\documentclass[12pt]{amsart}
\usepackage[a4paper, margin = 1in]{geometry}

\usepackage{xcolor}
\usepackage{soul}
\usepackage{xr-hyper}
\usepackage{graphicx, caption}
\usepackage{multicol}
\usepackage{hyperref}
\hypersetup{
     colorlinks=true,
     linkcolor=blue,
     filecolor=blue,
     citecolor = blue,      
     urlcolor=blue,
     }
\usepackage[
    natbib=true,
    style=numeric-comp,
    sorting=none,
    backend = bibtex
]{biblatex}
\newcommand{\pd}{\text{\sc pd}} % for persistence diagram
\newcommand{\wpd}{\text{\sc wpd}}
\newcommand{\stimulus}{\text{\small stim}}
\newcommand{\V}{\text{\small simp}}
\newcommand{\ori}{\text{\small ori}}
\newcommand{\dir}{\text{\small dir}}
\definecolor{forest}{RGB}{34, 139, 34}

\definecolor{WildStrawberry}{RGB}{242, 105, 142}

\begin{document}

\title{Tracking the topology of neural manifolds across populations}

\author[Yoon, Henselman-Petrusek, Yu, Ghrist, Smith, Giusti]{Iris H.R. Yoon \textsuperscript{a,1}, Gregory Henselman-Petrusek\textsuperscript{b}, Yiyi yu\textsuperscript{c},\\ Robert Ghrist\textsuperscript{d}, Spencer LaVere Smith\textsuperscript{c}, Chad Giusti\textsuperscript{e, 1}}

% \author[a,1]{Iris H.R. Yoon}
% \author[b]{Gregory Henselman-Petrusek}
% \author[c]{Yiyi Yu}
% \author[d]{Robert Ghrist}
% \author[c]{Spencer LaVere Smith}
% \author[e,2]{Chad Giusti}

% \curraddr[a]{Department of Mathematics and Computer Science, Wesleyan University, 265 Church Street
% Middletown, CT 06459}
% \affil[b]{Pacific Northwest National Laboratory, 902 Battelle Boulevard, Richland, WA 99352}

% \affil[c]{Department of Electrical and Computer Engineering, University of California Santa Barbara, Santa Barbara, CA 93106}

% \affil[d]{Department of Mathematics and Electrical \& Systems Engineering, University of Pennsylvania, Philadelphia, PA 19104}
% \affil[e]{Department of Mathematics, Oregon State University, 2000 SW Campus Way, Corvallis, Oregon 97331}

 \maketitle

{\small
\noindent
\textsuperscript{A}Department of Mathematics and Computer Science, Wesleyan University, 265 Church Street
Middletown, CT 06459. \\
\textsuperscript{B}Pacific Northwest National Laboratory, 902 Battelle Boulevard, Richland, WA 99352. \\
\textsuperscript{C}Department of Electrical and Computer Engineering, University of California Santa Barbara, Santa Barbara, CA 93106. \\
\textsuperscript{D}Department of Mathematics and Electrical \& Systems Engineering, University of Pennsylvania, Philadelphia, PA 19104. \\
\textsuperscript{E} Department of Mathematics, Oregon State University, 2000 SW Campus Way, Corvallis, Oregon 97331.\\
\textsuperscript{1} Corresponding authors. E-mail: hyoon@wesleyan.edu or chad.giusti@oregonstate.edu \\

%\dates{This manuscript was compiled on \today}
%\noindent This article is published in \url{https://www.pnas.org/doi/epub/10.1073/pnas.2407997121}

    \begin{abstract}
    Neural manifolds summarize the intrinsic structure of the information encoded by a population of neurons. Advances in experimental techniques have made simultaneous recordings from multiple brain regions increasingly commonplace, raising the possibility of studying how these manifolds relate across populations. However, when the manifolds are nonlinear and possibly code for multiple unknown variables, it is challenging to extract robust and falsifiable information about their relationships. We introduce a framework, called the method of analogous cycles, for matching topological features of neural manifolds using only observed dissimilarity matrices within and between neural populations. We demonstrate via analysis of simulations and \emph{in vivo} experimental data that this method can be used to correctly identify multiple shared circular coordinate systems across both stimuli and inferred neural manifolds. Conversely, the method rejects matching features that are not intrinsic to one of the systems. Further, as this method is deterministic and does not rely on dimensionality reduction or optimization methods, it is amenable to direct mathematical investigation and interpretation in terms of the underlying neural activity. We thus propose the method of analogous cycles as a suitable foundation for a theory of cross-population analysis via neural manifolds.
    \end{abstract}

Among the most successful models for coding in populations of biological or artificial neurons is the \emph{neural manifold} model. In this model, individual neurons correspond to \emph{receptive fields}, spatially localized regions in some metric space or manifold (Fig. \ref{fig:figure1}A). The aggregate input to and state of the population corresponds to a point in the neural manifold, and a neuron is active precisely when that point is within its corresponding receptive field; a common model for the firing rate of each neuron is the tuning curve, given by a bump function supported within the receptive field \cite{Hubel1959ReceptiveFO, okeefe_geometric_1996, niell_highly_2008}. 
Historically, neural manifolds were discovered by a direct comparison of the activity of individual neurons to known stimuli or behaviors which carried known geometric structure \cite{Hubel1959ReceptiveFO,OrientationBenYishai, OKeefe1971TheHA,Hafting2005MicrostructureOA}. Recent efforts have developed tools for detecting and studying the structure of neural manifolds intrinsically, without reference to external correlates \cite{rubinRevealingNeuralCorrelates2019, GridCellTorus, rybakkenDecodingNeuralData2019}. As we discover and characterize more intricate neural manifolds, the question now becomes understanding what they represent and how they are related to one another.
     
\smallskip

When a neural manifold is linear (Fig.~\ref{fig:figure1}A, left), there are many robust methods available for extracting its structure from observations of the system \cite{villetteInternallyRecurringHippocampal2015, rubinRevealingNeuralCorrelates2019}, and basic correlation analysis can be applied to compare such spaces. However, non-linear neural manifolds are common \cite{deChaudhuriNonlinear, chaudhuriIntrinsicAttractorManifold2019, GridCellTorus, zhangHippocampalSpatialRepresentations2023, zhouHyperbolicGeometryOlfactory2018}, and the most familiar among these are circular coordinates (Fig.~\ref{fig:figure1}A, right). For example, simple cells in primary visual cortex and head direction cells in the hippocampus encode a notion of angle using a circular coordinate system, while the quasi-periodic activity of pacemaker circuits encodes a circular coordinate  temporally. Machine learning approaches have been employed to detect these structures \emph{in vivo} and to compare them across populations \cite{degenhartStabilizationBraincomputerInterface2020, ganjaliUnsupervisedNeuralManifold2024, gallegoCorticalPopulationActivity2018, gallegoLongtermStabilityCortical2020, pandarinathInferringSingletrialNeural2018, pandarinathLatentFactorsDynamics2018}.
However, such approaches often require a good initial dimensionality estimate and reduction, and they aim to align local geometry \cite{degenhartStabilizationBraincomputerInterface2020, ganjaliUnsupervisedNeuralManifold2024, gallegoCorticalPopulationActivity2018, gallegoLongtermStabilityCortical2020}. For systems such as grid and conjunctive cells in entorhinal cortex, where different modules have incompatible geometries, such methods can be difficult to apply \cite{Hafting2005MicrostructureOA, stensolaEntorhinalGridMap2012}.
 The authors posit that the development of a mathematically well-founded theory of neural computation in and among neural manifolds requires a bottom-up approach, in which quantitative tools for detecting and interacting with these systems derive from the structure of neural manifolds themselves. Recent work has demonstrated that tools from applied algebraic topology \cite{Ghrist_barcodes, Carlsson2009TopologyAD, PH_survey} provide an effective mathematical and computational framework for studying the intrinsic shape of neural population activity (Fig.~\ref{fig:figure1}B) \cite{GridCellTorus, PlaceCellsCurtoItskov, DabaghianSpatialMap, CliqueTop}.

\begin{figure}[h!]%[\sidecaptionrelwidth][t]
\centering
\includegraphics[width=\textwidth]{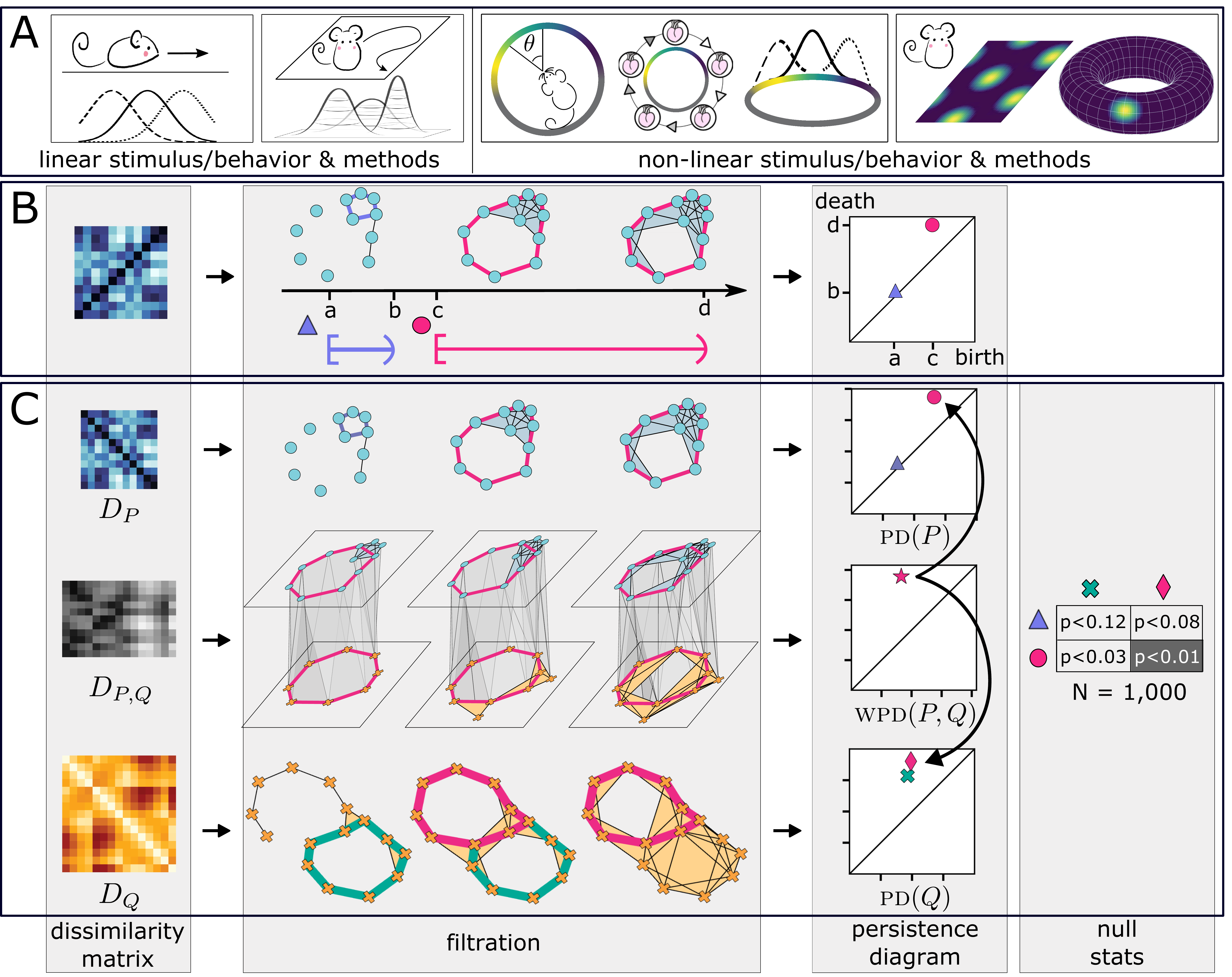}
\captionsetup{width=1\textwidth}
\caption{Topological comparison of neural manifolds across populations. \textbf{A.} Tuning curves assemble to represent linear (left) and non-linear(right) neural manifolds. \textbf{B.} Detection of circular features from a dissimilarity matrix via persistent homology. Given a dissimilarity matrix (left), a filtration of simplicial complexes (middle) encodes  the system at varying thresholds. The birth and death thresholds for topological features (middle) are summarized in a persistence diagram (right). \textbf{C.} The method of analogous cycles matches features. Given systems $P, Q$, with dissimilarity matrices $D_P, D_Q$, compute corresponding persistence diagrams $\pd(P), \pd(Q)$ (top, bottom). For a cross-system dissimilarity matrix $D_{P,Q}$ (middle row, left), create a witness filtration (middle row, center), with features summarized in the witness persistence diagram (middle row, right). The point (star) indicates one shared feature between the two systems. The analogous cycles method matches any representations of this feature in $\pd(P)$ (pink circle) and $\pd(Q)$ (pink diamond) consistent with $D_{P,Q}.$ 
The null model matching matrix (far right) gives the probability that the method returns a match between each pair of points in $\pd(P)$ and $\pd(Q)$ under the geometric null model (\emph{Methods and Materials}). Shading at the $(i,j)$ entry indicates that the corresponding pair is matched through the computed $\wpd(P,Q).$ 
}
\label{fig:figure1}
\end{figure}

Topological methods describe the structure of a neural manifold in terms of the intersections of receptive fields, providing a coarse summary of the shape of the space (SI Section 1A2). Such an encoding is natural in the context of population activity, where pairwise similarity of spike trains provides a proxy for overlap of receptive fields (Fig.~\ref{fig:figure1}B, SI Fig.~S1). This discards fine information about geometry in favor of robust representation of mesoscale, non-local information like connectivity. Finding precisely two  distinct paths from any point to any other in the neural manifold, for example, indicates the presence of a circular coordinate (Fig. \ref{fig:figure1}B, SI Fig.~S1). Encoding this structure as linear algebra, we obtain a tool, \emph{persistent homology} \cite{Ghrist_barcodes, PH_survey,Carlsson2009TopologyAD,ELZ_persistence}, for studying shape  that detects and characterizes non-linear \emph{cycles} in neural manifolds using only observations of population activity \cite{chaudhuriIntrinsicAttractorManifold2019, rybakkenDecodingNeuralData2019, singhTopologicalAnalysisPopulation2008, CliqueTop, PlaceCellsCurtoItskov, GridCellTorus}. These are represented as a \emph{persistence diagram}, denoted $\pd$, which summarizes the multi-scale structure of the cycles and their relative significance\footnote{While persistent homology can be used to study geometry of any dimension, in this paper we focus on the case of $1$-dimensional persistence diagrams, which reflect circular structure.} (Fig. \ref{fig:figure1}B, right).

Once we have detected a non-linear neural manifold, we are  left with the problem of assigning semantics. Without knowledge of the function of the population, how can we determine which, if any, external variables -- stimuli, behavior, or structured activity of other neurons -- are faithfully encoded by this structure? Our goal is not to assign semantics to individual neurons, but rather to use observations of population activity to determine which, if any, variables the population encodes. As the structure may be inherently high-dimensional, we cannot \emph{a priori} isolate  features to study. Thus, the methods must work simultaneously with all significant structure in both neural manifolds. 

To address this need, in \cite{analogous_bars} the authors introduced the mathematics underlying the \emph{method of analogous cycles} (Fig.~\ref{fig:figure1}C, SI Section 1A6). Given the persistent homology of two related systems and a measure of cross-dissimilarity between the two systems, the method of analogous cycles encodes how families of receptive fields in each system \emph{witness} those in the other (Fig. \ref{fig:figure1}C; SI Section 1A4). Leveraging this encoding, it enumerates all possible relations between their topological features consistent with the cross-dissimilarity measure. This method thus describes how representations of complex information in neural population activity evolve across brain regions and allows us to assign semantics from observations of stimuli and behaviors. These results are amenable to theoretical mathematical analysis using existing theory from topology and geometry. 

In this paper, we apply the method of analogous cycles to the study of coding in neural systems. We validate the tool in simulation studies against ground truth, establish statistical tests for significance of feature identification, and demonstrate how the method can be applied in both simulated and experimentally observed neural systems.

%-----------------------------------------------------
\section*{Results}
%-----------------------------------------------------

\subsection*{Assigning semantics to topology of neural manifolds in simulated visual systems}
Our fundamental question is whether the mathematical framework in \cite{analogous_bars}  can identify shared circular features in neural systems. To test this, we first consider a sequence of simulated neural populations for which we have access to ground truth. We investigate how topological structure in feed forward networks designed to represent or discard features of presented stimuli is matched across layers. We consider both networks that encode individual features of stimuli and those which synthesize inputs to describe more sophisticated structure.

\begin{figure}[h!]
\centering
\includegraphics[width=14cm]{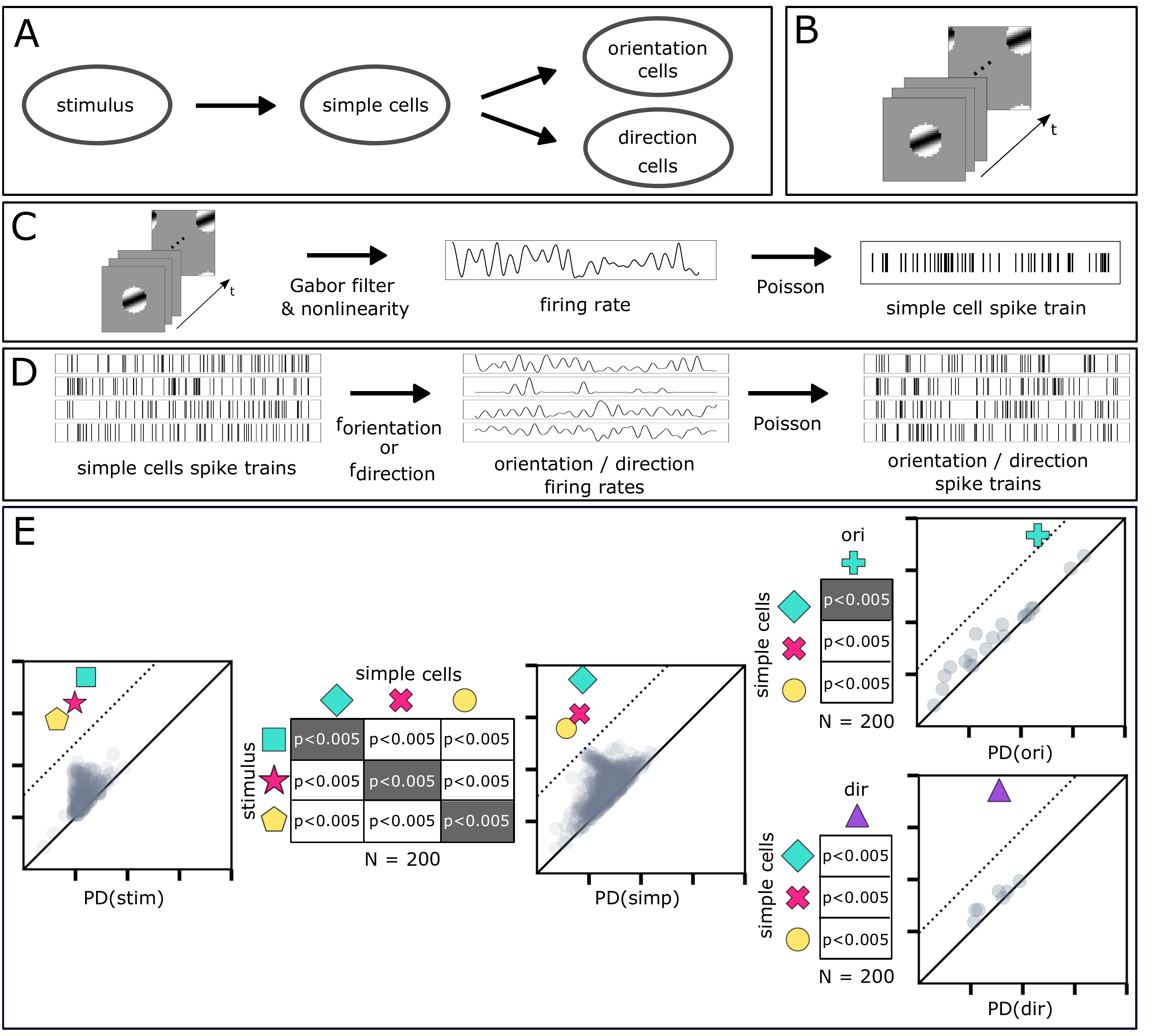}
\captionsetup{width=1\textwidth}
\caption{Analogous cycles correctly match features across a simulated visual system. \textbf{A.} Architecture of the simulation. Stimulus videos are presented to a population of simulated spiking simple cells. Spikes from these cells are presented to populations tuned to orientation and direction features of the stimulus. \textbf{B.} Stimuli are 40,000-frame videos of a circular grating with fixed orientation, with the center moving in a fixed direction with fixed speed, looping to the opposite edge at boundaries. Orientations, starting locations, and movement directions are sampled uniformly. \textbf{C.} Simulation of spiking simple cells. Firing rate is computed as the rectified dot product between a modified Gabor filter and frames of the stimulus video. Spikes are sampled from the resulting inhomogeneous Poisson process. 
\textbf{D.} Firing rates of orientation sensitive cells are given by $f_{\text{orientation}},$ which is described by a feed-forward neural network trained to approximate a unimodal tuning curve on orientations using the simple cell spike trains as input (see SI Section 2A3.) Output is obtained by sampling from the resulting firing rates using an inhomogeneous Poisson process. Direction cells are simulated similarly. \textbf{E. } Analogous cycles match cycles across  systems. (left) All three significant circular features of the stimulus are matched to those of the simple cells. Matches are indicated by color. All matches are statistically significant. (top right) One circular feature (teal diamond) in the simple cell neural manifold is matched to the unique significant feature (teal cross) of orientation sensitive population, implying a projection of encoded information from the simple cells. (bottom right) No matches are found between the simple cells and direction cells, indicating that the direction cells encode a novel feature.}
\label{fig:figure2}
\end{figure}

\subsubsection*{Identifying circular features in neural activity induced by structured stimuli}
%Question: Can our method identify circular features when multiple features are present in the systems we are studying? To answer this question, we developed a modified version of a standard grating stimulus and a computational model of simple cells in V1. 

Can the analogous cycles method identify the features of a stimulus that are encoded by a population of neurons? To address this question, we developed a modified version of a standard grating stimulus with interesting topology and a computational model of simple cells in V1 (Fig. \ref{fig:figure2}A1). Our stimuli are videos of a masked grating on a square torus\footnote{A square region becomes a square torus by identifying the left and right edges and by identifying the top and bottom edges, as in the video game \emph{Asteroids}.}, (Fig. \ref{fig:figure2}B). The location of the grating and the orientation of the grating vary continuously (SI Section~2A1), creating three intrinsic circular coordinates. We simulated simple cells via samples from an inhomogeneous Poisson process with rate given by a rectified dot product between a modified Gabor filter and the stimulus video  (Fig. \ref{fig:figure2}C, SI Section~2A2). This population of simple cells, by design, encodes all of the significant topological features of the stimulus videos, providing a collection of expected ground-truth matches among encoded features against which we can compare the output of the analogous cycles method.

% PH on each system
For the video stimuli, we randomly sampled 400 out of 40,000 stimulus images, and we computed a pairwise dissimilarity matrix $D_{\text{stim}}$ via the $L_2$ distance on the orientation and location of the circular mask (see SI Section~1B3). For the simple cells, we computed a dissimilarity matrix $D_{\text{simp}}$ via a windowed cross-correlation dissimilarity (\textit{Materials and Methods}). From these two matrices, we computed persistence diagrams $\pd(\stimulus)$ and $\pd(\V)$ (Fig. \ref{fig:figure2}E left, center, without colors). The persistence diagrams indicated that both the  stimuli and the neural manifold for the simulated simple cell population have multiple circular features. In both cases, three of the features are significant, quantified using a quantile test on distance from the birth equals death line (\emph{Materials and Methods}). However, simply observing that both populations carry circular features does not provide us with a method for comparing them. Are the circular structures observed in the activity of the simulated simple cells driven by the stimulus? If so, which feature of the stimulus is encoded by which circular feature of the simple cells? 

The method of analogous cycles requires a measure of dissimilarity between the stimuli and the activity of each simple cell. To each video, we associated a binary vector with one entry per frame of the video, with value $1$ at frame $t$ if the presented stimulus video is displaying an image within a threshold distance in the space of images (SI Section~1B4). To compute the dissimilarity matrix $D_{\text{stim, simp}},$ for each image and simulated cell, we applied the windowed cross-correlation dissimilarity (\textit{Materials and Methods}) to the corresponding binary vector and spike train. The method of analogous cycles computed with $D_{\text{stim}}, \, D_{\text{simp}}$, and $D_{\text{stim, simp}}$ as inputs matched three pairs of points between $\pd(\stimulus)$ and $\pd(\V)$ (Fig. \ref{fig:figure2}E left, center, with colors). To validate, we compared the results to a geometric null model (\textit{Materials and Methods}, Fig. \ref{fig:figure2}E left matrix) and computed the likelihood of observing the teal, pink, and yellow pairs ($p<0.005$, $N = 200$ trials). 

We can verify the results of our computations by visualizing the matches. Visualizing the cycles detected in $\pd(\stimulus)$ indicates that the points represented by the teal square, yellow pentagon, and pink star respectively represent orientation, $x$-coordinates, and $y$-coordinates in the stimulus video (SI Fig.~S25). Comparing the matched cycles of $\pd(\V)$ against the geometry induced by the stimuli provides verification of their semantics (SI Fig.~S26).

\subsubsection*{Tracking and falsifying feature propagation across populations}

The previous experiment validates the method of analogous cycles for populations which faithfully encode the topology of the input space. However, neural computation across populations involves selection and synthesis of features. In real neural systems, we therefore expect that only a subset of features would propagate between populations, while new features that are not intrinsic to neural manifolds of upstream populations would be generated. This led us to ask whether this method correctly identifies which, if any, topological features are shared with other populations and which are novel.

To address this question, we extended our simulated visual system to include two additional simulated neural populations, each receiving input from the population of simulated simple cells (Fig.~\ref{fig:figure2}A). The  neurons in these populations were trained to carry unimodal tuning curves for grating orientation and motion direction in the stimulus movies, respectively (Fig.~\ref{fig:figure2}D, SI Sections 2A3 and 2A4). In both cases, the coding properties of the population are well-described by circular neural manifolds. The orientation-sensitive population selects a single feature from those represented by the upstream simple cell population, while the motion sensitive population synthesizes position information across a short time window, and thus does not reflect any existing feature described by the simple cell neural manifold.
Persistence diagrams computed from the windowed cross-correlation dissimilarity matrices $D_{\text{ori}}$ and $D_{\text{dir}}$ for both populations consisted of a single point (Fig. \ref{fig:figure2}E), confirming that the simulated population activity was well-constrained to the expected neural manifolds.

We applied the method of analogous cycles to compare the neural manifold for the simulated simple cell population and both of these new circular neural manifolds. The cross-system dissimilarity matrices $D_{\text{simp, ori}}$, $D_{\text{simp, dir}}$ were computed by the windowed cross-correlation dissimilarity of spike trains (\textit{Materials and Methods}). The method identified one pair of analogous points between $\pd(\V)$ and $\pd(\ori)$ (Fig. \ref{fig:figure2}E, top row, teal), correctly identifying the teal diamond point in $\pd(\V)$ as the feature of the simple cell neural manifold that corresponds to the circular feature in neural manifold of the new orientation-selective population. Comparison with the geometric null model (Fig. \ref{fig:figure2}E, top right) indicates that this match is statistically significant ($p < 0.005, N = 200$ trials). On the other hand, the method of analogous cycles correctly identified no matches between $\pd(\V)$ and $\pd(\dir)$ (Fig. \ref{fig:figure2}E, bottom right), indicating that the circular feature encoded by the direction-sensitive population must either be synthesized from upstream information but not intrinsically described by the earlier neural manifold, or projected from some unobserved population. 

\subsection*{Disentangling conjunctive cell coding in simulated entorhinal cortex}
While the method of analogous cycles was effective in feature identification and selection in the simulated visual system, it is possible that there is some aspect of those simulations that was particularly well-suited to topological analysis. As the method is intended to be a general tool, we next asked whether the choice of the model or the architecture of the simulation impacts the performance of the method. In order to address this concern, we applied the method to simulated populations of entorhinal cortex neurons (Fig.~\ref{fig:figure3}A). Using experimental recordings of location and head directions from rats engaging in foraging behavior taken from \cite{GridCellTorus}, we simulated populations of grid cells, head-direction cells, and conjunctive cells. Grid cells, located in the dorsocaudal medial entorhinal cortex, are activated when an animal's position coincides with a regular grid that spans the environment \cite{Hafting2005MicrostructureOA}. We simulated  grid cell firing rates using the continuous attractor network model, as in \cite{CoueySimulation}. Head direction (HD) cells, which fire when an animal's head is aligned with a specific direction, were simulated using selected tuning curves (SI Section~2B). Conjunctive grid cells, a population of grid cells whose firing is modulated by a preferred head direction, can be found in the deep layers of medial entorhinal cortex\cite{GerleiGrid, MoserConjunctive}. We computed the firing rate of each such conjunctive cell as the minimum firing rate of a model grid cell and a model HD cell to a given stimulus (Fig. \ref{fig:figure4}B).

\begin{figure}[h!]
\includegraphics[width=11cm]{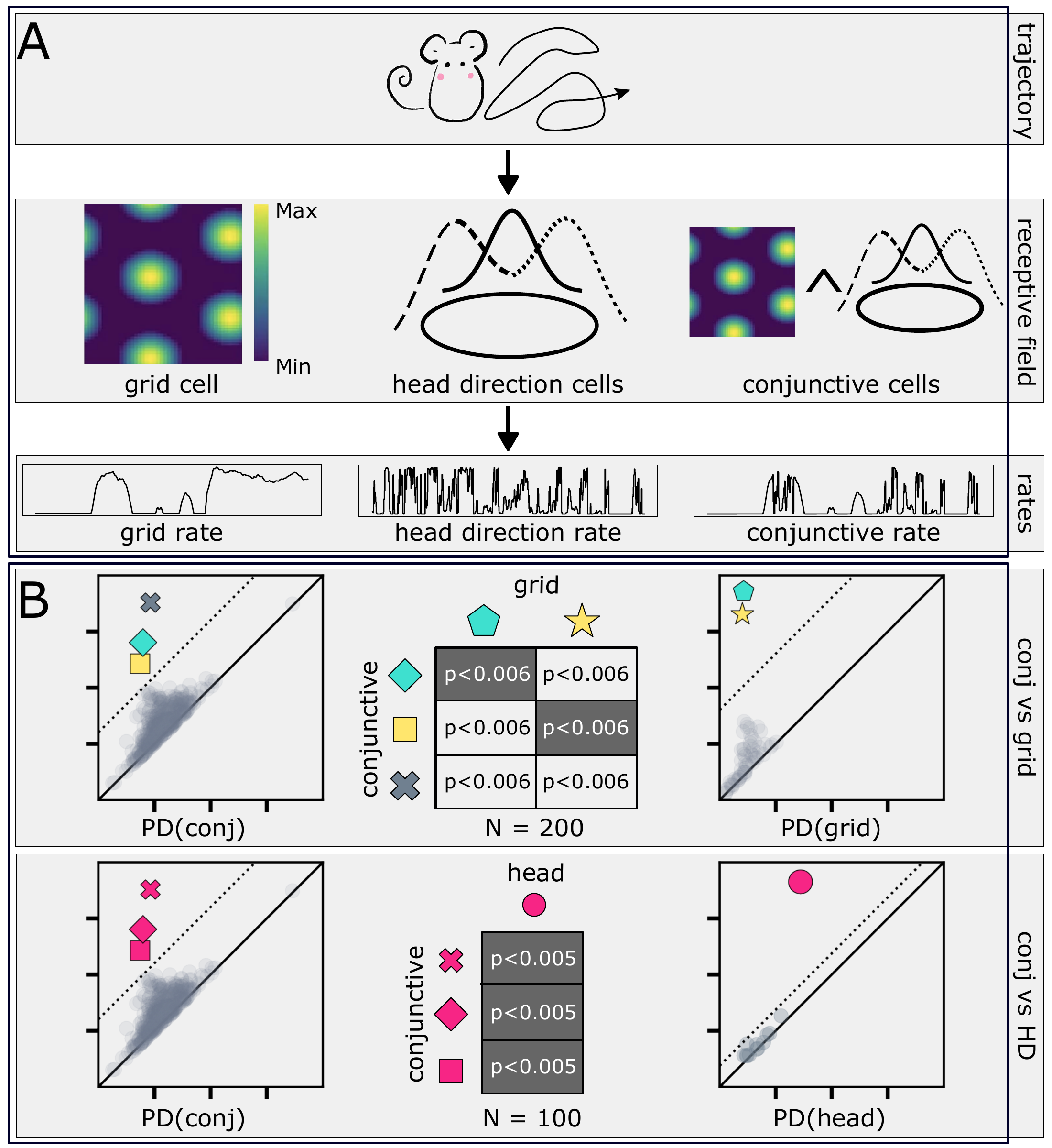}
\captionsetup{width=1\textwidth}
\caption{Analogous cycles assign semantics to simulated navigational system. \textbf{A.} Firing rate simulation. Given a rat trajectory, the grid cells are simulated using a continuous attractor networks model. The head-direction(HD) cells are simulated by tuning curves. The conjunctive cells firing rates are computed by the minimum firing rates of the grid cells and HD cells. \textbf{B.} (Top) Analogous cycles between the grid and conjunctive cells indicated by the colors. The teal and yellow points in the conjunctive PD encode the torus arising from the grid cell organization. (Bottom) Analogous cycles between the HD and conjunctive cells. The single significant feature in $\pd(\text{head})$ is analogous to a combination of the significant features in $\pd(\text{conj})$. Since the diamond and square points of the conjunctive PD are analogous to the two points in the grid PD, we conclude that the cross point of the conjunctive PD must encode the cyclicity of HD cells. }
\label{fig:figure3}
\end{figure}

It has recently been experimentally verified that grid cells carry a toroidal neural manifold \cite{GridCellTorus, GridTwisted}, well-parameterized by two independent circular coordinates, while HD cells have a circular neural manifold \cite{HD_topography}. We thus expect conjunctive cells to have a neural manifold parameterized by three independent circular features. We computed  internal dissimilarity matrices $D_{\text{grid}}$, $D_{\text{HD}}$, $D_{\text{conj}}$, and cross-system dissimilarity matrices $D_{\text{grid, conj}}$, $D_{\text{HD, conj}}$ using the windowed cross-correlation dissimilarity (\textit{Materials and Methods}), and obtained persistence diagrams $\pd(\text{grid}),$ $\pd(\text{head}),$ and $\pd(\text{conj})$ that  verify our expected count of circular features in the three simulated populations (Fig.~\ref{fig:figure3}B). %We applied the analogous cycles method between the grid cells and the conjunctive cells, and we also applied the method between the grid cells and the HD cells. 

We applied the analogous cycles method to the populations of grid and conjunctive cells to test whether the method correctly matches the two circular features of the conjunctive cells that arise from the grid cells. The method identifies two pairs of analogous cycles between the grid and conjunctive persistence diagrams, indicated by the pairs of teal and yellow points in Figure \ref{fig:figure3}B. Comparison with the geometric null model (Fig.~\ref{fig:figure3}B) indicates that a randomly arising match between either the teal points or the yellow points are both highly unlikely ($p< 0.006$, $N=200$ trials). We thus conclude that these two features of the conjunctive cell stimulus space represent the toroidal coordinates described by the constituent grid cells.

We then applied the analogous cycles method between the conjunctive and the HD cells. As illustrated in Figure \ref{fig:figure3}B, the method matches the significant feature (pink circle) in $\pd(\text{head})$ to the three significant features in the $\pd(\text{conj})$. Such a many-to-one match involves a linear combination of the corresponding cycles, indicating that there is some interaction among the three circular coordinates in the chosen cycle basis for the conjunctive cells (SI Fig.~S27, S28). Here, we conclude that the chosen cycle basis for $\pd(\text{conj})$ correlated the head direction and direction of motion. One can verify this deduction by constructing a change of basis for the cycles in $\pd(\text{conj})$ which isolates the match to the cycle which is unmatched to points in $\pd(\text{grid})$. As before, comparing with the geometric null model (Fig. \ref{fig:figure3}B) indicates that these matches are statistically significant ($p< 0.005$, $N = 200$ trials). 
%Analogous cycles between conjunctive cells and HD cells tells us that the orientation-cyclicity of the HD cells are encoded in the conjunctive cells. \IY{Need to add interpretation of Figure 3 being analogous to three points}
%The point is, we know how grid cells, HD cells, and conjunctive cells are related, and our experiment verifies the relations.

\subsection*{Matching \text{in vivo} neural manifolds for primary visual cortex and anterolateral region}
Finally, we asked whether the analogous cycles 
method is robust enough to identify shared features of neural manifolds in experimental data. To address this question, we considered \emph{in vivo} recordings from two mouse neural systems that are known to share coding properties: the primary visual cortex (V1) and the anterolateral region (AL) (Fig.~\ref{fig:figure4}A) \cite{yu2022selective, andermann2011functional, wang2007area}. 

We extracted single cell spike trains from two-photon calcium imaging traces in both regions from a head-fixed mouse (Fig.~\ref{fig:figure4}A). Recordings were made while presenting black-white full contrast drifting gratings. Four different orientations were presented twice with opposite drifting directions for twenty trials (Fig.~\ref{fig:figure4}B, \textit{Materials and Methods}). We aggregated the spike trains across the trials (Fig.~\ref{fig:figure4}B) and performed preprocessing steps to remove neurons that fire uniformly and unreliably (SI Section 2C1). All internal and cross-system dissimilarity matrices were computed using the windowed cross-correlation dissimilarity (\textit{Materials and Methods}). We computed persistence diagrams $\pd(\text{V1})$ and $\pd(\text{AL})$ for the two regions, both of which contained only two points (Fig.~\ref{fig:figure4}C). For such small persistence diagrams, our usual method of selecting a significance threshold for points in a persistence diagrams cannot be applied. In this case, we use an alternative significance threshold given by the largest lifetime from a sample of persistence diagrams computed from synthetic spike trains that match the observed population statistics (\emph{Materials and Methods}).

We applied the analogous cycles method to these two diagrams, and identified a matched pair of points as shown in Figure \ref{fig:figure4}C. Comparison to the geometric null model (Figure \ref{fig:figure4}C) indicates that the identified analogous pair is statistically significant ($p < 0.005$, $N=200$ trials). The fact that V1 and AL neurons encode a matched feature is consistent with prior observations that neurons in V1 and AL share coding properties for orientation of visual stimuli \cite{Marshel2011FunctionalSO, CorticoProjection}. 
Visualizing the spike trains for neurons in V1 and AL that constitute the matched cycles shows these subpopulations of neurons in V1 and AL have similar spiking profiles, without reference to stimuli (SI Fig.~S29).

\begin{figure}[h!]
\centering
\includegraphics[width=9.5cm]{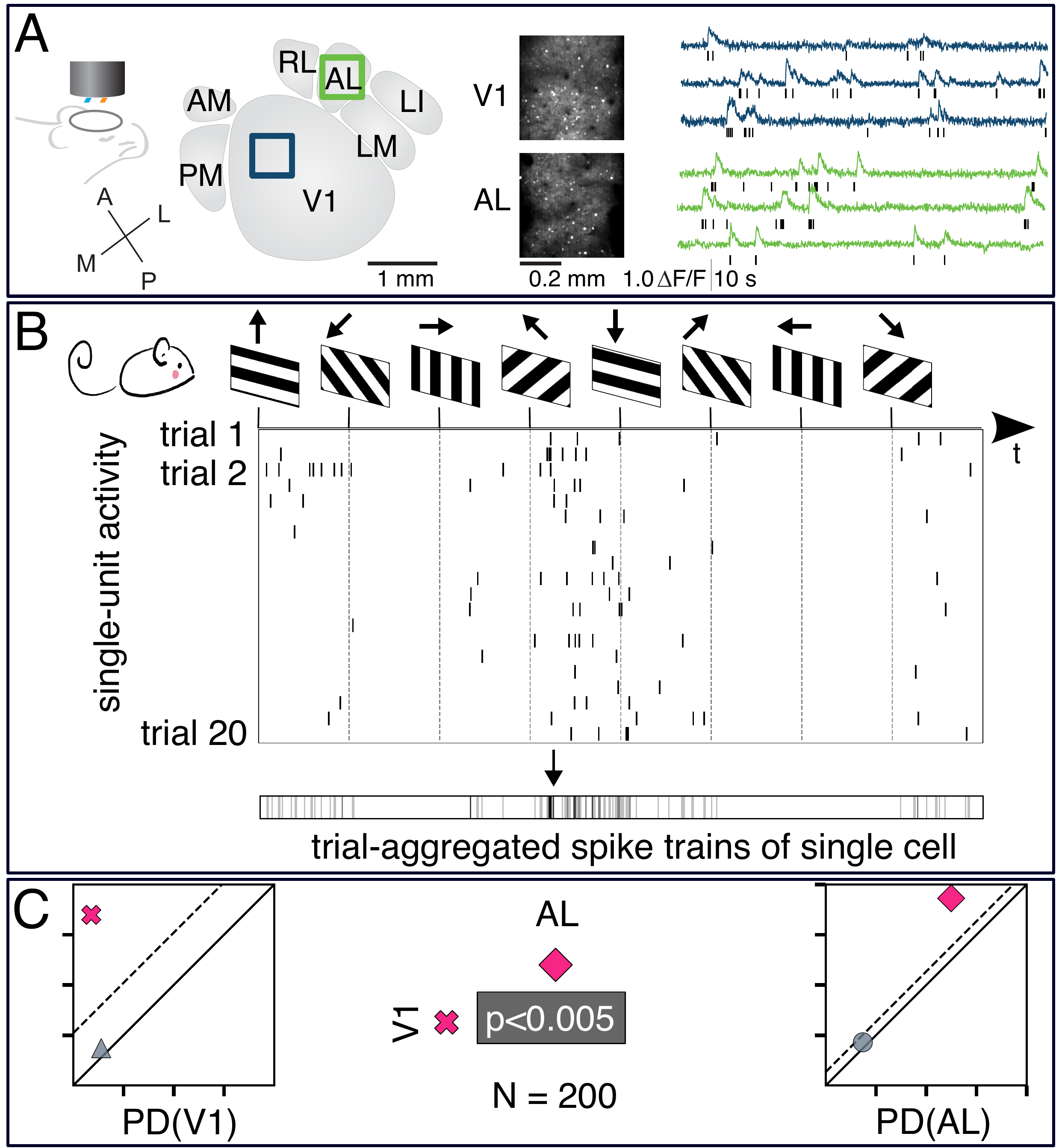}
\captionsetup{width=1\textwidth}
\caption{Analogous cycles for neural coding propagation on experimental data. \textbf{A.} Schematic (top) of dual region two-photon calcium imaging in primary (V1) and anterolateral (AL) visual areas (left). Spike trains were inferred from the calcium sensor dynamics (right). \ \textbf{B.} The mouse was presented with a video consisting of drifting gratings of four orientations. Each stimulus was repeated 20 times. For each cell, the trial-aggregated spike train was computed. \textbf{C.} The analogous cycles method identifies a pair of points in the V1 and AL persistence diagram that encodes the same circular information. }
\label{fig:figure4}
\end{figure}

%-----------------------------------------------------
\section*{Discussion}
%-----------------------------------------------------
% 4 paragraphs. %1st paragraph: recall what we did in broad terms and state its significance; middle paragraphs, call up details that were interesting or surprising and briefly address questions or suggest future directions, compare to other work/methods; last paragraph, indicate how the method can be used and what we expect it will provide to practitioners. very briefly discuss possible directions

% 1st paragraph: Summary of what we did (in broad terms and state significance)
In this study, we introduced and validated the method of analogous cycles, a method for comparing topological structure in neural manifolds across populations. We demonstrated that the method reliably matches related circular coordinates between neural manifolds and rejects matching cycles that are not related. The method utilizes only the data of a cross-dissimilarity matrix to deterministically identify matches. Furthermore, the method works in the context of multiple simultaneously represented features, providing tools for comparing complex neural representations without descending to the level of activity of individual neurons. While we did not investigate such data in this study, these methods can be applied to other high-dimensional neural manifolds with non-trivial topology, such as spheres. Together, these features provide a robust tool set for assigning semantics to nonlinear structure within neural manifolds, and understanding how structures interact across populations.

In contrast to some prior studies that compare neural coding across populations, the method does not require any \emph{a priori} knowledge of encoded variables or the use of optimization methods \cite{dabagiaAligningLatentRepresentations2023, degenhartStabilizationBraincomputerInterface2020, ganjaliUnsupervisedNeuralManifold2024, gallegoCorticalPopulationActivity2018}. While some existing studies on information transmission across neural populations focus on anatomical connections \cite{tracingcentury, orientedAxon, CorticoProjection} and spike train propagation \cite{SignalPropLogic, stablePropagation, corticalPropagation, FeedbackPropagation}, this method provides mathematically robust tools for studying and falsifying the propagation of non-linear structure in codes at the population-level. Unsupervised approaches such as manifold learning have been successful in comparing neural manifolds across populations, but such analyses are hindered when alignment of local geometry is difficult or such an alignment does not exist \cite{degenhartStabilizationBraincomputerInterface2020, ganjaliUnsupervisedNeuralManifold2024, gallegoCorticalPopulationActivity2018, gallegoLongtermStabilityCortical2020, Hafting2005MicrostructureOA, stensolaEntorhinalGridMap2012}. Prior studies using topological methods \cite{chaudhuriIntrinsicAttractorManifold2019, rybakkenDecodingNeuralData2019, singhTopologicalAnalysisPopulation2008, CliqueTop, PlaceCellsCurtoItskov, GridCellTorus} have been effective at describing structure of neural manifolds, but these did not provide tools for studying multi-system structure.

Looking forward, many existing methods perform dimensionality reduction on population vectors to mitigate computations and the effects of noise. Our method deals with dissimilarity between neural units, providing interpretability but leaving in place these concerns. Extending the method to work with  dimensionality-reduced population vectors, as used  in \cite{GridCellTorus}, would allow us to integrate strengths of existing methods with analogous cycles. For example, many studies identify network states in aggregate population vectors, assign meanings to each state using external stimuli/behavior, and then interpret neural systems by via network state. A suitably augmented analogous cycles method could identify families of network states more intricate than clusters. Furthermore, while the method presented in this paper identifies geometrically independent features across neural manifolds, in some cases where those features have complex interactions, the algorithmically selected match may not align with intuition, and applying significance thresholding may return false negatives for matching. We give an example of such a pair of neural manifolds in SI Section 3E and describe an \emph{ad hoc} method for correcting such misalignments. Developing formally justified methods for ameliorating these issues will require further research. 

In summary, the analogous cycles method allows us to study how nonlinear coordinate systems encoded by distinct populations of neurons are related, without estimating tuning curves or relying on known stimuli or behavior. Complementing recent developments in multi-region imaging techniques \cite{junFullyIntegratedSilicon2017, kim_integration_2017}, the method provides new avenues for the study of information flow and computation across brain regions and for comparison of coding properties between homologous regions in distinct organisms.

% \subsection*{Supporting Information Appendix (SI)}

 % \subsubsection*{SI Datasets} 

% Supply .xlsx, .csv, .txt, .rtf, or .pdf files. This file type will be published in raw format and will not be edited or composed.

% \subsubsection*{SI Movies}
% The stimulus videos used in the simulated visual system and the mouse experiment are attached as SI movies. 

\section*{Materials and Methods}
%Please describe your materials and methods here. This can be more than one paragraph, and may contain subsections and equations as required. 

% reference: https://digitalcommons.unl.edu/cgi/viewcontent.cgi?article=1083&context=mathfacpub

\subsection*{Dissimilarity between spike trains and firing rates}
Let $\vec{x}$ and $\vec{y}$ be vectors of the same dimensionality that each represent either a spike train or a firing rate. 
% The cross-correlation between $\vec{x}$ and $\vec{y}$ for a displacement value of $n$ is 
% \begin{equation*}
% (\vec{x} * \vec{y})_n = \sum_{m} \vec{x}_m \vec{y}_{m+n}.
% \end{equation*}
We define the similarity between $\vec{x}$ and $\vec{y}$ to be the normalized sum of shifted convolutions, for a limited range of displacement values specified by the shift parameter $\ell: $  
\begin{equation*}
Sim_\ell(\vec{x}, \vec{y}) = \frac{\sum_{n=-\ell}^\ell \sum_{m} \vec{x}_m \vec{y}_{m+n}.
}{ \| \vec{x} \|_2 \, \|\vec{y} \|_2 } .
\end{equation*}
Given a collection of time series $P = \{\vec{x}_i\}_{i=0}^{K}$, we compute  pairwise dissimilarity as \textit{windowed cross-correlation dissimilarity}
\begin{equation*}
Dis_\ell(\vec{x}, \vec{y}) = \begin{cases} 1 - \frac{Sim_\ell(\vec{x}, \vec{y})}{M} \quad \text{if } \vec{x} \neq \vec{y} \\
0 \quad \text{if } \vec{x} = \vec{y} \\
\end{cases}
\end{equation*}
where $M = \max_{\{\vec{s}, \vec{t} \in P\}} Sim_{\ell}(\vec{s}, \vec{t})$. See SI Section 1B2 for a thorough discussion on dissimilarity. A comparison of this dissimilarity measure for varying $\ell$ and a comparison to existing spike train dissimilarities can be found in SI Section 3C.

\subsection*{Persistent homology}
For a neural population or system $P$ with $N$ elements, we perform topological data analysis using an $N \times N$ pairwise dissimilarity matrix $D_P$. Let $\{\varepsilon_i\}_{i=1}^m$ be a collection of parameters. We construct a filtration of simplicial complexes 
\[X_{\varepsilon_1} \subseteq X_{\varepsilon_2} \subseteq \cdots \subseteq X_{\varepsilon_m}\]
where $X_{\varepsilon_k}$ consists of $N$ vertices and has $n$-simplex $[v_0, \dots, v_{n}]$ precisely when all pairwise dissimilarity among the listed elements is at most $\varepsilon_k$. Computing homology in dimension $1$ with $\mathbb{Z}_2$ coefficients, we obtain a sequence of vector spaces summarizing the circular features in each simplicial complex at each parameter. We track the birth and death of such circular features via a \textit{persistence diagram}, denoted $\pd(P)$, where a feature that is born at parameter $b$ and dies at $d$ is represented by a point in the plane with coordinates $(b, d)$. See SI Section 1A3 for a gentler exposition and SI Section 3D for experiments establishing robustness of persistence diagrams against various noise in simulations.

\subsection*{Significant points on persistence diagrams}
The points on a persistence diagram that are far from the diagonal line represent mesoscale features of the system, while the remaining points are often considered noise. To identify important features, we compute the empirical distribution of the lifetimes $d - b$ of all points, and threshold at $Q_3 + 3 (Q_3 - Q_1),$ where $Q_i$ is the $i$th quartile. We indicate this threshold value as a dotted line on the persistence diagram, taking any point  above the threshold to be significant. A comparison of methods for identifying significant points in persistence diagrams is given in SI Section 3A. % IY: I double-checked that we use k = 3 and not 1.5

If a persistence diagram contains insufficiently many points to apply the non-parametric statistical test to the distributions of lifetimes, we can apply an alternative test for significance developed in \cite{GridCellTorus}. We sample a population of shuffled spike trains by binning our data spike trains into 0.0751s bins and randomly permuting the entries of the resulting vector. We compute a persistence diagram for this population as for the data. Applying this process $N=10,000$ times, we obtain a distribution of lifetimes, and take our significance threshold to be the largest observed lifetime. In SI Section 3B we detail alternative tests for significance one can employ when dealing with small persistence diagrams.

\subsection*{Analogous cycles}
%One paragraph on Dowker / Witness
Let $P$, $Q$ be two systems. The inputs to the analogous cycles method are the internal dissimilarity matrices $D_P$, $D_Q$, and the cross-system dissimilarity matrix $D_{P,Q}$. Let $\pd(P)$ and $\pd(Q)$ denote the dimension-1 persistence diagrams of $D_P$ and $D_Q$. From $D_{P,Q}$, we construct a sequence of witness complexes 
\[W_{P, Q}^{\psi_1} \subseteq W_{P,Q}^{\psi_2} \subseteq \cdots \subseteq W_{P,Q}^{\psi_t}. \]
The witness complex $W_{P,Q}^{\psi_k}$ is a simplicial complex with vertices $P$ and an $n$-simplex $[p_0, \dots, p_{n}]$ precisely when there exists a point $q \in Q$ such that $D_{P,Q}[p_j, q] \leq \psi_k$ for all $j \in \{0, \dots, n \}$. We compute the homology in dimension $1$ with $\mathbb{Z}_2$ coefficients and summarize the results as the  \textit{witness persistence diagram}, denoted $\wpd(P,Q)$ \cite{Witness}. By the Dowker duality theorem, $\wpd(P,Q)$ and $\wpd(Q,P)$ are identical \cite{Dowker, functorial_dowker}. See SI Section 1A4 for details. 

% One paragraph on the similarity-centric analogous cycles
For each significant point $w \in \wpd(P,Q)$, we construct auxiliary complexes which admit inclusions to the filtration of complexes on $P$ (resp. $Q$) and the witness complex at a maximal parameter $\psi(w)$. We leverage induced maps on homology to construct a linear system whose solutions enumerate matches $p_w \subseteq \pd(P)$ and $q_w \subseteq \pd(Q)$ through $w$. For details, see SI Section 1A6 and \cite{analogous_bars}. These (possibly empty) collections of significant points in the persistence diagrams are referred to as matched, or analogous, cycles. The output is illustrated via color-coordination of points in $\pd(P)$ and $\pd(Q)$. For a visualization of the output including the witness persistence diagrams, see SI Fig.~S24.

\subsection*{Significance of analogous cycles from geometric null models}
Random or shuffled dissimilarity matrices provide a null model inconsistent with any geometry and empirically always produce no significant matches. A better null model would assume an underlying shared geometry consistent with the number of cycles observed in each system, in which features are randomly aligned. 

To construct such a null model in the relevant $1$-dimensional case, let $P, Q$ be observed systems with $n_P$ and $n_Q$ elements, and let $D_P$, $D_Q$, and $D_{P,Q}$ be the dissimilarity matrices. Let $m_P$ and $m_Q$ denote the number of significant points in $\pd(P)$ and $\pd(Q)$, and assume $m_P \geq m_Q$. Construct a $m_P$-dimensional torus $T_P$\footnote{An $m$-dimensional torus has $m$ independent circular coordinates, and can be obtained by taking an $m$-dimensional cube and identifying its opposite faces.} and let $P'$ be a uniform sample of $n_P$ points from that torus. Select an $m_Q$-dimensional torus embedded within $T_P$, and let $Q'$ denote a sample of $n_Q$ points from this subspace. Take $D_{P',Q'}$ to be the cross-distance matrix for these points. Let $\{(p_w, q_w) | w \in \wpd(P', Q') \text{ significant}\}$ be the output of the method of analogous cycles applied to $D_P, D_Q$ and $D_{P',Q'}$ for a sample from this null model. For each $p_i \in \pd(P)$ and $q_j \in \pd(Q)$, let $F_{p,q}$ be the number of significant $w \in \wpd(P', Q')$ such that $p_i \in p_w$ and $q_j \in q_w$, and summarize the outputs in a frequency matrix $F$ whose $(i,j)$-entry is $F_{p_i, q_j}$. We repeat this sampling procedure $N \geq 100$ times and report the empirical probability matrix $\frac{F}{N}$ to provide non-parametric $p$-values for matches.

\subsection*{Experimental data}
All animal procedures and experiments were approved by the Institutional Animal Care and Use Committee of the University of North Carolina at Chapel Hill or the University of California Santa Barbara and performed in accordance with the regulation of the US Department of Health and Human Services. Cranial windows were surgically implanted in mice expressing the fluorescent calcium sensor protein GCaMP6s \cite{chen2013ultrasensitive} in layer 2/3 pyramidal neurons in primary visual cortex (V1) and higher visual area anterolateral (AL) \cite{YuSmith, Yu_2024}. Large field-of-view two-photon calcium imaging \cite{yu2021diesel2p,StirmanSmith} measured neuronal activity responses to visual stimuli. Fluorescence dynamics were converted to inferred spike trains \cite{pnevmatikakis2016simultaneous}. Visual stimuli were displayed on a 60 Hz LCD monitor (9.2 cm x 15 cm). All stimuli were displayed in full contrast. Stimuli consisted of drifting square-wave gratings. We repeated the stimulus presentation 20 times. The experimental data is presented in \cite{Yu_2024}. See SI Section~2C for data preprocessing steps.

\subsection*{Computations}
Code and experimental data can be found at the following Github repository \href{https://github.com/irishryoon/analogous_neural}{https://github.com/irishryoon/analogous\_neural}. All persistent homology computations are performed via Eirene \cite{henselmanghristl6}.
}

\section*{Acknowledgements}This work was supported by NSF DMS 1854683 (IY, CG), NSF DMS 1854748 (GHP), AFOSR FA9550-21-1-0334 (RG), and AFOSR FA9550-21-1-0266 (CG). The authors thank Carina Curto and Vladimir Itskov for helpful conversations.

 % Display the acknowledgments section

%\bibliography{paper}
\printbibliography

\end{document}

%% Comment out or remove this line before generating final copy for submission; this will also remove the warning re: "Consecutive odd pages found".
%\instructionspage  

\maketitle

%% Adds the main heading for the SI text. Comment out this line if you do not have any supporting information text.
%\SItext

\tableofcontents

\section{Supplementary methods}
%--------------------------------------------
\subsection{Topological methods}

\subsubsection{Simplicial complex and simplicial homology}
An \textit{(abstract) simplicial complex} $K = (V, F)$ consists of the vertex set $V$ and the collection of simplices $F$, where a simplex is an unordered subset of $V$. Given a simplex $\sigma \in F$, every non-empty subset of $\sigma$ is also in $F$. Here, the vertex set $V$ will often correspond to individual neurons or spike trains, and $F$ will consist of collections of neurons with similar spike trains. A collection of $n+1$ vertex elements, say $(v_0, \dots, v_n)$, that is in $F$ is called an \textit{$n$-simplex}. 

While there are many definitions of homology groups, here we compute the homology of a simplicial complex $K$ with field coefficients. As a result, all homology computations involve vector spaces and linear maps. We direct the reader to \cite{hatcherAlgebraicTopology} for a more general introduction to homology. \\

\noindent{\textbf{Chains and boundary homomorphisms.}}
Let $F_n$ be the number of $n$-simplices in $K$. An $n$-chain is a finite formal sum of $n$-dimensional simplices
\[ \sum_{i}^{F_n} c_i \sigma_i^n\]
where $c_i \in \mathbb{R}$ and $\sigma_i^n$ refers to an $n$-simplex in $K$. The collection of all such $n$-chains are denoted $C_n(K)$. Note that $C_n(K)$ is the vector space $\mathbb{R}^{F_n}$.

A \textit{boundary homomorphism} $\partial_n: C_n(K) \to C_{n-1}(K)$ is constructed as the following. Given an $n$-simplex $\sigma = (v_0, \dots, v_{n})$, we let $(v_0, \dots, \hat{v}_i, \dots, v_{n})$ denote the $(n-1)$-simplex obtained by removing $v_i$ from the original collection. We define the boundary homomorphism on $\sigma$ to be 
\[ \partial_n(\sigma) = \sum_{i=0}^n (-1)^i (v_0, \dots, \hat{v}_i, \dots, v_n).\]
We then extend this map linearly to all $n$-chains. That is, we define $\partial_n: C_n(K) \to C_{n-1}(K)$ as 
\[\partial_n\Big(\sum_{i}^{F_n} c_i \sigma_i^n\Big) = \sum_{i}^{F_n} c_i \, \partial_n(\sigma_i^n).\]

This results in the following sequence of vector spaces and linear maps 
\[\cdots \to C_{n+1}(K) \xrightarrow{\partial_{n+1}} C_n(K) \xrightarrow{\partial_n} C_{n-1}(K) \cdots \xrightarrow{\partial_1} C_0(K) \to 0 \]

One can check that the composition of the boundary homomorphisms is the zero map, i.e., $\partial_{n} \circ \partial_{n+1} = 0$ for all $n$. \\

\noindent{\textbf{Homology.}}

The fact that $\partial_{n} \circ \partial_{n+1} = 0$ implies that $\text{im}\, \partial_{n+1} \subseteq \ker \partial_n$. We refer to elements of $\ker \partial_n$ as \textit{cycles} and elements of $\textit{im}\, \partial_{n+1}$ as \textit{boundaries}. An element in $\ker \partial_n$ represents a collection of $n$-simplices that form an $n$-dimensional hole. If there are $(n+1)$-simplices that can ``fill in" this $n$-dimensional hole, then we do not consider such hole as being valid. We thus consider only the $n$-dimensional holes that cannot be filled in via $(n+1)$-simplices. Such information is computed via the $n^{\text{th}}$ $\textit{homology }$:
\[ \displaystyle H_n(K) = \frac{\ker \partial_n}{\text{im}\, \partial_{n+1}}.\]

$H_n(K)$ is a vector space whose dimension is the number of $n$-dimensional holes that are not filled in.

\subsubsection{Nerve Theorem}
\label{SI:Nerve}

% https://journals.plos.org/ploscompbiol/article?id=10.1371/journal.pcbi.1000205#s5 for reference

A fundamental result in algebraic topology, called the Nerve Theorem, enables the study of a neural manifold in terms of the intersections of receptive fields. We first introduce the relevant terminology. 

% definition and lemma
Given a topological space $X$, let $\mathcal{U} = \{ U_{\alpha} \}$ be a collection of open sets such that every point $ x \in X$ is in some $U_{\alpha}$. We refer to such $\mathcal{U}$ as an \textit{open cover}.

Given an open cover $\mathcal{U}$, we construct a simplicial complex $N\mathcal{U}$ called the \textit{nerve} of $\mathcal{U}$. The vertex set of $N\mathcal{U}$ consists of the elements $U_{\alpha}$ in the open cover. A simplex $(v_{\alpha_1}, \dots, v_{\alpha_k})$ is in $N\mathcal{U}$ if $U_{\alpha_1} \cap \cdots U_{\alpha_k} \neq \emptyset$.  

% version that is relevant to us
\begin{theorem}[Nerve Theorem, paraphrased]
Let $\mathcal{U} = \{ U_{\alpha}\}$ be an open cover of a compact space $X \subset \mathbb{R}^n$ such that every $U_{\alpha}$ is convex. Then, the homology of $X$, denoted $H_1(X)$ is equal to the homology of the nerve, denoted $H_1(N\mathcal{U})$.
\end{theorem}

\begin{figure}[H]
\centering
\includegraphics[width=0.6\textwidth]{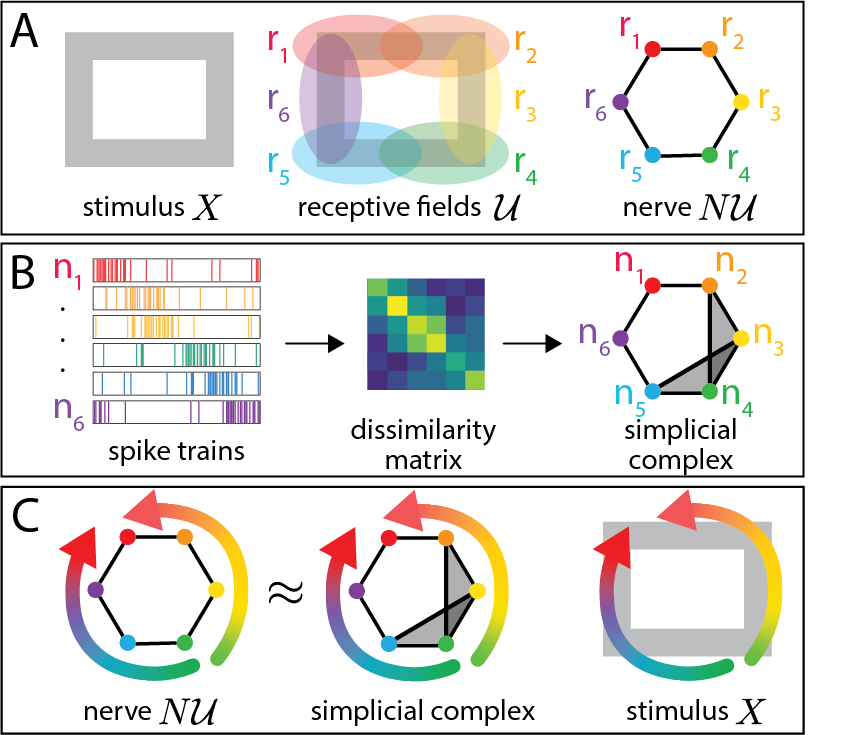}
\caption{\textbf{Nerve Theorem in navigational system.} \textbf{A.} The stimulus space $X$ is a two-dimensional environment explored by an animal. A collection of convex receptive fields form a good cover $\mathcal{U}$. The nerve $N\mathcal{U}$ is a simplicial complex representing the intersection patterns of the receptive fields. \textbf{B.} The intersections of the receptive fields can be approximated from spike train similarities. One can build a simplicial complex from the spike train dissimilarities. 
\textbf{C.} The nerve $N\mathcal{U}$ is approximated by the simplicial complex built from the spike train dissimilarities. Finding two distinct paths from any point to any other in the simplicial complex indicates the presence of a circular
coordinate. Such circular coordinate is detected by first homology $H_1$. Here, the circular coordinate reflects the hole in the environment. }
\label{SIfig:nerve}
\end{figure}

% what this means in our context
Here, the space $X$ can be the space of visual stimuli or a two-dimensional environment explored by an animal. Each $U_{\alpha}$ corresponds to a single neuron with convex receptive field. If $X$ corresponds to visual stimuli, then $\mathcal{U}$ consists of neurons responding to specific visual cues. If $X$ corresponds to a two-dimensional environment, then the collection of place fields is a valid option for $\mathcal{U}$ (Figure~\ref{SIfig:nerve}A).

The theorem guarantees that the nerve $N\mathcal{U}$ reflects the structure of the stimulus space $X$ (Figure~\ref{SIfig:nerve}A). Note that $N\mathcal{U}$ is constructed from the intersection patterns of receptive fields. One can instead use similarities between spike trains as a proxy for the intersection of the corresponding receptive fields and construct another simplicial complex that approximates $N\mathcal{U}$. To do so, we first compute the pairwise dissimilarities among all spike trains (Figure~\ref{SIfig:nerve}B). We then fix a dissimilarity threshold $\varepsilon$ and build a simplicial complex that reflects the dissimilarities among the spike trains; Each spike train is a vertex in the simplicial complex and an $n$-simplex corresponds to $(n+1)$ spike trains whose pairwise dissimilarities are all less than $\varepsilon$ (Figure~\ref{SIfig:nerve}B). 

% some explanation of H1 in this context. 
The first homology $H_1$, whether it is computed from the stimulus space $X$, the nerve $N\mathcal{U}$, or the approximated simplicial complex (Figure~\ref{SIfig:nerve}C), is a vector space whose dimension equals the number of circular features. When $X$ is a visual stimulus space, the circular features reflect a collection of visual stimuli whose variance is explained by one parameter that repeatedly revisits the stimuli as one varies the parameter. Examples include orientation and movement direction. When $X$ is a two-dimensional environment, then the circular features correspond to regions in the environment that are unavailable for the animal (Figure~\ref{SIfig:nerve}C).
%\IY{should talk about neural manifolds that are intrinsically circular, like head direction or grid cells. }

In practice, the similarities between spike trains do not always perfectly capture the homology of $X$. This is partly due to the noise and stochasticity of spike trains. In addition, it is unclear how one should choose the dissimilarity threshold $\varepsilon$ at which two neurons are considered to have overlapping receptive fields. To address both concerns, we use persistent homology (see the following section), which computes the homology of the nerve $N\mathcal{U}$ for varying levels of dissimilarities among neurons. 

While the above Theorem is stated for first homology $H_1$, the theorem can be stated for higher-dimensional homology. The Theorem follows from the following Corollary \cite{hatcherAlgebraicTopology}.

% the nerve lemma, as stated in HAtcher
\begin{theorem}[Nerve Theorem \cite{hatcherAlgebraicTopology} Corollary 4G.3]
Let $\mathcal{U}$ be an open cover of a compact space $X$ such that every nonempty intersection of finitely many sets in $\mathcal{U}$ is contractible. Then, $X$ is homotopy equivalent to the nerve $N\mathcal{U}$.
\end{theorem}

% discussion of generalizations
For a thorough discussion of the various versions of the Nerve Theorem, we direct the reader to \cite{bauer_unified_2023}.

%--------------------------------------------
\subsubsection{Persistent homology}
\label{methods:PH}
We provide a brief introduction to persistent homology. Given a population of interest, we represent the system via a simplicial complex, which is a generalization of networks that incorporates higher-order interactions. 
A simplicial complex $X = (V, F)$ consists of the vertex set $V$ and the collection of simplices $F$, where a simplex is an unordered subset of $V$. Given a simplex $\sigma \in F$, every non-empty subset of $\sigma$ is also in $F$.

Let $P = \{ p_1, \dots, p_n \}$ denote a population of interest. Assume that we know the dissimilarity between every pair $(p_i, p_j)$. To study $P$ with respect to the similarity of its constituents, one can choose a dissimilarity threshold $\varepsilon$ and construct the simplicial complex $X_P^{\varepsilon} = (P, F_{\varepsilon})$ whose vertex set equals $P$ and a simplex $\sigma \in F_{\varepsilon}$ consists of elements of $P$ whose pairwise dissimilarity is at most $\varepsilon$. One can then look for cyclic features in $X_P^{\varepsilon}$ by computing its homology $H_1(X_P^{\varepsilon})$, which is a vector space whose dimension equals the number of cycles in the simplicial complex. 

% paragraph on persistent homology
So far, we constructed the simplicial complex by studying pairwise dissimilarities upto some parameter $\varepsilon$. It is difficult to choose the  appropriate parameter $\varepsilon$ without prior knowledge. We thus turn to persistent homology, which examines the homology $H_1(X_P^{\varepsilon})$ as the parameter $\varepsilon$ varies. Given a collection of parameters $\{\varepsilon_1, \varepsilon_2, \cdots, \varepsilon_N \}$, the simplicial complexes at various parameters form a filtration
\[X_P^{\varepsilon_1} \subseteq X_P^{\varepsilon_2} \subseteq \cdots \subseteq X_P^{\varepsilon_N}.\]
Persistent homology applies homology $H_1( \cdot)$ to this sequence of nested simplicial complexes to obtain the following sequence of vector spaces and linear maps between them:
\begin{equation}
\label{eq:persistence}
H_1(X^\bullet_P): \, H_1(X_P^{\varepsilon_1}) \to H_1(X_P^{\varepsilon_2}) \to  \cdots \to H_1(X_P^{\varepsilon_N}).
\end{equation}
The maps arise from the fact that if $X_P^{\epsilon_i}$ is a subcomplex of $X_P^{\epsilon_j}$, then there is an induced linear map $f: H_1(X_P^{\epsilon_i}) \to H_1(X_P^{\epsilon_j})$. The map $f$ identifies cycles in $X_P^{\epsilon_i}$ with cycles in $X_P^{\epsilon_j}$. Persistent homology then extracts the birth and death parameters of cyclic features in Equation \ref{eq:persistence}. 
% paragraph on persistence diagram 
Given a cyclic feature with birth parameter $b$ and death parameter $d$, one can summarize the evolution of cycles in Equation \ref{eq:persistence} by plotting a point at coordinates $(b,d)$. The resulting plot is referred to as the persistence diagram. For details, see \cite{Ghrist_barcodes, PH_survey, Carlsson2009TopologyAD, ELZ_persistence},

\subsubsection{Witness persistent homology} 
\label{methods:WPH}
Let $P = \{p_1, \dots, p_n\}$ and $Q = \{q_1, \dots, q_m\}$ be two populations of interest, and let $M_{P,Q}$ denote the $n \times m$ cross-system dissimilarity matrix. We fix a dissimilarity threshold $\psi$ and construct the witness complex $W_{P,Q}^{\psi} = (P_{\psi}, F_{\psi})$, which is a simplicial complex whose vertex set $P_{\psi}$ consists of points whose dissimilarity to the collection $Q$ is at most $\psi$. That is, 
\[P_{\psi} = \{p_i \in P \, | \, \exists q_j \in Q  \text{ such that } M_{P,Q}(p_i, q_j) \leq \psi \}. \]
The simplices $F_{\psi}$ consist of subsets of $P$ whose dissimilarity to a single element of $Q$ is at most $\psi$. In other words, 
\[F_{\psi} = \{ (p_{i_1}, \dots, p_{i_k}) \, | \, \exists q_j \in Q \text{ such that } \max \{ \, M_{P,Q}(p_{i_1}, q_j), \cdots,  M_{P,Q}(p_{i_k}, q_j)  \, \}\leq \psi \}. \]

In order to overcome the challenge of choosing an appropriate parameter $\psi$, we construct the witness complex at various parameters $\{\psi_1, \psi_2, \cdots, \psi_N \}$:
\[ W_{P, Q}^{\psi_1} \subseteq W_{P,Q}^{\psi_2} \subseteq \cdots \subseteq W_{P,Q}^{\psi_N}.\]

Applying homology $H_1(\cdot)$ to this sequence of nested simplicial complexes leads to the following sequence of vector spaces and linear maps between them:

\[H_1(W^\bullet_{P,Q}): \, H_1(W_{P, Q}^{\psi_1}) \to H_1(W_{P,Q}^{\psi_2}) \to \cdots \to H_1(W_{P,Q}^{\psi_N}).\]

Persistent homology then extracts the birth and death times of cyclic structures as the parameter evolves. The result is summarized in a witness persistence diagram denoted by $\pd_1(W_{P,Q})$. 
The points on the witness persistence diagram that are far from the diagonal indicates cyclic structures in $P$ and $Q$ that are shared. 

% using the transposed matrix.
Note that given the cross-system dissimilarity matrix $M_{P,Q}$, one could take the transpose of this matrix $M_{Q,P} = (M_{P,Q})^T$ and build the witness complex $W_{Q,P}$ that has $Q$ as its possible vertex set. The functorial Dowker's Theorem \cite{Dowker, functorial_dowker} tells us that the resulting witness persistence diagrams $\pd_1(W_{P,Q})$ and $\pd_1(W_{Q,P})$ are identical. Throughout, we use the notation $\wpd(P,Q)$ to refer to $\pd_1(W_{P,Q})$ and $\pd_1(W_{Q,P})$.

\subsubsection{Significant points on a persistence diagram}
\label{SI:significant_PD}
%---------------------------------------------------------------
Given a point on a persistence diagram, we refer to the difference between the death and birth parameters as the lifetime of the point. Given a persistence diagram, we consider points whose lifetimes lie outside the interquartile range as significant points. Let $Q_1$ denote the lower quartile, let $Q_2$ denote the median, and let $Q_3$ denote the upper quartile of the collection of lifetimes of a persistence diagram. A standard way for determining the outlier is to compute the following thresholds
\begin{align*}
o_U = Q_3 + k * (Q_3 - Q_1) \\
o_L = Q_1 - k * (Q_3 - Q_1).
\end{align*}
Usually, any value above $o_U$ or $o_L$ is considered an outlier. A typical choice of $k$ is $k=1.5$ or $k=3$. Here, we use $k=3$. %checked. k = 3

When identifying the significant points in a persistence diagram, we compute the interquartile range of lifetimes of persistence points. The lifetimes are left-skewed, so we consider points whose lifetimes are at least $o_U$ to be significant points. Given a persistence diagram $\pd$, we use $\pd_*$ to denote the collection of significant points. 
Note that the outlier-detection method is viable only when the persistence diagram contains enough points. When the persistence diagram doesn't have enough points, we take all points as significant. 

%\IY{A brief discussion on the lack of appropriate null model for this? Should we discuss our various experiments (with relative length, confidence sets, etc?)} -- maybe if a reviewer asks for it. 
See Section \ref{sec:significance_pd} for a comparison of alternative methods of determining significant points on a persistence diagram. 

\subsubsection{Analogous cycles}
\label{methods:analogous_cycles}

% motivation 
In order to identify cycles in two systems, we developed the method of analogous cycles. The method uses Witness persistent homology to identify shared cycles between the two systems. We present a pseudocode of the analogous cycles method in Algorithm \ref{alg:analogous_cycles}.

Let $P = \{p_1, \dots, p_n \}$ and $Q = \{q_1, \dots, q_m \}$ denote two systems of interest. Assume that we are given a cross-system dissimilarity matrix $D_{P,Q}$ where the entry at row $i$ and column $j$ denotes dissimilarity between $p_i$ and $q_j$. We compute the persistence diagrams $\pd(P), \pd(Q), \wpd(P,Q)$ from the dissimilarity matrices $D_P, D_Q$, and the cross-system dissimilarity matrix $D_{P,Q}$.

In step 2, we find the significant points $\pd_*(P), \pd_*(Q), \wpd_*(P,Q)$ of the persistence diagrams (see SI Section \ref{SI:significant_PD}). We consider $\wpd_*(P,Q)$ as an indication of shared circular features between $P$ and $Q$.

For each point $w \in \wpd_*(P,Q)$, we aim to find their representations $p_w \subseteq \pd(P)$ and $q_w \subseteq \pd(Q)$. To do so, in step 3A, we first find homology classes in Witness complexes that represent $w$. Fix $\psi$ to be the largest parameter smaller than the death time of $w$. Then, one can find homology classes $[w_P] \in W_{P,Q}^{\psi}$ and $[w_Q] \in W_{Q,P}^{\psi}$ that corresponds to $w$\footnote{In \cite{analogous_bars}, it is shown that such homology class is unique when homology is computed with $\mathbb{Z}/2\mathbb{Z}$ coefficients}.

%Using Dowker's Theorem we find the point $k_Q$ in $PD_1(W_{Q,P})$ that is dual to $k_P$. Repeat the above process to find the cycle $[w_Q] \in W_{Q,P}^{\psi}$ that represents $k_Q$.%\IY{comment that the persistence diagrams and $w_P$, $w_Q$ are actually the same.}

% Persistent extension
The technicality of the analogous cycles method lies in the persistent extension method in steps 3b and 3c. We now briefly describe the persistent extension method. Given a homology class $[w_P] \in W^{\psi}_{P,Q}$, the (class-to-class) persistent extension (\cite{analogous_bars} Algorithm 1) finds a collection of homology classes $[x_P] \in H_1(X^{\varepsilon}_P)$ that represent $[w_P]$ at various parameters $\varepsilon$. It does so by finding all possible representations of $[w_P]$ in $H_1(X_P^{\varepsilon})$ for various $\varepsilon$ via the intersected complex $W^{\psi}_{P,Q} \cap X^{\varepsilon}_P$. That is, consider the following sequence of vector spaces and induced linear maps
\[ H_1(W_{P,Q}^{\psi}) \xleftarrow{\chi_{\varepsilon}} H_1(W_{P,Q}^{\psi} \cap X_P^{\varepsilon}) \xrightarrow{\Upsilon_{\varepsilon}} H_1(X_P^{\varepsilon}).  \]
If there exists a homology class $[r] \in H_1(W^\psi_{P,Q} \cap X^\varepsilon_P)$ such that $\chi_{\varepsilon}[r] = [w_P]$, then we consider its image $\Upsilon_{\varepsilon}[r]$ to be a representation of $[w_P]$ in $H_1(X^\varepsilon_P)$. We call such image $\Upsilon_{\varepsilon}[r]$ a class extension. Note that there can be many class extensions. 

For a complete understanding of how $[w_P] \in H_1(W^\psi_{P,Q})$ is represented in $H_1(X_P^{\varepsilon})$, one should report all class extensions at various parameter $\varepsilon$. Here, we only report one class extension for conciseness. We select the parameter $\varepsilon_0$ to be the minimum parameter at which one can find a class extension. We then report one homology $[x_P] \in H_1(X_P^{\varepsilon_0})$ that is reported by the algorithm (see Algorithm 1 and \cite{analogous_bars}) \footnote{In the full analogous cycles method, we report a collection of class extensions. As such, the output does not depend on the choice of basis of the persistence module $H_1(W^{\psi}_{P,Q} \cap X^\bullet_P)$. Here, we report only one class extension, and such output depends on the choice of the basis.}. In step 3b, we repeat the same process to find the homology class $[x_Q] \in H_1(X^\varepsilon_Q)$ that represents $[w_Q]$ via persistent extension. 

%Once we select the baseline cycle extension $[x_P] \in H_1(X_P^{\varepsilon_0})$, we then find the points on the persistence diagram $\pd_1(X_P)$ that represent $[x_P]$. Such points $x_P$ are the outputs of step (2). One considers the $x_P$ as the points on the persistence diagram $\pd_1(X_P)$ that represent $[w_P]$.  
 
In step 3d, we find collections of points $p_w \subseteq \pd_*(P)$ that represent $[x_P]$ and $q_w \subseteq \pd_*(Q)$ that represent $[x_Q]$. Given $[x_P] \in H_1(X^\varepsilon_P)$, we first find the points in $\pd(P)$ that represent $[x_P]$. We then consider the subset that consists of the significant points $\pd_*(P)$ and denote it by $p_w$\footnote{In \cite{analogous_bars}, the analogous cycles method returns all representations of $[x_P]$ under different choices of basis of the persistence module $H_1(X^\bullet_P)$. Here, the output $p_w$ depends on the choice of basis of $H_1(X^\bullet_P)$}. We find $p_q \in \pd_*(Q)$ in a similar manner. We consider $p_w$ and $p_q$ as representing analogous circular features in $P$ and $Q$, and we refer to $(p_w, q_w)$ as an analogous pair via $w$. 

We direct readers to \cite{analogous_bars} for details of the method.

\begin{algorithm}
\caption{Analogous cycles}
\label{alg:analogous_cycles}
\begin{flushleft}
\textbf{Inputs}: Dissimilarity matrices $D_P, D_Q$, and cross-system dissimilarity matrix $D_{P,Q}$. \\
\end{flushleft}
\begin{flushleft}
\textbf{Outputs}: 
\end{flushleft}
\begin{itemize}
\item persistence diagrams $\pd(P), \pd(Q)$,
\item witness persistence diagram $\wpd(P,Q)$, and 
\item a collection $\{(p_w,q_w) \, | \, w \in \wpd_*(P,Q) \}$ where $p_w \subseteq \pd_*(P)$, $q_w \subseteq \pd_*(Q)$.
\end{itemize}
\begin{flushleft}
\textbf{Algorithm}:
\end{flushleft}
\begin{enumerate}
\item Compute persistence diagrams $\pd(P)$, $\pd(Q)$, and witness persistence diagrams $\wpd(P,Q)$.
\item Compute the significant points $\pd_*(P), \pd_*(Q)$, and $\wpd_*(P,Q)$
\item For every $w \in \wpd_*(P,Q)$:
\begin{enumerate}
\item Find homology classes $[w_P] \in W_{P,Q}^{\psi}$ and $[w_Q] \in W_{Q,P}^{\psi}$ that correspond to $w$ for some $\psi$. Typically, we choose $\psi$ to be the largest parameter smaller than the death time of $w$. 
\item Find homology class $[x_P] \in H_1(X_P^{\varepsilon})$  for some $\varepsilon$ that represent $[w_P]$ via persistent extension
\item Find homology class $[x_Q] \in H_1(X_Q^{\varepsilon})$  for some $\varepsilon$ that represent $[w_Q]$ via persistent extension
\item Find points $p_w \subseteq \pd_*(P)$ and $q_w \subseteq \pd_*(Q)$ that each represent $[x_P]$ and $[x_Q]$.
\end{enumerate}
\item return $\{(p_w, q_w) \, | \, w \in \wpd_*(P,Q)\}$
\end{enumerate}

\end{algorithm}

\subsection{Methods for computing dissimilarities among spike trains and stimulus images}
\subsubsection{Circular distances}
An interval $[0,I]$ whose boundaries are identified is considered a circle. Given points $u,v \in [0, I]$, we define its circular distance as the distance between $u,v$ on the circle obtained by identifying the boundary of $[0,I]$. That is, let $A = \min \{u, v \}$ and $B = \max \{u,v\}$. We denote the circular distance between $u,v$ by
\begin{equation}
\label{eq:angular_distance}
d_I(u,v) = \min \{B -A, A + I - B\}. 
\end{equation}

%--------------------------------------------------------------
\subsubsection{Dissimilarity between spike trains and firing rates}
\label{sec:SI_timeseries_dissimilarity}
%--------------------------------------------------------------
We represent a spike train or firing rate with $L$ bins as $\vec{x}$ of dimension $L$. For spike trains, the entries of $\vec{x}$ can be binary, where $\vec{x}_i = 1$ indicates that a spike has occurred at $i^{\text{th}}$ bin:

\begin{equation*}
\vec{x}_i = 
\begin{cases}
1 \quad \text{ if } x \text{ has a spike at bin } i \\
0 \quad \text{ otherwise.} \\
\end{cases}
\end{equation*}
The entries of $\vec{x}$ can also be integer-valued, where $\vec{x}_i$ represents the number of spikes that occurred at bin $i$. When representing a firing rate, the entries of $\vec{x}$ can be real-valued. In this study, we demonstrate the analogous cycles method on binary-, integer-, and real-valued vectors representing neural activity.

% checked: 1. simulations are binary rasters (from the first output of "simulate_rasters"). 2. UCSB Smith results in integer-valued (maximum 3) vectors 3. grid cell simulation are real-valued rates. 

Let $\vec{x}, \vec{y}$ be two vectors representing neural activity. The cross-correlation between $\vec{x}$ and $\vec{y}$ for a displacement value of $n$ is 
\begin{equation*}
(\vec{x} * \vec{y})_n = \sum_{m} \vec{x}_m \vec{y}_{m+n} .
\end{equation*}
We define the similarity between $\vec{x}$ and $\vec{y}$ to be the normalized sum of $(\vec{x} *\vec{y})_n$ for a limited range of displacement values specified by the shift parameter $\ell: $  
\begin{equation}
\label{eq:spiketrain_similarity}
Sim_\ell(\vec{x}, \vec{y}) = \frac{\sum_{n=-\ell}^\ell (\vec{x} * \vec{y})_n}{\sqrt{(\vec{x} \cdot \vec{x}) (\vec{y} \cdot \vec{y}) }}. 
\end{equation} % this formula for similarity was also double-checked, both for the simulations (julia code) and for the Smith data (python code "functions.py"). 

Given a collection of vectors $P = \{\vec{x}_i\}_{i=0}^{K}$ representing time series, our goal is to compute the pairwise dissimilarity between every pair of vectors in $P$. We first compute the pairwise similarity between every pair. We then compute the maximum pairwise similarity value 
\[M = \max_{\{\vec{x}, \vec{y} \in P\}} Sim_{\ell}(\vec{x}, \vec{y}),\]
and normalize all similarity by $M$ so that the similarity values live in $[0,1]$. We then define the dissimilarity between $\vec{x}$ and $\vec{y}$ as 
\begin{equation}
\label{eq:spike_train_dissimilarity}
Dis_\ell(\vec{x}, \vec{y}) = 1 - \frac{Sim_\ell(\vec{x}, \vec{y})}{M}. 
\end{equation}
if $\vec{x} \neq \vec{y}$. If $\vec{x} = \vec{y}$, we assign $Dis_{\ell}(\vec{x}, \vec{x}) =0$. We refer to such dissimilarity as the \textit{windowed cross-correlation dissimilarity}. Note that cross-correlation measures similarity of two time series as a function of the displacement of one time series relative to the other. The resulting dissimilarity, then, assigns a low dissimilarity score if the two time series are similar within the displacement bound $\ell$. One can consider the displacement bound $\ell$ as being akin to the time scale parameter of the Victor-Purpura distance \cite{VPdistance} and the van Rossum distance \cite{VRdistance}.

%While we discussed the dissimilarity in the context of spike trains, we can use the same construction to compute dissimilarity between two firing rates. 

%---------------------------------------------------------------
\subsubsection{Dissimilarity between images on a square torus}
\label{SI:dissimialrity_images}
Consider a window of size $s \times s$ considered as a square torus by identifying the left and right edges and the top and bottom edges. Let $S_{x, y, \theta}$ denote an image with a grating of orientation $\theta$ and a circular mask centered at $(x,y)$. The collection of such images constitutes the stimulus in the simulation studies. See Equation \ref{eq:sampled_image} for the equation of $S_{x, y, \theta}$ and SI Fig. \ref{fig:stimulus_summary} for an example images.%\IY{might need to explain why we compute dissimilarity this way and not pixel-wise}

Let $S_{x_i, y_i, \theta_i}$ and $S_{x_j, y_j, \theta_j}$ be two such images. Let $d_{s}(x_i, x_j)$, $d_{s}(y_i, y_j)$, and $d_\pi(\theta_i, \theta_j)$ denote the circular distances (Equation \ref{eq:angular_distance}). We define the dissimilarity between the two stimulus images $S_{x_i, y_i, \theta_i}$ and $S_{x_j, y_j, \theta_j}$ as 
\begin{equation}
\label{eq:image_dissimilarity}
d(S_{x_i, y_i, \theta_i}, S_{x_j, y_j, \theta_j}) = \sqrt{ \Big(\frac{d_s(x_i, x_j)}{s}\Big)^2 + \Big( \frac{d_s(y_i, y_j)}{s} \Big)^2 + \Big( \frac{d_\pi(\theta_i, \theta_j)}{\pi} \Big)^2}.
\end{equation}

%\IY{This is a "intrinsic" dissimilarity based on the circular coordinates of x, y, and theta}

\subsubsection{Dissimilarity between a spike train and images in a video}
\label{sec:dissimilarity_spiketrain_image}
% corresponding notebook: `1_generate_stimulus_parameters.ipynb`

Let $S$ denote a stimulus video consisting of $L_S$ images (frames), and let $\vec{x}$ be a spike train vector representing the firings of a neuron in response to the video $S$. Let $L_{x}$ be the length of $\vec{x}$. One should consider $\vec{x}$ as either an experimental or simulated spike train that is measured simultaneously as the video $S$.  

% \IY{comment on the length, etc.. new notation for stimulus video}  Do we have a new notation for the stimulus video? 

We compute the dissimilarities between an image  $S_{x, y, \theta}$ in $S$ and a spike train $\vec{x}$ as follows. We first construct a vector $u_{x, y, \theta}$ of length $L_S$ indicating the frames at which an image similar to $S_{x, y, \theta}$ occurs in the video $S$. Given an image frame $t$, let $(x_t, y_t)$ denote the center location of the circular mask of the stimulus video at frame $t$. Similarly, let $\theta_t$ denote the orientation of the stimulus video at frame $t$. We constructed a binary vector $\vec{u}^{x, y, \theta}$ of length $L_S$ as %L = 100,000? 
\begin{equation}
\label{eq:sampled_stimulus_binary_array}
\vec{u}^{x, y, \theta}_t = 
\begin{cases}
1 \quad \text{ if } \theta = \theta_t  \text{ and } \sqrt{d_s(x, x_t)^2 + d_s(y, y_t)^2 } \leq 6 , \\
0 \quad \text{ otherwise} \\
\end{cases}
\end{equation}
where $d_s$ is the circular distance (Equation \ref{eq:angular_distance}). That is, $\vec{u}^i_t = 1$ if the stimulus video $S(x_t, y_t, \theta_t)$ at frame $t$ and the image $S_{x, y, \theta}$ have the same orientation and if their circular masks are close in location. 

We then converted $\vec{u}^{x, y, \theta}$ into a vector of length $L_x$. Here, each frame in $S$ corresponded to twentyfive time bins in $\vec{x}$, i.e., $L_x = 25 * L_S$. We thus lengthened $\vec{u}^{x, y, \theta}$ by repeating each entry $25$ times, which results in a vector $\vec{U}^{x, y, \theta}$ of length $L_x$. We then compute the dissimilarity between $\vec{U}^{x, y, \theta}$ and $\vec{x}$ using the windowed cross-correlation dissimilarity (Equation \ref{eq:spike_train_dissimilarity}). We consider the resulting dissimilarity as the dissimilarity between image $S_{x, y, \theta}$ and $\vec{x}$. 

%\IY{details of how many such images we sample. Does this go here or somewhere else} -- I put it in later section when we discuss analogous cycles between stimulus and V1 simple cell.

%==============================================
\newpage
\section{Data}
\label{SI:data}
%==============================================

%==============================================
\subsection{Simulated visual system}
%==============================================

\subsubsection{Stimulus} 
\label{sec:stimulus_generation}
\begin{figure}[H]
\centering
\includegraphics[width=0.5\textwidth]{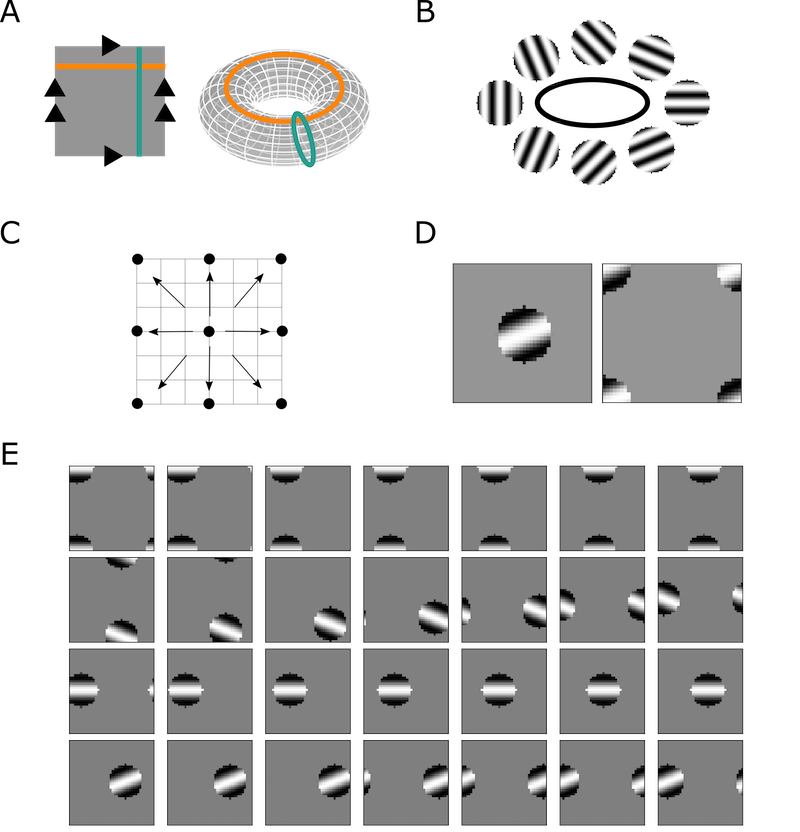}
\caption{Illustration of the design of stimulus space. \textbf{A} The stimulus window as a torus. \textbf{B} Eight orientations of the stimulus. \textbf{C} Eight movement directions of the stimulus. \textbf{D} Example stimulus images $S_{s/2, s/2, \pi/8}$ and $S_{0, 0, \pi/8}$. \textbf{E} First 28 frames of the stimulus video.}
\label{fig:stimulus_summary}
\end{figure}

We designed the stimulus to have non-trivial topology. Given a window of size $s \times s$, we considered the window as a square torus by identifying the left and right edges and the top and bottom edges (Figure \ref{fig:stimulus_summary}A). In particular, the $x$-coordinate and the $y$-coordinates of this window each lived in an interval $[0,s]$ with their boundaries identified. We created a video of a grating with a circular mask (Figure \ref{fig:stimulus_summary}D) that moves continuously on this square torus.

%--------------------------------------------------
% Figure: stimulus
%--------------------------------------------------

\noindent\textbf{Stimulus image.} 
Given a window of size $s \times s$, a grayscale sinusoidal grating with orientation $\theta$ centered at $(s/2, s/2)$ can be described by 
\[ B_{\theta}(x,y) = \cos \Big( 2 \pi \frac{(x-s/2) \cos \theta + (y -s/2) \sin \theta}{\lambda} + \phi\Big),  \]
where $\lambda$ is the wavelength and $\phi$ is the phase. % note: coordinates start at (0,0) on top left corner
We fixed a window size of $s = 40$ pixels, a wavelength of $\lambda = 12$ and a phase of $\phi = \pi$. We then applied a circular mask of radius $r$ centered at $(s/2, s/2)$ by
\begin{equation}
\label{eq:stimulus_centered}
M_\theta(x,y) = 
\begin{cases}
B_\theta(x,y) \quad \text{ if } d(x, s/2)^2 + d(y, s/2)^2 \leq r^2, \\
0 \quad \text{ otherwise} \\
\end{cases}
\end{equation}
with $r = 8$. Figure \ref{fig:stimulus_summary}D illustrates an example $M_\theta$ with $\theta = \pi/8$. The stimulus image $S_{x_0, y_0, \theta}$ with orientation $\theta$ and center location $(x_0, y_0)$ is obtained by
\begin{equation}
\label{eq:sampled_image}
S_{x_0, y_0, \theta}(x,y) = M_\theta(x-d_x \mod s, \quad y-d_y \mod s) ,
\end{equation}
where $d_x = x_0 - s/2$ and $d_y = y_0 - s/2$.
Figure \ref{fig:stimulus_summary}D illustrates an example stimulus image $S_{0, 0, \pi/8}$. One can consider $S_{x_0, y_0, \theta}$ as the image obtained from $M_{\theta}$ by moving the location of the circular mask and the sinusoidal grating from $(s/2, s/2)$ to $(x_0, y_0)$ on the square 
torus.  \\

\noindent\textbf{Stimulus video.}
Given an initial stimulus image $S_{x_0, y_0, \theta}$, we created a stimulus video of 40,000 frames in which the orientations and the center locations changed continuously. 

The stimulus orientations could take one of eight values $\{ 0, \pi/8, \dots, 7 \pi/ 8 \}$. Figure \ref{fig:stimulus_summary}B visualizes the eight orientations. Note that one can consider the space of all possible orientations as an interval $[0, \pi]$ with their boundaries identified. Thus, the collection of orientations has a circular feature. Given an orientation $\theta$, the neighboring orientations $N_{\text{orientation}}(\theta) = \{ \theta - \pi/8 \mod \pi, \theta + \pi/8 \mod \pi \}$ refers to the two orientations that are immediately to the left and right of $\theta$ in Figure \ref{fig:stimulus_summary}B. For example, $N_{\text{orientation}}(0) = \{\pi/8, 7\pi/8 \}$.

Similarly, throughout the video, the centers of the stimulus images could change by the following coordinates: 
\[\{(3,0), (3,3), (0,3), (3,0), (3,3), (0,3), (-3, 3), (-3,0), (-3,-3), (0,-3), (3, -3) \} \]
Figure \ref{fig:stimulus_summary}C illustrates the eight possible movement directions. Note that the collection of movement directions also has a circular feature. Given a direction $m$, its neighboring directions $N_{\text{direction}}(m)$ refers to the two directions that are immediately to the left and right of direction $m$ in Figure \ref{fig:stimulus_summary}D. For example, $N_{\text{direction}}((3,0)) = \{(3,3), (3, -3) \}$.

We created an initial stimulus image $S_{0,0,0}$ with orientation $\theta =0$ and center $(0,0)$. We assigned an initial movement direction of $m =(3,0)$. For $7$ frames, we moved the center of the stimulus image by $(3,0)$. That is, the first $8$ frames of the video consisted of the images $S_{0,0,0}, S_{3,0,0}, S_{6,0,0}$, and so on. 

Every 7 frames, we randomly chose a new orientation from its neighbors $N_{\text{orientation}}(\theta)$ and a new movement direction from its neighbors $N_{\text{movement}}(m)$. Given the new orientation $\theta$ and movement direction $m$, we created new stimulus images of orientation $\theta$ whose center locations changed by $m$ in every following frame. We repeated this process until we obtained a video of 40,000 frames. We assume that each frame is shown for $1$ second in an experiment. \\

\subsubsection{Simulated simple cells in V1}
\label{sec:V1_simulation}

We simulated V1 simple cells in response to the stimulus video using the linear-nonlinear-Poisson (LNP) model \cite{PANINSKI2007, doya2007bayesian}. A simple cell was first represented by some filter. We then computed the dot product between the stimulus and the filter, applied a non-linear function to obtain the firing rate, and simulated the spike trains via the Poisson process. We chose Gabor filters on a square torus as the filters, and we chose the hyperbolic tangent function as the nonlinear function. \\

\noindent\textbf{Modified Gabor filters}. 
Simple cells are commonly represented by Gabor filters \cite{abbottDayan, Jones1987AnEO}. A two-dimensional Gabor filter with orientation $\theta$ centered at $(x_0, y_0)$ is given by 
 \begin{equation}
\label{standard_Gabor}
G_{x_0, y_0, \theta}(x, y) = \exp\Big(-\frac{x'^2 + \gamma^2 y'^2}{2 \sigma^2} \Big) \cos \Big( 2 \pi \frac{x'}{\lambda }+ \phi \Big)
\end{equation}
where $x' = (x-x_0) \cos \theta + (y-y_0) \sin \theta$ and $y' = -(x-x_0) \sin \theta + (y-y_0) \cos \theta$. Here, $\lambda$ represents wavelength, $\sigma$ is the standard deviation of the Gaussian envelope, $\gamma$ is the spatial aspect ratio, and $\phi$ is the phase offset. 

We used a variation of the standard Gabor filters by creating Gabor filters that live on a square torus of size $s \times s$. We first created Gabor filters $G_{x_c, y_c, \theta}(x,y)$ centered at $(x_c = s/2, y_c = s/2)$ with fixed parameters $\lambda = 16$, $\phi = \pi$, $\sigma = 5$ and $\gamma = 1$. Note that $\lambda$ and $\phi$ were chosen to be identical to the stimulus parameters. We defined the Gabor filter on a torus with orientation $\theta$ centered at $(x_0, y_0)$ as the following
\[G^{\text{T}}_{x_0, y_0, \theta}(x, y) = G_{x_c, y_, \theta}(x - d_x \mod s, \, y - d_y \mod s)  \]
where $d_x = x_0-x_c$ and $d_y = y_0 - y_c$. That is, the Gabor filter on a torus with center $(x_0, y_0)$ is obtained by moving the center location of the Gabor filter $G_{x_c, y_c, \theta}$ from $(x_c, y_c)$ to $(x_0, y_0)$ while maintaining the properties of the square torus. The final filter we used is the negative of the real part of $G^T$: 
\begin{equation}
\label{eq:V1_filters}
V_{x_0, y_0, \theta}(x,y) = - \text{Re}(G^{\text{T}}_{x_0, y_0, \theta}(x, y) ). 
\end{equation}

\begin{figure}[H]
\centering
\includegraphics[width=0.9\textwidth]{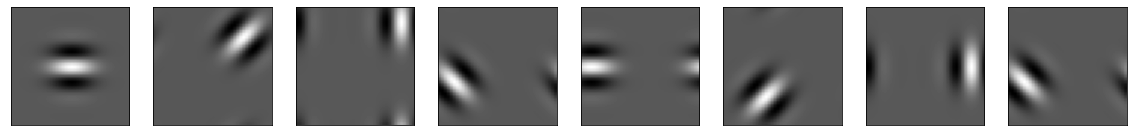}
\caption{Example modified Gabor filters on a torus.}
\label{fig:example_Gabor_filters}
\end{figure}

We generated a total of 800 filters $V_{x_0, y_0, \theta}$ with varying orientations and center locations. The orientations $\theta$ could be one of the following eight values $\{ 0, \pi/8, \dots, 7 \pi/ 8 \}$, the same set of orientations as the stimulus images. For each of the eight orientations, we sampled 100 filters on a torus with random centers. The center locations $x_0$ and $y_0$ could vary between $0$ and $s$. See Figure \ref{fig:example_Gabor_filters} for examples.

\noindent\textbf{Firing rates.}
Recall that we simulated the simple cells via a linear-nonlinear-Poisson model. For each filter $V_{x_0, y_0, \theta}$, we computed its dot product with each of the $40,000$ stimulus images. %Figure \ref{fig:histogram_dotproduct} shows the histogram of the dot products. 

The non-linear functions convert the dot product into firing rates. Some common choices include the rectified linear function, sigmoidal function, hyperbolic tangent, and Heeger normalization \cite{Heeger1992NormalizationOC, abbottDayan}. We used the hyperbolic tangent function
\[ F(L) = r_0 + r_{max} [\tanh(g_2(L-L_0))]_+\] 
where $L$ is the value of the dot product, $L_0$ is the threshold, $r_0$ is the baseline firing rate, and $r_{max}$ is the above-baseline maximum firing rate, and $g_2$ determines how rapidly the firing rate increases. The firing rates of simple cells vary widely across experiments and across animals \cite{Chance4785, Niell7520, Carandini470, Prince2002QuantitativeAO}.  We fixed values of $r_0 = 1$, $L_0 = 10$, $g_2 = 0.05$, and $r_{\max} = 30$. Once we applied the non-linear function, we ended up with the firing rates of 800 simulated simple cells in response to each of the 40,000 stimulus images. \\

\noindent\textbf{Spike train simulation.}
Given a time-varying firing rate $\theta = \{ \lambda_t \}$ of a single neuron where $\lambda_t$ denotes the firing rate at time $t$, we generated the spike train via the inhomogeneous Poisson process \cite{Shlens2014NotesOG}. Consider the spike train $Y$ as a vector of spike counts $\{y_t\}$ binned at a time resolution $\Delta$. We fixed $\Delta = 0.04$ s \footnote{Recall that the simulation video was designed to show each static image for 1 second. }.
Let $y_t$ denote the number of spike counts at bin $t$. The likelihood of the spike train $Y$ given the firing rate $\theta$ is 
\[P(Y | \theta) = \prod_t \frac{(\lambda_t \Delta)^{y_t}}{y_t!}  \exp(- \lambda_t \Delta).\]

We sampled from the inhomogeneous Poisson process using \cite{tick_Python}, which results in a vector $\vec{y}$ of spike counts for each bin. We converted into a binary spike train $\vec{x}$ where 
\begin{equation*}
\vec{x}_t = 
\begin{cases}
1 \quad \text{ if } \quad  \vec{y}_t > 0, \\
0 \quad \text{ if } \quad \vec{y}_t = 0. \\
\end{cases}
\end{equation*}
The binary spike train $\vec{x}$ had size $100,000$. 

We repeated the spike train simulation for all $800$ simple cells. As a result, we ended up with a binary raster of size $100,000$ by $800$, where each bin corresponded to $0.04$ seconds. \\

\subsubsection{Orientation cells}
\label{data:orientation}
A standard process for simulating orientation-sensitive neurons is to specify the tuning curve, compute the firing rate, and sample spike times via inhomogeneous Poisson process. Our goal is to model a collection of orientation cells and the propagation of neural encoding from simple cells to orientation cells. We thus represented the orientation cells via a function $\fOri$ whose input is the binary spike train of simple cell neural population and whose output is the firing rates of the orientation cells. The process is broken into the following three steps: 
\begin{enumerate}
\item specify tuning curves of orientation cells
\item compute firing rates of orientation cells by approximating the function $\fOri$ using a trained neural network, and
\item simulate spike trains via inhomogeneous Poisson process.
\end{enumerate}

\noindent\textbf{Tuning curves.}
The orientation cells were specified by their preferred firing orientation and their tuning curve. We sampled 64 preferred orientations from $[0, \pi]$. We specified the tuning curve using the Gaussian distribution wrapped on a circle \cite{orientation_direction_selectivity} given by
\begin{equation}
\label{eq:tuning_curve}
T_{\theta_{P}}(\theta) = C + R_P \exp \Big( - \frac{d_\pi(\theta, \theta_{P})^2}{2 \sigma^2}\Big),
\end{equation}
where $C$ represents the baseline firing rate, $R_P$ is the above-baseline rate at the preferred orientation, $\theta$ is the orientation of the stimulus, $\theta_{P}$ is the preferred orientation of the neuron, $d_{\pi}(\theta, \theta_{P})$ is the circular distance between $\theta$ and $\theta_{P}$ (Equation \ref{eq:angular_distance}), and $\sigma$ is the tuning width. We fixed the parameters at $C = 0.25$ Hz, $R_P = 2$ Hz, and $\sigma = 0.2$. %Figure \ref{fig:tuning_curve} shows an example tuning curve. 
Equation \ref{eq:tuning_curve} provides the ``ground-truth'' firing rates of the orientation cells and it is a variation of the von Mises distribution on a circle \cite{Swindale1998}.  \\
%\IY{I didn't think it was necessary to include an example plot of a tuning curve, but we can add if needed}  

\noindent\textbf{Firing rates.} We modeled neural propagation from V1 simple cells to orientation cells via a function $\fOri$ that takes as input the binary vector representing the firing of V1 simple cells and outputs the firing rates of orientation-sensitive neurons. To be precise, recall that there were 800 simulated simple cells and 64 simulated orientation-sensitive neurons. The input of $\fOri$ was a binary vector of size 800 whose $i^{\text{th}}$ entry is 1 if the $i^{\text{th}}$ simple cell fired and 0 otherwise. The output was a vector of size 64 whose $i^{\text{th}}$ entry indicated the firing rate of the $i^{\text{th}}$ orientation-sensitive neurons. 

We approximated $\fOri$ using a fully-connected three-layer feed-forward neural network. The input layer had 800 nodes, the hidden layer had 128 nodes, and the output layer had 64 nodes. To create the training set, we created a training stimulus video and simulated the simple cell spike trains in response to this stimulus video as in SI Section \ref{sec:stimulus_generation}. Using the tuning curve in Equation \ref{eq:tuning_curve}, we computed the ground-truth firing rates of the orientation-sensitive neurons in response to the training video. We used the simple cell spike trains and the ground-truth firing rates as the training data. We trained the neural network using mean squared error as the loss function. At the end of the training, the mean squared error on the training set was 0.045, and the mean squared error on the test set was 0.047. 

Let $\fOriApprox$ denote the function represented by the trained neural network. During simulation, we considered the output of $\fOriApprox$ as the firing rates of orientation cells. Note that by design, the firing rates were driven by the spike trains of the simple cells. 

\noindent\textbf{Spike train simulation.} Once we approximated $\fOriApprox$, we simulated the spike trains of orientation-sensitive neurons as the following. We computed the spikes trains of simulated simple cells and used it as input to $\fOriApprox$. The output, $\tilde{T}_{\theta_P}$ is the approximated firing rate for each orientation-sensitive neuron at every time point. We then constructed the firing rate of each neuron by adding Gaussian noise and rectifying the negative values to zero as following \cite{Butts2006TuningCN}
\begin{equation}
\label{eq:TC_to_rate}
r_{\theta_P}(\theta) = [\tilde{T}_{\theta_P}(\theta) + \eta_{\sigma}]_{+},
\end{equation}
where $\eta$ represents the Gaussian noise with zero mean and standard deviation $\sigma$, and $[ \quad ]_+$ represents the rectification of negative values to zero. 

% noise from reliability
% function: `add_unreliable_intervals`
We added another layer of noise by introducing intervals in which the simulated neurons fire at baseline firing rate. For each non-overlapping unit of interval $U$, the neuron fired at the predicted rate $r_{\theta_P}(\theta)$ with probability $p$ and at baseline firing rate $C$ with probability $1-p$. Here, we used $U =$ 7 frames $= 7 $ seconds, $p = 0.8$, and $C = 0.25$ Hz. See SI section \ref{Supp:experiments_messy_barcodes} for a discussion of how the parameters in the simulation of orientation-sensitive neurons affect the persistence diagrams. 

\begin{figure}[H]
\centering
\includegraphics[width=0.9\textwidth]{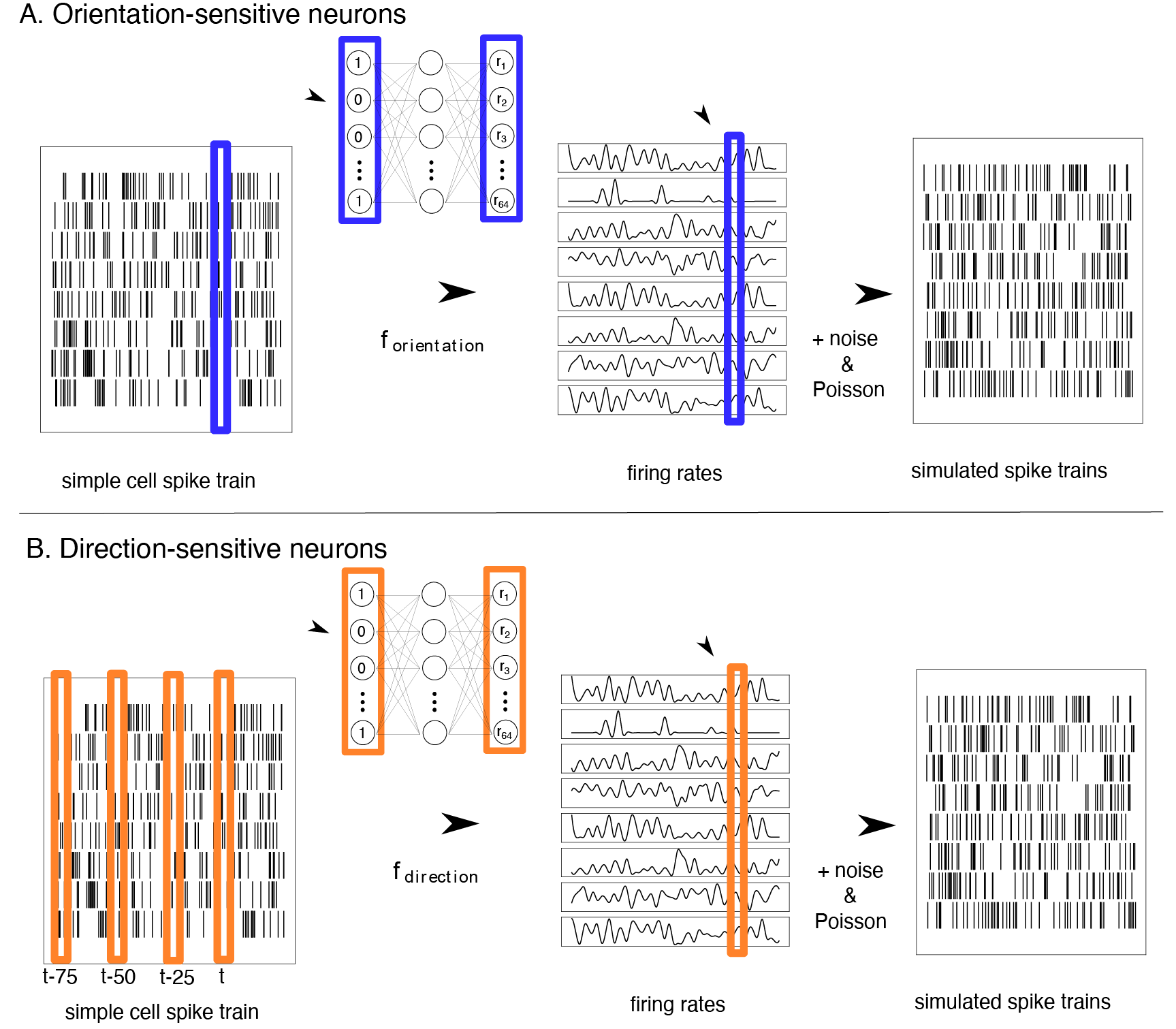}
\caption{Spike train simulation of orientation and direction cells. \textbf{A.} A feed-forward neural network predicts the firing rates of orientation cells at bin $t$ from the simple cell spike trains at bin $t$. We then use the inhomogeneous Poisson process to simulate the spike trains.
\textbf{B.} A neural network predicts the firing rates of direction cells at bin $t$ from the simple spike trains at bins $t, t-25, t-50$, and $t-75$. Spike trains are simulated via an inhomogeneous Poisson process. } % For direction cells, it is taking as input the spike trains at 4 different points, but not consecutive. It's taking the spike trains at a given frame, and the 3 preceding frames (so 25 bins ahead). 
\label{fig:downstream_neurons_NN}
\end{figure}

Once we computed the firing rates, we then simulated the spike trains via inhomogeneous Poisson process. Figure \ref{fig:downstream_neurons_NN}A summarizes the simulation process.

%---------------------------------------------------
\subsubsection{Direction cells}
\label{data:direction}
We simulated the direction-sensitive neurons in a manner similar to the orientation cells with minor adjustments in the design of the neural network. In contrast to the orientation-sensitive neurons, direction-sensitive neurons need to observe the behavior of simple cells over an interval of time in order to compute the movement direction of a stimulus during that interval. Thus, the firing rates of direction cells were computed from a neural network that takes as input the spike trains of simple cells over some specified interval. Again, the process is broken into the following three steps: 
\begin{enumerate}
    \item specify tuning curves of direction cells
    \item compute firing rates by approximating the function $\fDir$ using a neural network, and
    \item simulate spike trains via inhomogeneous Poisson process.
\end{enumerate}

\noindent\textbf{Tuning curves.} The direction-sensitive neurons were specified by their preferred movement direction and their tuning curve. We sampled 64 movement directions and specified the tuning curve using Equation \ref{eq:tuning_curve} with parameters $C= 0.25, R_P = 2$, and $\sigma = 0.4$. The tuning curve provided the ground-truth firing rates of the direction cells. 
% for the training data, the direction value was determined at a single point (not an average in the preceding interval) 

\noindent\textbf{Firing rates.} The firing rates of direction cells were modeled via a function $\fDir$ whose input is the spike trains of simple cells. Since the movement direction is a dynamic feature, the direction cells needed access to the simple cell spike trains over multiple time points in order to calculate the movement direction. We modeled $\fDir$ using a neural network that takes as input the concatenation of binary vectors of simple cell spike trains at bins $t$, $t-25$, $t-50$, and $t-75$\footnote{Ideally, the input would be a concatenation of spike trains at all bins for an interval of time leading up to $t$. This is computationally infeasible}. Since there are 800 V1 neurons, the input of $\fDir$ was a vector of size 3200. 

We approximated $\fDir$ using a fully-connected three-layer feed-forward neural network. The input layer had 3200 nodes, the hidden layer had 128 nodes, and the output layer had 64 nodes. We used the same training set stimulus video and spike trains as that of the orientation cells. At the end of the training, the mean squared error on the training set was 0.036 and the mean squared error on the test set was 0.043.

\noindent\textbf{Spike train simulation.} Given the firing rate of a direction-sensitive neuron, we follow the ``spike train simulation" outlined in SI Section \ref{data:orientation} to simulate the spike trains of direction-sensitive neurons. Figure \ref{fig:downstream_neurons_NN}B illustrates the simulation of direction-sensitive neurons.

%==============================================
\subsection{Simulated navigational system}
\label{SI:simulated_navigation}
%==============================================
%\IY{note! Here we use rates instead of spike trains, I think. Double check the data: Yes. We use firing rates here, and not spike trains to compute dissimilarity. Argument: Method works for both firing rates and spike trains}
We simulated grid, head-direction, and conjunctive cells as follows. We simulated grid cells using the data and methods from \cite{GridCellTorus}. In \cite{GridCellTorus}, authors performed an experiment in which rats engaged in foraging behavior in a square open-field enclosure. To simulate the grid cells, they take the first 1,000 seconds of the recorded trajectory (of rat R). While the original was sampled at 10 ms, they interpolated to 2-ms time steps. The authors then generated a 56 $\times$ 44 grid-cell network based on the continuous attractor networks (CAN) model. In particular, they use only the lateral inhibition in the connectivity matrix as described in \cite{CoueySimulation}. The activity $\vec{s}$ is updated as 

\[ \vec{s}_{i+1} = \vec{s}_i + \frac{1}{\tau} (- \vec{s}_i + (l + \vec{s}_i \cdot W + \alpha v(t) \cos(\theta(t) - \tilde{\theta}))_+) \]
using parameters $I = 1$, $\alpha = 0.15$, $l =2$, $W_0 = -0.01$, $R = 20$, and $\tau = 10$. They start with random activity and perform 2,000 updates, allowing the activity pattern to stabilize. For computational reasons, activity was set to 0 if $\vec{s}_i < 0.0001$, and the simulation was down-sampled to keep only every $5^{\text{th}}$ time frame. %This is the downsampling done in GridCellTorus, and we take their code for simulations. 
We refer the readers to \cite{GridCellTorus, CoueySimulation} for details. Using the same trajectory data and methods, we simulated the firing rates of 200 grid cells. The result is a matrix of size 200 by 598,999, where 598,999 is the number of timebins. For computation reasons, we limit to the first 100,000 timebins. 
%\IY{What is the output? Grid cell firing rates? for how many time bins? what is the size of the matrix here}

Next, the firing rates of 200 head-direction (HD) cells were first simulated using tuning curves 
\begin{equation}
\label{eq:HD_sim}
T_{\theta_{P}}(\theta) = C + R_P \exp \Big( - \frac{d_\pi(\theta, \theta_{P})^2}{2 \sigma^2}\Big)
\end{equation}
with parameters $R_P = 0.4$, $\sigma = 0.4$, $C = 0$. The 200 preferred directions were sampled randomly.

% unreliability - intervals of a spike train at which firing rates are forced to be at baseline. % maybe call these "unreliable intervals" 
We then introduced another layer of noise by forcing the neurons to fire at baseline firing rates at random intervals. For each non-overlapping interval of length $U$ timebins, the neuron fires at the predicted rate $T_{\theta_P}(\theta)$ with probability $p$ and at baseline firing rate $C$ with probability $1-p$. Here, we used $U =100$ timebins, $p = 0.5$, and $C = 0$ Hz. See SI section \ref{Supp:experiments_messy_barcodes} for a discussion of how the parameters in the simulation of orientation-sensitive neurons affect the persistence diagrams.

%We then added unreliable intervals to the simulated firing rates of HD cells in order to introduce noise to the persistence diagrams. \IY{Need equation describing unreliability. State parameters used}

Lastly, we simulated 800 conjunctive cells. We first simulated 800 pure grid cells as described above. We then simulated 800 head-direction cells using the tuning curve in Equation \ref{eq:HD_sim} with the same parameters. We computed the firing rates of the $i^{\text{th}}$ conjunctive cell as the minimum of the firing rate of the $i^{\text{th}}$ grid cell and the $i^{\text{th}}$ head-direction cell at each time point.

%==============================================
\subsection{Experimental data}
\label{sec:data_experimental}

The experimental data was collected as described in \textit{Materials and Methods} in the main text.

\subsubsection{Spike train preprocessing for experimental data}
\label{sec:spike_train_preprocessing}

Experimental spike trains often contain spike trains that are rarely firing and those that are uniformly firing. While topological methods are robust to noise, such outlier spike trains can obfuscate the output of topological methods. We thus preprocess the experimental data to select neurons with reliable spike patterns. 

%Note to self: in UCSB Smith data, we are reporting "results 2". 
\noindent\textbf{i. Conversion to binned spike train.}

The experiment results in a list of spike times for neurons. For each neuron, we convert the spike times into a binned spike train as follows. We first partitioned the duration of the experiment (32 seconds) into 426 bins of equal length. Each bin then corresponds to $32/426 \approx 0.0751$ seconds.

%IS this necessary?: Each stimuli direction was presented for $4$ seconds, which corresponds to $53.25$ bins.    

For a given neuron, its binned spike train is a vector of length $426$, whose $i^{\text{th}}$ entry is the number of spikes at the $i^{\text{th}}$ bin. This resulted in 352 binned spike trains in V1 and 163 binned spike trains in AL.  \\

\noindent\textbf{ii. Neuron selection by consistency across trials} % originally, I called this unreliability. Changed to "consistency" 

We selected neurons that fired consistently across the 20 trials. Given a neuron, let $\vec{x}_i$ denote the spike train from trial $i$. Recall from Equation \ref{eq:spiketrain_similarity} the similarity between $\vec{x}_i$ and $\vec{x}_j$. 

For each neuron and for each pair of repeats, we computed the similarity using a shift parameter of 25 bins\footnote{The shift parameter corresponds to approximately 1.9 seconds. Recall that the stimuli consisted of various grating directions, and each direction was presented for 4 seconds.}. We defined a neuron's reliability as the average of such similarity over all pairs of 20 trials
\[\sum_{i \leq 20} \sum_{i < j \leq 20} Sim_{25}(\vec{x}_i, \vec{x}_j).\]

\begin{figure}[h]
\centering
\includegraphics[width=1\textwidth]{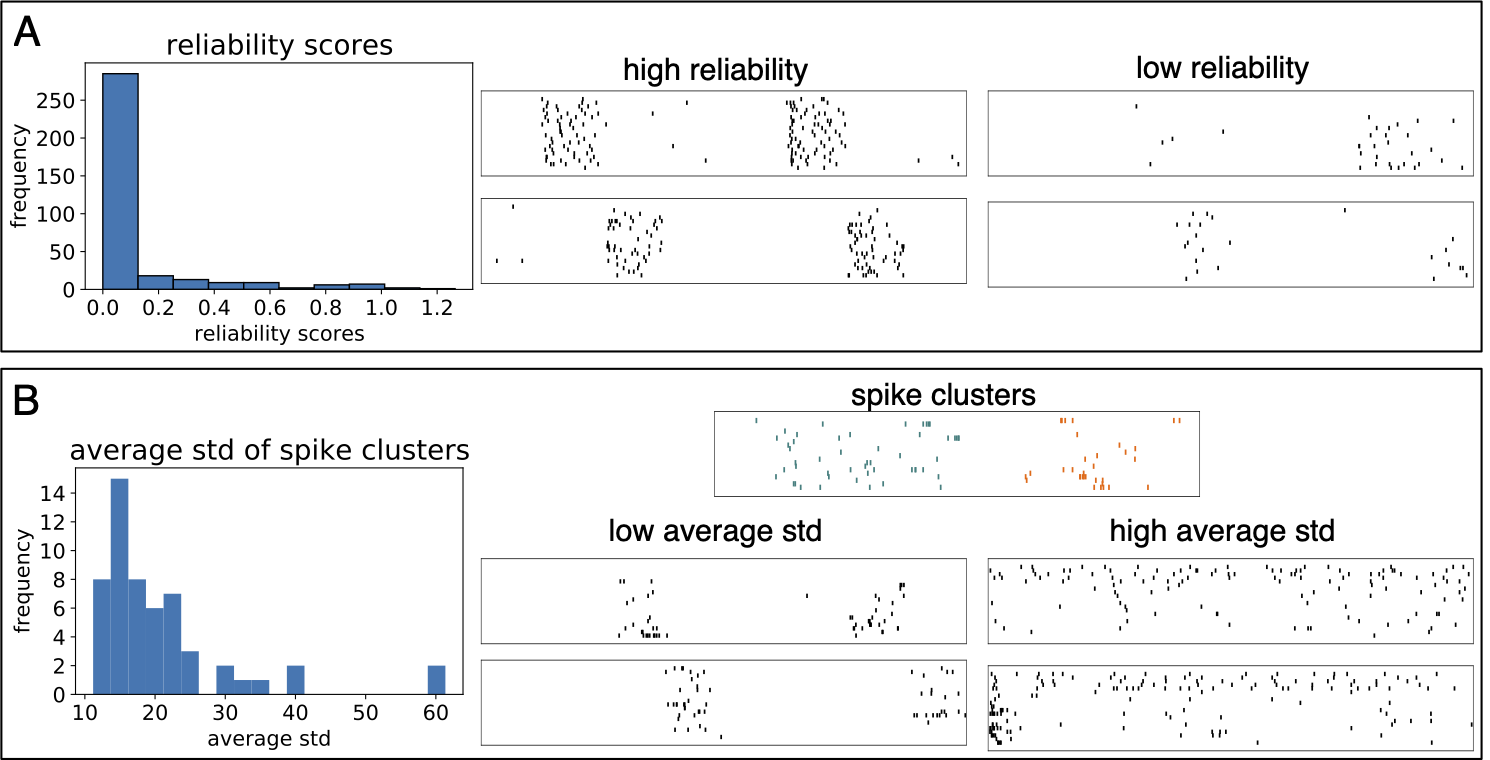}
\caption{Illustration of the preprocessing steps on V1 neurons. \textbf{A} (Left) Histogram of reliability scores of V1 neurons. (Right) Example neurons with high and low reliability scores. \textbf{B}. (Top) Example clusters of spike times. The green spikes form one cluster and the orange spikes form another cluster. (Left) Histogram of the average standard deviation of spike clusters. (Right) Example neurons with high and low average standard deviations. }
\label{fig:UCSB_V1_preprocess}
\end{figure}

% Selecting reliability
For each system, say V1, we computed the reliability scores of each neuron. We then used the histogram of the reliability scores to determine the minimum reliability scores required for each neuron. Figure \ref{fig:UCSB_V1_preprocess} A illustrates the histogram for V1 neurons and example neurons with high and low reliability scores. We chose a cutoff value of $0.2$ and selected neurons that whose reliability scores were higher than $0.2$. A total of 55 neurons were selected.

We performed a similar selection process for AL neurons. A total of 44 AL neurons were selected. \\

\noindent\textbf{iii. Neuron selection by the average standard deviation of spike clusters}

We then removed neurons that had a tendency to fire uniformly throughout the experiment. We measured such uniform firing behaviors via the average standard deviation of spike clusters explained below. 

For each neuron, we clustered its spike times using DBSCAN \cite{DBSCAN, scikit-learn} with parameters `eps' = 40 s and `min\_samples'  = 20\footnote{`eps'=20 indicates that two spikes must occur within 40 seconds of each other to be considered as being in the same neighborhood. `min\_samples' = 20 indicates that a spike must have at least 20 nearby spikes in order to be considered a core point.}. Figure \ref{fig:UCSB_V1_preprocess}B shows an example raster with spikes clustered. 

For each cluster, we computed the standard deviation of the spike times for all spikes belonging to that cluster. We then computed the average of such standard deviation over all clusters. Neurons with low average standard deviation have well-defined spike time clusters, while neurons with high average standard deviation tend to have uniform spike times. Figure \ref{fig:UCSB_V1_preprocess}B provides example neurons with high and low average standard deviations. 

For each system, we computed the average standard deviation of spike clusters for each neuron. We then examined their histogram to decide on a threshold for the average standard deviation score. Figure \ref{fig:UCSB_V1_preprocess}B illustrates the histogram for V1 neurons. We decided to exclude two neurons whose average standard deviation score was above 55, resulting in a total of 53 V1 neurons. We performed a similar process for the AL neurons, which resulted in 41 neurons.

%===============================================================
\section{Supplementary Experiments}

\subsection{Experiments for identifying significant points on a persistence diagrams}
\label{sec:significance_pd}
We present a variety of methods that can be used to determine the significant points on a persistence diagram. We implemented the methods on simulated orientation cells (SI Section \ref{data:orientation}) and summarize the output in Fig. \ref{fig:significance_experiments}. 

\subsubsection*{Features of points}
Given a persistence diagram, we consider three features that measure the ``importance" of a point. Let $p$ denote a point in a persistence diagram $\pd$ with birth parameter $b$ and death parameter $d$. The three possible features of $p$ are:. 
\begin{enumerate}
\item \textbf{lifetime}: $d-b$.
\item \textbf{relative lifetime}: $\frac{d-b}{m}$, where $m$ is the median lifetime of all points in $\pd$.
\item \textbf{persistence ratio}: $\frac{d}{b}$. 
\end{enumerate}

\subsubsection*{Methods} We experiment with seven different ways of identifying significant points on a persistence diagram. The first three methods are applied to the three features listed above. 
\begin{figure}[h!]
\centering
\includegraphics[width=0.63\textwidth]{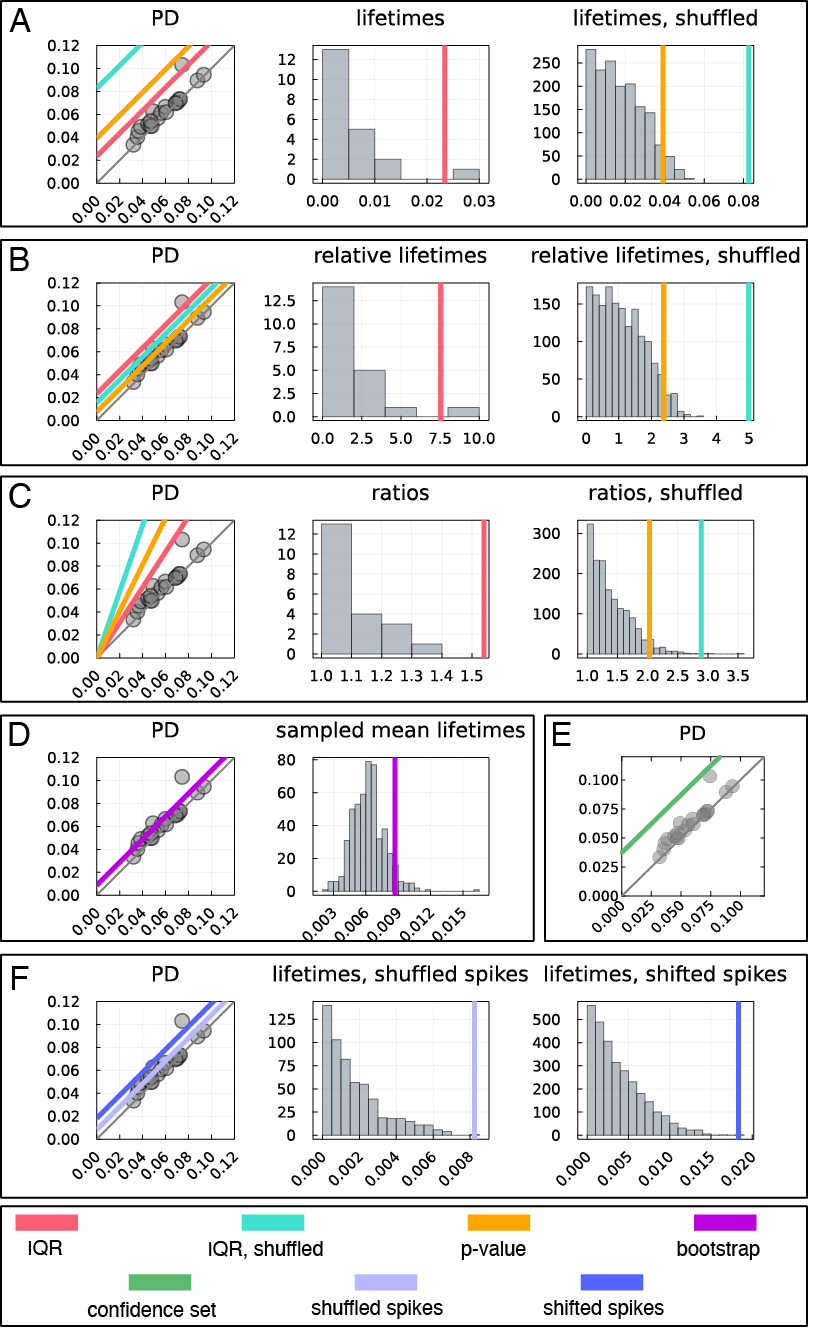}
\caption{A comparison of the significance thresholds obtained from various methods and features. \textbf{A.} Significance thresholds using lifetime features. (left) Three thresholds from IQR (red), shuffled IQR (teal), and $p$-value (orange) methods are shown on a persistence diagram. (center) Histogram of lifetimes of points in $\pd$ and the threshold from IQR method. (right) Histogram of lifetimes of points in $\pd$ from shuffled dissimilarity matrices. The orange and teal vertical lines indicate the significance threshold. \textbf{B.} Significance thresholds using relative lifetime features. (left) Three thresholds from IQR (red), shuffled IQR (teal), and $p$-value (orange) methods are shown on a persistence diagram. (center) Histogram of relative lifetimes from PD and the threshold from IQR method. (right) Histogram of relative lifetimes from shuffled dissimilarity matrix and significance thresholds from shuffled IQR and $p$-value methods. \textbf{C.} Significance thresholds using persistence ratio features. (left) Three thresholds from IQR (red), shuffled IQR (teal), and $p$-value (orange) methods are shown on a persistence diagram. (center) Histogram of persistence ratios from PD and the threshold from IQR method. (right) Histogram of persistence ratios from shuffled dissimilarity matrix and significance thresholds from shuffled IQR and $p$-value methods. \textbf{D.} (left) Threshold from bootstrap method. (right) Histogram of mean lifetimes and threshold. \textbf{E.} Significance threshold from confidence sets.  \textbf{F.} Significance thresholds from shuffled (lavender) and shifted (navy) spike trains. (left) Significance thresholds shown on $\pd$. (center) Histogram of lifetimes from shuffled spike trains. (right) Histogram of lifetimes from shifted spike trains. }
\label{fig:significance_experiments}
\end{figure}

\begin{enumerate}
\item\textbf{IQR}: We use the interquartile range (IQR) of features to identify the significant points (SI Section \ref{SI:significant_PD}). The significance threshold is shown in red in Fig. \ref{fig:significance_experiments}. 

\item\textbf{IQR from shuffled dissimilarity matrix}: We shuffle the entries of the dissimilarity matrix while maintaining the symmetry. We perform such shuffling 10 times and obtain an empirical distribution of the features. We then compute the IQR. Any point whose feature lies above the IQR (with $k=3$) is considered significant. The significance threshold is shown in teal in Fig. \ref{fig:significance_experiments}. % PD_significance.jl select_from_shuffled_IQR

\item\textbf{$p$-value from shuffled dissimilarity matrix}: Similarly to the above method, we shuffle the entries of the dissimilarity matrix 10 times while maintaining symmetry. For we shuffled matrix, we compute the persistence diagram and extract the three features. From the empirical distribution of the features, we selected a threshold $t$ such that $(1-\alpha)$ percent of the features are below $t$. We chose $\alpha = 0.05$. The significance threshold is shown in orange in Fig. \ref{fig:significance_experiments}. 
% PD_significance.jl select_from_shuffled_dissimilarity

\item\textbf{bootstrap}: Given a dissimilarity matrix $D$ of size $n$, we sample $n$ rows and columns with replacement. Let $D_b$ be the resulting dissimilarity matrix. We compute the persistence diagram $\pd(D_b)$ and compute the mean lifetime of points in $\pd(D_b)$. We repeat the process $5,000$ times and obtain an empirical distribution of the mean lifetimes (see Fig. \ref{fig:significance_experiments}D, right). We then select a threshold $t$ such that $(1-\alpha)$ percent of the mean lifetimes are below $t$. We chose $\alpha =0.001$. The significance threshold is shown in Fig. \ref{fig:significance_experiments}D.

\item\textbf{confidence sets / topological bootstrapping}: We identified the significant points on a persistence diagram using the confidence sets (Method I: subsampling in \cite{confidencesetsPD}). The method was later refined in {\cite{Chazal_topological_inference}} and was called "Bottleneck bootstrap method". We implemented the refined version in \cite{Chazal_topological_inference}. As mentioned in the original paper, the method is very conservative (Fig. \ref{fig:significance_experiments}E). Note that the method only considers the lifetime feature of a point.  

\item\textbf{shuffled spike trains}: The following two methods are specific to persistence diagrams computed from dissimilarity matrices among spike trains. The first method is motivated by \cite{GridCellTorus}. Given a collection of spike trains, we randomly shuffle the spike times of every neuron. We compute the dissimilarity matrix among the shuffled spike trains, compute the persistence diagram $\pd$, and compute the lifetime feature of every point on the $\pd$. We repeat this process 50 times to get an empirical distribution of the lifetimes from shuffled spike trains (Fig. \ref{fig:significance_experiments}F center). We then use the maximum lifetime observed as the threshold for determining significant points on the original persistence diagram (Fig. \ref{fig:significance_experiments}F, left in lavender).

% in notebook "analysis/Experiment_shuffle_spike_times.ipynb" need to move to experiments

\item\textbf{shifted spike trains}: Given a collection of spike trains, instead of randomly shuffling the spike times, we shifted the entire spike train by a random time interval. Given a spike train of length $\ell$, we randomly sampled some shift length $s$ and delayed every spike time by $s$. That is, a spike that originally occurred at time $t$ now occurs at time $t + s$. If $t + s \geq \ell$, then the new spike occurs at $t + s \mod \ell$. From the collection of shifted spike trains, we compute the dissimilarity matrix, compute the persistence diagram, and compute the lifetime features of every point on the $\pd$. We repeat this process $50$ times to get an empirical distribution of the lifetimes from the shifted spike trains (Fig. \ref{fig:significance_experiments}F, right). We then use the maximum lifetime observed as the threshold for determining significant points on the original persistence diagram (Fig. \ref{fig:significance_experiments}F, left in navy).
\end{enumerate}

Figure \ref{fig:significance_experiments} summarizes the result of implementing the above methods on the simulated orientation cells (SI Section \ref{data:orientation}). For the analysis of this paper, we chose the IQR method on lifetime features due to its simplicity and computational speed.

%-----------------------------------------------------------------

\subsection{Identifying significant points on persistence diagrams with few points}

Since the persistence diagrams $\pd(V1)$ and $\pd(AL)$ contain only a few points (see Fig.~4C of main text), we cannot use the proposed non-parametric statistical test (see \textit{Materials} \textit{and} \textit{Methods}) to distinguish significant features from noise. In such situations, we use randomly generated spike trains to identify the significant points on a persistence diagram. We discuss the details in this section (SI Section~{\ref{sec:significant_points_random}}). 
%In the main text, we make the conservative choice to consider all points on the persistence diagrams as candidate matches in the analogous cycles method. We emphasize that bypassing the test for significant points on the persistence diagram doesn't imply that all points should be considered as true topological features of a neural manifold. Rather, we lack the statistical power to \emph{a} \emph{priori} reject their significance, and will require \emph{post} \emph{hoc} investigation of matches to interpret their meaning in terms of the underlying topology. 

There is an alternative statistical test, called topological bootstrapping, that can be used for persistence diagrams consisting only of a few points. However, we caution the reader that this method can lead to a high threshold for significant points, and such thresholds can be especially problematic in experimental data where the topological structure may be difficult to detect and thus appear to have short lifetimes. In the interest of completeness, we include a discussion of topological bootstrapping (SI Section~{\ref{sec:PI_significance_bootstrapping}}).

\subsubsection{Identifying significant points on persistence diagrams via random spike trains with matched statistics}
\label{sec:significant_points_random}

% "shuffled" data" 
\begin{figure}[h]
    \centering
    \includegraphics[width=0.9\linewidth]{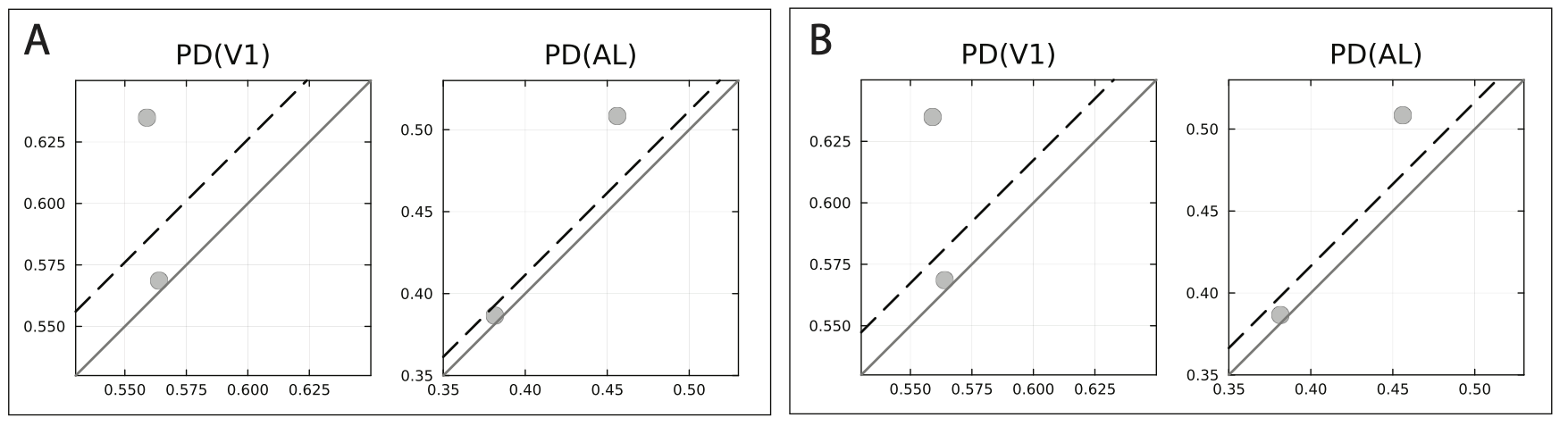}
    \caption{The persistence diagrams $\pd(V1)$ and $\pd(AL)$ with the significance threshold computed from random spike trains. \textbf{A.} The dotted lines indicate the significance thresholds computed from randomly shuffled aggregate spikes. \textbf{B.} The dotted lines indicate the significance thresholds computed from spike trains generated from homogeneous Poisson processes.}
    \label{fig:PD_significance_shuffled_spiketrains}
\end{figure}

In \cite{GridCellTorus}, the authors apply a version of a standard non-parametric test for significance of features in persistence diagrams. This test is performed by generating synthetic spike-train data matching spiking statistics of the target neural populations but discarding any intrinsic geometry, passing this data through the same persistent homology pipeline as the data, and using the empirical distribution of lifetimes in the resulting persistence diagrams to test for significance. We demonstrate this method using the V1 and AL data analyzed in the main text for Fig.~4.

% In \cite{GridCellTorus}, 
Such synthetic spike trains can be constructed by (1) binning the population spike trains in 0.0751s bins and randomly permuting their entries, or (2) simulating the spike trains via a homogeneous Poisson process whose rate matches the observed rates. Once the synthetic spike trains are generated, we compute dissimilarity using the same measure as for the experimental data, and we normalize by the maximum observed pairwise similarity $M$ (Equation~{\ref{eq:spike_train_dissimilarity}}). For consistency of scale of dissimilarity values, we take $M$ to be the maximum observed pairwise similarity value from the original data (the original $V1$ spike trains). We then use the dissimilarity matrices to compute the persistence diagrams. We repeat the process $n = 10,000$ times and use the maximum observed lifetime as the threshold for identifying significant points on the original persistence diagram. SI Figure~{\ref{fig:PD_significance_shuffled_spiketrains}} illustrate the significance thresholds obtained from shuffled spike trains (SI Fig.~{\ref{fig:PD_significance_shuffled_spiketrains}}A) and spike trains generated from homogeneous Poisson processes (SI Fig.~{\ref{fig:PD_significance_shuffled_spiketrains}}B).

\subsubsection{Identifying significant points on persistence diagrams via topological bootstrapping}
\label{sec:PI_significance_bootstrapping}

Topological bootstrapping,  initially introduced in {\cite{Fasy_confidence_set}} as a ``subsampling method" and was later refined in {\cite{Chazal_topological_inference}} as the ``bottleneck bootstrap method", is a bootstrap method for generating synthetic datasets, as follows:

\begin{enumerate}
    \item Compute the persistence diagram $\pd$ from the input dissimilarity matrix $D$.
    \item Sample from the original data $N$ times with replacement, i.e., sample the rows and columns of $D$ with replacement $N$ times. For each sample, compute the persistence diagram, which results in $N$ persistence diagrams $\pd_1, \pd_2, \dots, \pd_N$. 
    \item Compute the Bottleneck distance $d_{\infty}(\pd_i, \pd)$ between the sampled persistence diagram and the original persistence diagram. % Here, we only consider dimension 1.
    \item Let $c$ be the $1- \alpha$ percentile of the Bottleneck distances computed in step 3.
    \item Only points on the original persistence diagram $\pd$ whose lifetime is greater than $2c$ are considered significant. The significance threshold is indicated on $\pd$ by a band of width $2c$ above the diagonal line. 
\end{enumerate}

\begin{figure}[h!]
    \centering
    \includegraphics[width=8cm]{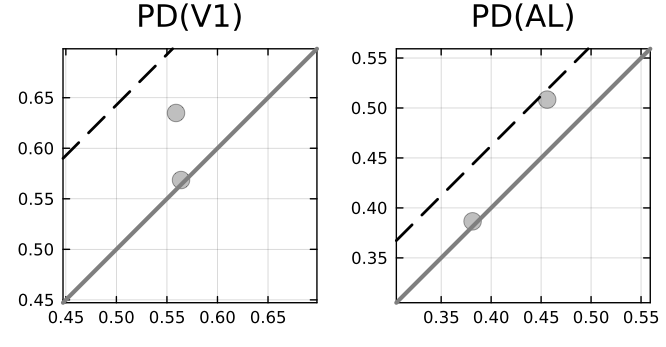}
    \caption{The persistence diagrams $\pd(V1)$ and $\pd(AL)$ with the significance threshold computed from topological bootstrapping. Note that the scales of the birth ($x$-axis) and the death ($y$-axis) parameters have been modified from SI Fig.~\ref{fig:PD_significance_shuffled_spiketrains} in order to show the threshold from bootstrapping. Bootstrapping provides a conservative threshold for separating signal from noise.}
    \label{fig:bootstrap_thresholds}
\end{figure}

We implemented the topological bootstrap method {\cite{Brown2024_TDApplied_R}} on the 53 VI neurons and 41 AL neurons studied in Fig~4 in the main text, using parameters $N = 1,000$ and $\alpha = 0.05$. As shown in Fig.~{\ref{fig:bootstrap_thresholds}}, the method resulted in thresholds for significance strong enough that all points on both persistence diagrams were rejected.

% threshold, for V1, according to bootstrapping: 0.1428
% threshold, for AL, according to bootstrapping: 0.0622

%Note that a common method of separating signal from noise in a persistence diagram uses kernel density estimation  

%-----------------------------------------------------------------
\subsection{Experiments with spike train dissimilarities}
\label{SI:dissimilarity_experiment}
%-----------------------------------------------------------------

In the remainder of this section, we 
show the robustness of the persistence diagram to the choice of the shift parameter $\ell$ (SI Section~\ref{sec:SI_timeseries_dissimilarity}), and we compare the windowed cross-correlation dissimilarity to other existing methods for computing spike train dissimilarity. We show that the windowed cross-correlation method is robust against the presence of unreliably-firing intervals. We also show that the windowed cross-correlation method assigns a high dissimilarity score whenever a sparse spike train is involved, which is a desirable trait for topological analysis. 

\subsubsection*{Varying the shift parameter}
\label{sec:displacement_limit}

The dissimilarity between time series in Equation \ref{eq:spike_train_dissimilarity} depends on the shift parameter $\ell$. The shift parameter represents the time scale at which one considers two time series as being similar. We performed an experiment illustrating how different choice of the shift parameter affects the persistent homology analysis. %Recall  that the shift limit parameter $\ell$ affects the dissimilarity between spike trains, which then affects the persistence diagram computed from the dissimilarity matrix. 

%In order to understand the effect of the shift parameter, we studied a simulated system for which we knew the time scale at which change occurred. \IY{language - "change occurred"? } We thus experimented with the simulated orientation cells (SI Section \ref{data:orientation}). 

Recall from the simulated visual system (SI Section \ref{sec:stimulus_generation}) that we designed a stimulus video that changed orientations every 7 frames, and we simulated spike trains of 64 orientation-sensitive neurons. In the simulation, every frame of the video corresponded to 25 bins in a spike train. Thus, the stimulus video changed its orientation every 175 bins, and the spike trains also exhibit changes in its firing rate every 175 bins. 

\begin{figure}[H]
\centering
\includegraphics[width=0.9\textwidth]{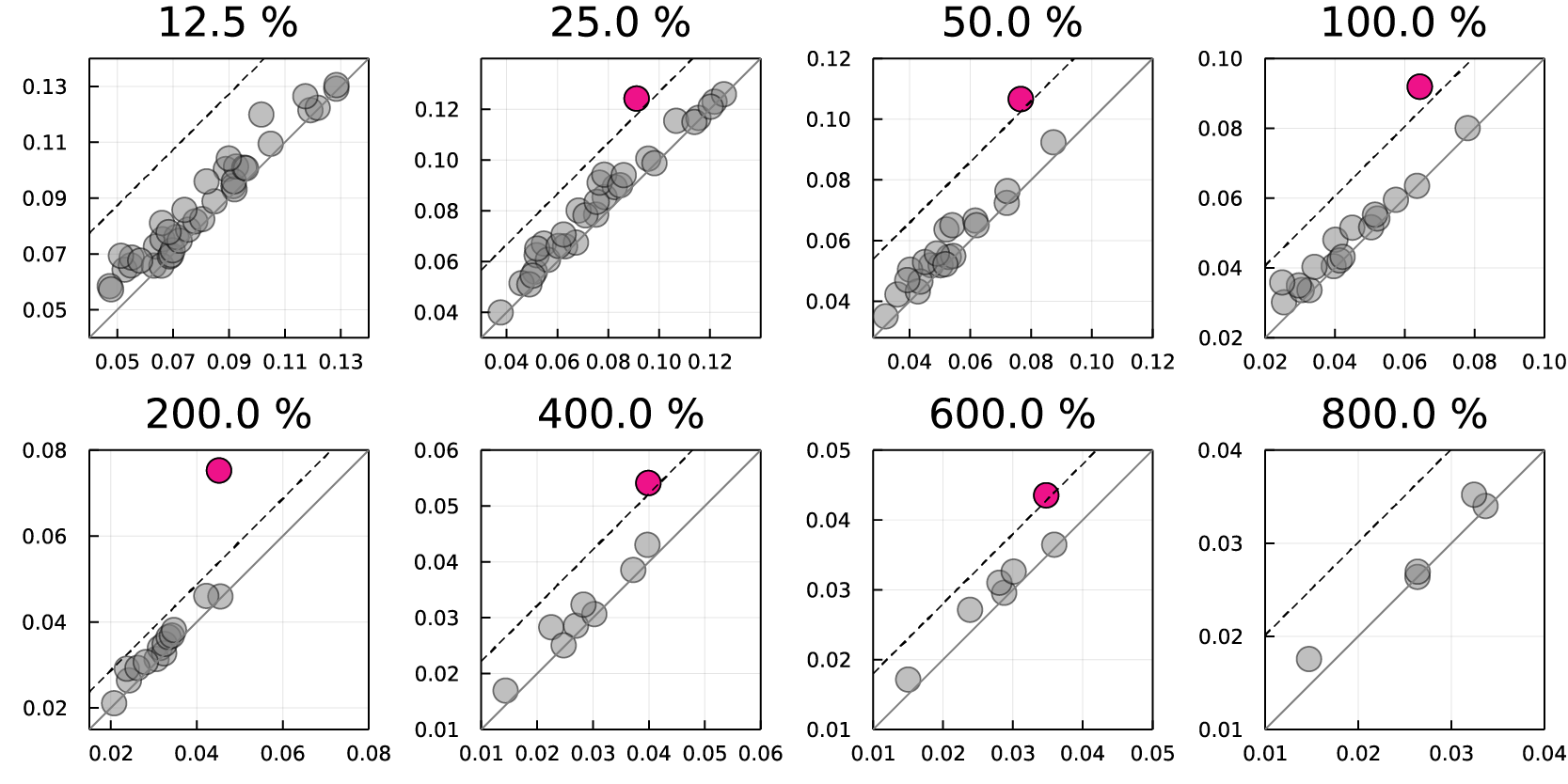}
\caption{Persistence diagrams computed from dissimilarity matrices obtained from varying shift parameter $\ell$. The percentage indicates the ratio of the shift parameter $\ell$ (in bins) to the unit of change of stimulus orientation (175 bins). The persistence diagrams capture the expected cyclicity for a range of shift parameters (from $25.0 \%$ to $600.0 \%$ of the change unit).}
\label{fig:displacement_limit_experiments}
\end{figure}

We computed the spike train dissimilarity matrices from the orientation neurons for the following range of shift parameters: $22, 44, 88, 175, 700, 1050, 1400$, which each corresponds to $12.5 \%, 25.0 \%, 50.0 \%, 100.0 \%, 200.0 \%$, $400.0 \%, 600.0 \%,$ and $800.0 \%$ of the unit of change in orientation (175 bins). We then computed the dimension-1 persistence diagrams. We expect the dimension-1 persistence diagram to contain one significant point which captures the orientation cyclicity. As summarized in Figure \ref{fig:displacement_limit_experiments}, the persistence diagrams capture the expected cyclicity for a range of shift parameters (from $25.0 \%$ to $600.0 \%$ of the change unit).  

\subsubsection*{Comparison to other spike train dissimilarities}
\label{sec:spiketrain_dissimilarities}

We compare the windowed cross-correlation dissimilarity to the Victor-Purpura distance \cite{VPdistance}, van Rossum distance \cite{VRdistance}, ISI-distance \cite{KreuzISI}, SPIKE-distance \cite{KreuzSPIKE}, and SPIKE-synchronization distance \cite{KreuzSPIKEsync}.

The Victor-Purpura distance between two spike trains measures the minimal cost of transforming one spike train to another via deletion, insertion, and shifting of spikes. The van Rossum distance first smooths the spike train by applying an exponentially decaying kernel to each spike. It then measures the resulting differences in the smoothed waveforms. Victor-Purpura distance and van Rossum distance are considered spike-resolved distances because they are based on matching spikes between two spike trains. 

While the Victor-Purpura distance and the van Rossum distances depend on a time scale parameter, there are a collection of time-resolved spike train distances that does not require a choice of the time parameter. The ISI-distance measures the dissmilarity between two spike trains by first computing the instantaneous rates from the inverse of the local interspike intervals (ISIs). It then computes the average of the rate dissimilarity over the total length of the spike train. Spike trains with similar firing rate profiles will be assigned a low dissimilarity. The SPIKE-distance incorporates both the local rate dissimilarity and the spike timing. SPIKE-synchronization measures similarities between spike trains based on the coincidence of spikes. It is easily converted to a distance. A detailed discussion of the above-mentioned spike train distances can be found at \cite{Satuvuori2018WhichST, Mulansky2015AGT}.

To compare the different dissimilarities, we computed six dissimilarity matrices on the simulated orientation neurons (SI section \ref{data:orientation}) based on the windowed cross-correlation distance, Victor-Pupura distance, van Rossum distance, ISI-distance, SPIKE-distance, and SPIKE-synchronization distance. We used Elephant \cite{elephant18} and PySpike \cite{Mulansky2016PySpikeA} to compute the dissimilarities. For each of the six dissimilarity matrix, we computed the persistence diagram in dimension 1. Since orientations are arranged in a circular fashion, we expect the persistence diagrams to consist of one significant point far from the diagonal. 

\begin{figure}[H]
\centering
\includegraphics[width=0.7\textwidth]{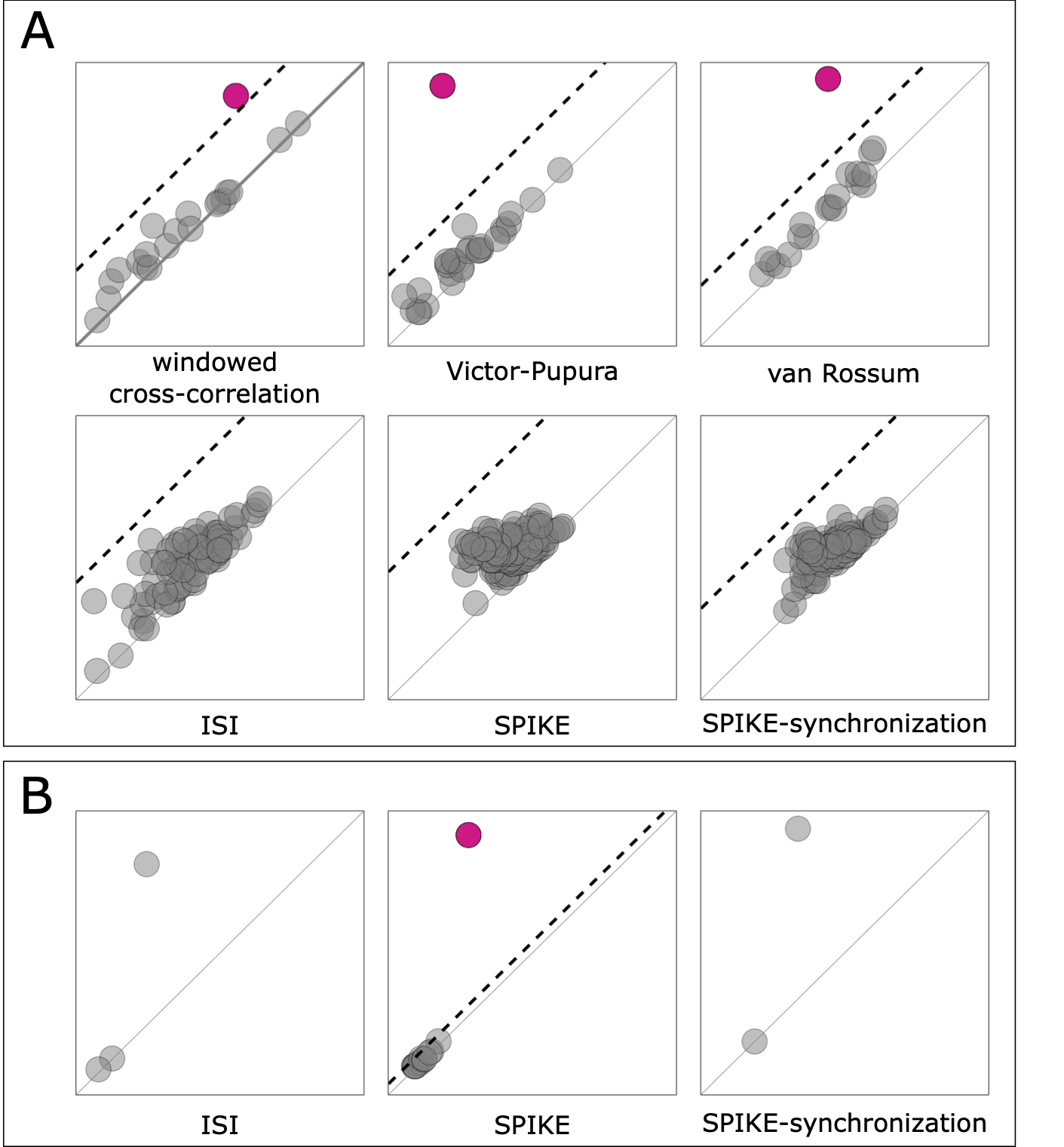}
\caption{A comparison of six spike train dissimilarities: windowed cross-correlation dissimilarity, Victor-Pupura distance, van Rossum distance, ISI-distance, SPIKE-distance, and SPIKE-synchronization distance. We computed six dissimilarity matrices of the simulated orientation-sensitive neurons using the six dissimilarity methods. A: Dimension-1 persistence diagrams of the six dissimilarity matrices. The persistence diagrams from the windowed cross-correlation dissimilarity, Victor-Pupura distance, and van Rossum distance contains one significant point (pink), indicating that the three dissimilarity methods successfully captures the circular arrangement of the orientation-sensitive neurons. The persistence diagrams from the ISI-, SPIKE-, and SPIKE-synchronization distances do not contain any significant points, indicating that the three distance methods fail to capture the circular arrangements. B: Dimension-1 persistence diagrams from ISI-, SPIKE-, and SPIKE-synchronization distances applied to the simulated orientation-sensitive neurons without the reliability noise.}
\label{fig:spiketrain_distances}
\end{figure}

Figure \ref{fig:spiketrain_distances} panel A illustrates the six persistence diagrams in dimension 1. When we use the windowed cross-correlation, Victor-Purpura, and van Rossum dissimilarity as inputs, the resulting persistence diagrams contains one significant point far from the diagonal, successfully capturing the circular organization of orientation neurons. On the other hand, the ISI-distance, SPIKE-distance, and SPIKE-synchronization distances fail to capture such circular organization. 

One possible explanation for the difference in performance is the robustness of the various dissimilarities against different types of noise. As mentioned in SI Section \ref{data:orientation}, we added noise to the simulated orientation-sensitive neurons by randomly picking intervals at which the neurons fire only at baseline rate. In the presence of such noise, the windowed cross-correlation, Victor-Purpura, and van Rossum dissimilarities were more robust. 

In order to test this hypothesis, we simulated another collection of orientation neurons without adding such noise. In the absence of such noise, the persistence diagrams from ISI-, SPIKE-, and SPIKE-synchronization distances detect the circular structure of the orientation neurons (Figure \ref{fig:spiketrain_distances}B)\footnote{Due to the small number of points in the persistence diagram, we do not compute the significance threshold for the persistence values.}.

\begin{figure}[H]
\centering
\includegraphics[width=1\textwidth]{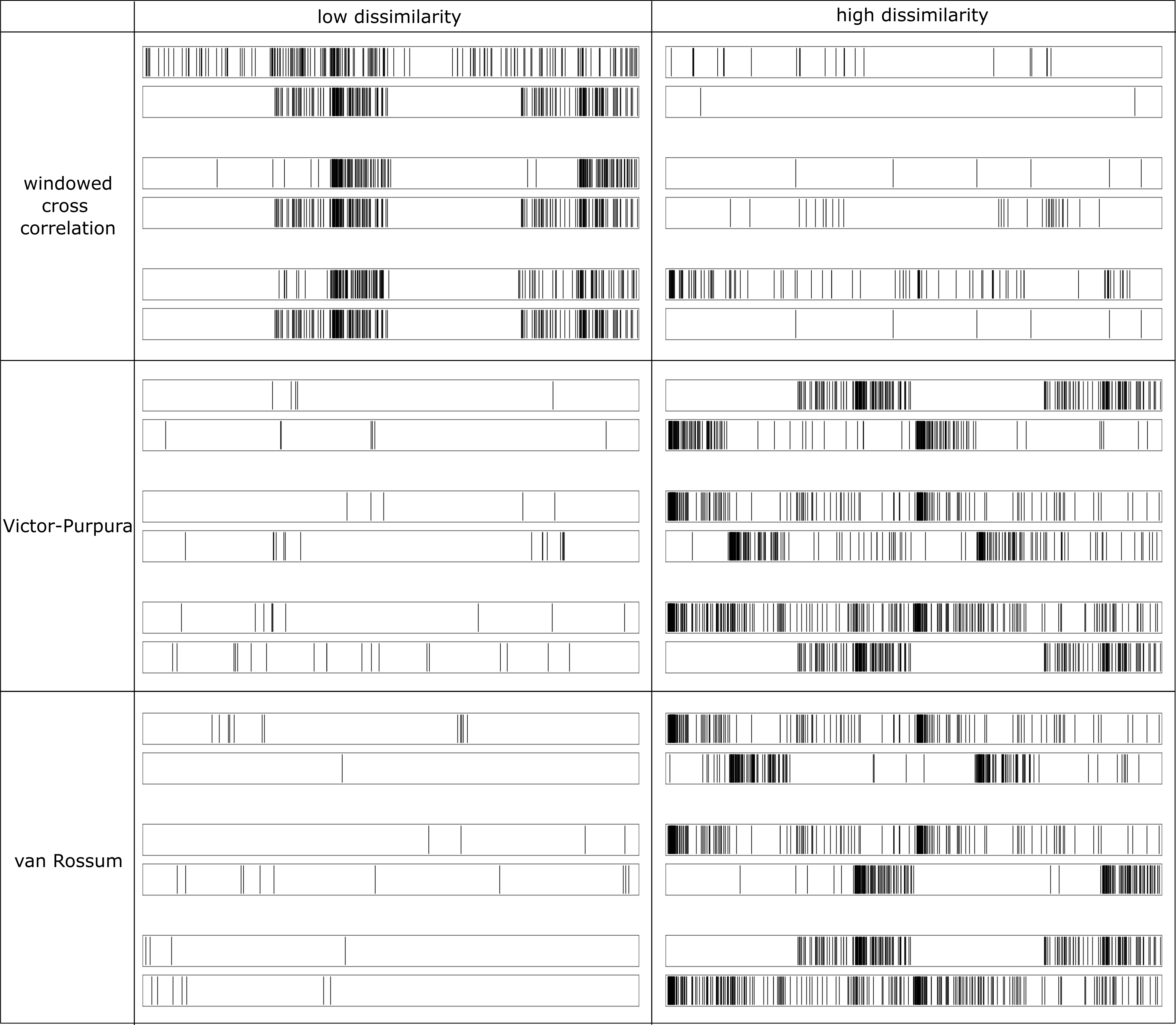}
\caption{Example pairs of spike trains with high and low dissimilarity according to the windowed cross-correlation dissimilarity (top), Victor-Purpura distance (middle), and the van Rossum distance (bottom). The windowed cross-correlation dissimilarity assigns high dissimilarity when a sparse spike train is involved, whereas the other two methods assign a low dissimilarity for sparse spike trains. }
\label{fig:distance_comparison}
\end{figure}

Even though both Victor-Pupura and van Rossum distances successfully encode the expected topological features, we used the windowed cross-correlation dissimilarity in our analysis because it assigns a high dissimilarity value for sparse spike trains, while the Victor-Pupura and Rossum distances tend to assign a low dissimilarity value whenever sparse spike trains are involved (Fig.\ref{fig:distance_comparison}). Figure \ref{fig:distance_comparison} illustrates randomly selected pairs of spike trains from the experimental visual data with low and high dissimilarities for each method. 
%We used the pre-processed spike trains from the experimental AL spike trains \ref{sec:spike_train_preprocessing}. We computed three dissimilarity matrices using the windowed cross-correlation dissimilarity, Victor-Pupura distance, and the van Rossum distances. For each distance, we randomly selected three pairs of spike trains with low distance values (bottom 10\% of total dissimilarity alues), and we selected three pairs of spike trains with high distance values (top 10 \% of total dissimilarity values).
Both Victor-Pupura and van Rossum distances assign a low dissimilarity score to a pair of sparse spike trains while the windowed cross-correlation method assigns a high dissimilarity to a pair of sparse spike trains. When combining the dissimilarity computation with topological analysis, assigning a high dissimilarity to sparse spike train is more favorable than assigning a low dissimilarity. Assume that a dataset contains a very sparse spike train $\vec{s}$. Regardless of the dissimilarity method, the spike train $\vec{s}$ will be (more or less) equidistant to the rest of the spike trains. %That is, $\vec{s}$ represents the apex of a cone. 
If $\vec{s}$ is assigned a low dissimilarity to all other spike trains, then any topological feature of the population will be killed early on in the persistence module because of $\vec{s}$. However, if $\vec{s}$ is assigned a high dissimilarity to all other spike trains, then the topological feature will die later, thus maintaining the topological feature in the persistence module. For such reason, we prefer the windowed cross-correlation dissimilarity over the Victor-Pupura and van Rossum distance in this analysis. %Note that the Victor-Pupura and the van Rossum distance can still be used for topological analysis. It may just require more careful preprocessing steps to remove sparse spike trains. 

In summary, the windowed cross-correlation dissimilarity is robust against the presence of unreliably-firing intervals (Fig.~\ref{fig:spiketrain_distances}) and assigns a high dissimilarity score when a sparse spike train is involved (Fig.~\ref{fig:distance_comparison}), which is desirable for topological analysis. Furthermore, the windowed cross-correlation dissimilarity can be used to compute dissimilarities for both spike trains and firing rates.

%-----------------------------------------------------------------
\subsection{Experiments with the robustness of persistence diagrams against variations and noise}
\label{Supp:experiments_messy_barcodes}
%-----------------------------------------------------------------
In our analysis, we studied various simulated and experimental spike trains through the lens of persistence homology. In this section, we examine how the persistence diagram is affected by certain variations and noise. We expect that many of these variations and noise will be present in experimental spike trains. 

To study the robustness of the persistence diagrams, we varied different aspects of the simulation of the orientation-sensitive neurons (SI Section \ref{data:orientation}) and examined the impact on the persistence diagrams. In particular, we varied the parameters of the tuning curve and added various forms of noise.

Recall the simulation process. We first created a stimulus video, for which the image orientation changed "continuously" every 7 frames. We then followed the steps:

\begin{enumerate}
 \setlength\itemsep{0em}
    \item Create 64 orientation-sensitive neurons using the tuning curve 
    \begin{equation}
\label{eq:experiment_tuning}
T_{\theta_{P}}(\theta) = C + R_P \exp \Big( - \frac{d_\pi(\theta, \theta_{P})^2}{2 \sigma^2}\Big).
\end{equation}
    \item Calculate the firing rate in response to the stimulus video
    \item Simulate spike trains using inhomogeneous Poisson process
    \item Compute dissimilarity matrix among spike trains
    \item Compute persistence diagrams. 
\end{enumerate}

Recall that the persistence diagram of orientation cells contained one significant point, which is consistent with the expectation that orientation-sensitive neurons are organized circularly. 

We examined the robustness of the persistence diagram against variations to the parameters in Equation \ref{eq:experiment_tuning} and to different levels of noise. The following summarizes the variations that we studied: 
\begin{itemize}
    \item maximum firing rate,
    \item sparsity of spike trains,
    \item Gaussian noise,
    \item percentage of spike trains firing uniformly, and
    \item percentage of unreliable intervals.
\end{itemize}

\subsubsection*{Maximum firing rate} 
\begin{figure}[H]
\centering
\includegraphics[width=1\textwidth]{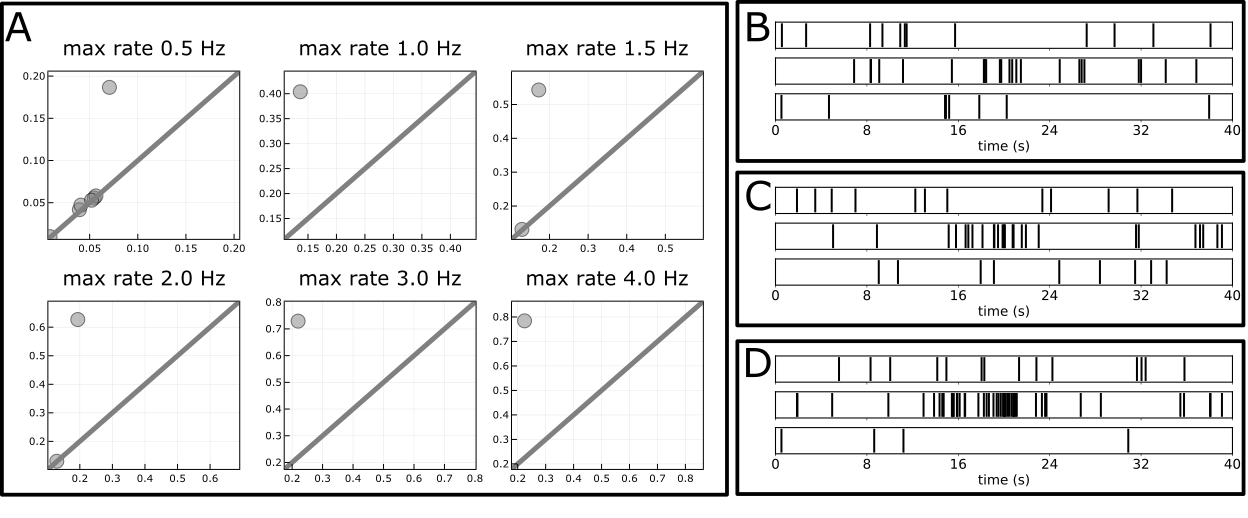}
\caption{Experiments while varying the maximum firing rate while keeping the baseline firing rate constant at $0.25$ Hz. \textbf{A.} Dimension-1 persistence diagrams. All persistence diagrams contain one point far from the diagonal. \textbf{B.} Example spike trains with maximum rate $0.5$Hz. \textbf{C.} Example spike trains with maximum rate $2$ Hz. \textbf{D.} Example spike trains with maximum rate $4$ Hz. }
\label{fig:maxrate_experiment}
\end{figure}

We examined changes to the dimension-1 persistence diagram as we varied the maximum firing rate $R_P$ of the tuning curve in Equation \ref{eq:experiment_tuning}. The maximum above-baseline firing rates varied from $0.5$ Hz, $1.0$ Hz, $1.5$ Hz, $2$ Hz, $3$ Hz, and $4$ Hz, while we fixed the baseline firing rate $C$ at $0.25$ Hz and the tuning width $\sigma$ at 0.2. For each candidate maximum firing rate $R_P$, we created 64 orientation-sensitive neurons using the tuning curve in Equation \ref{eq:experiment_tuning}.  
As shown in Figure \ref{fig:maxrate_experiment}, all persistence diagrams contain one point far from the diagonal. A low maximum firing rate of $0.5$ Hz leads to slightly messier persistence diagrams than the others. 

\subsubsection*{Sparsity}
\begin{figure}[H]
\centering
\includegraphics[width=1\textwidth]{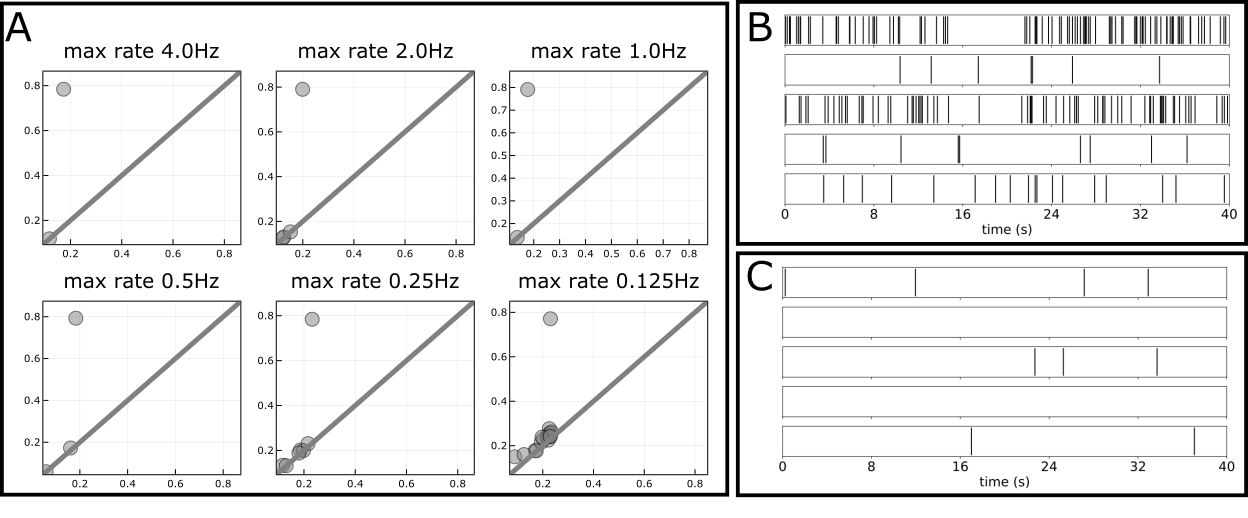}
\caption{Experiments while increasing sparsity of spike trains. \textbf{A.} Dimension-1 persistence diagrams for various maximum firing rates. The persistence diagrams consistently show one significant feature even as we increase the sparsity. \textbf{B.} Example raster of five neurons with maximum firing rate 4 Hz. \textbf{C.} Example raster of five neurons with maximum firing rate 0.125 Hz.  }
\label{fig:sparsity}
\end{figure}

We examined the changes to the dimension-1 persistence diagram as the spike trains became sparser. We decreased the neurons' firing rates while keeping the ratio between the maximum firing rate and the baseline firing rate at a constant 6.25 \%. The above-baseline maximum firing rates varied from $4$ Hz, $2$Hz, $1$Hz, $0.5$Hz, $0.25$Hz, and $0.125$Hz. The corresponding baseline firing rates were $0.25$ Hz, $0.125$ Hz, $0.063$ Hz, $0.031$ Hz, $0.016$ Hz, and $0.008$ Hz. Experiments with low maximum and baseline firing rates will have sparse spike trains. As illustrated in Figure \ref{fig:sparsity}, all persistence diagrams contain one point far from the diagonal even as we increase the sparsity of spike trains.

\subsubsection*{Presence of noise}
\begin{figure}[H]
\centering
\includegraphics[width=1\textwidth]{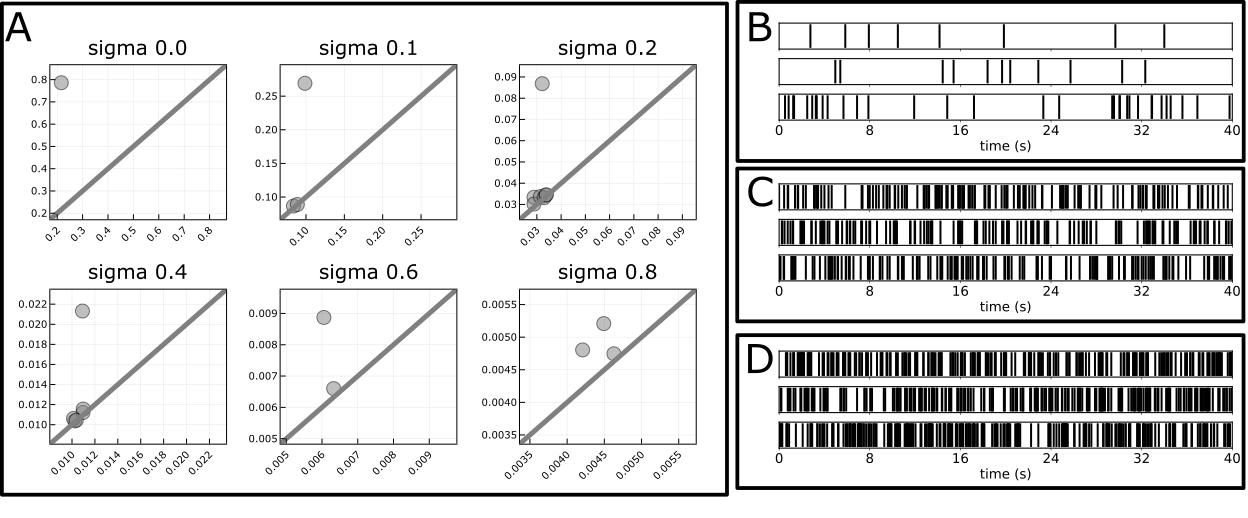}
\caption{Experiment with varying levels of Gaussian noise. \textbf{A.} Dimension-1 persistence diagram for various levels of Gaussian noise. The addition of Gaussian noise adds points with short persistence and decreases the lifetime of the point far from the diagonal. At $\sigma = 0.8$, we don't observe any point with significantly different persistence. \textbf{B.} Example spike trains with noise level $\sigma =0$, $\sigma = 0.4$, and $\sigma = 0.8$.}
\label{fig:noise_experiment}
\end{figure}

Recall from SI Section \ref{data:orientation} that Gaussian noise is added to a tuning curve, resulting in a firing rate function of the form 
\begin{equation}
r_{\theta_{P}}(\theta) = \Big[C + R_P \exp \Big( - \frac{d_\pi(\theta, \theta_{P})^2}{2 \sigma^2}\Big) + \epsilon \Big]_+,
\end{equation}
where $\epsilon$ represents Gaussian noise with mean 0 and some standard deviation $\sigma_{\epsilon}$, and $[ \, ]_+$ denotes rectification of negative values to zero \cite{Butts2006TuningCN}. We examined how the persistence diagram changes as we increase the standard deviation $\sigma_{\epsilon}$ of the noise term. As shown in Figure \ref{fig:noise_experiment}, increasing $\sigma_{\epsilon}$ adds some noisy points to the persistence diagram. It also  decreases the persistence of the points until one can no longer see a clear separation between significant and non-significant points.

\subsubsection*{Presence of uniformly-firing neurons}

We then studied how the presence of neurons firing at a uniform rate affects the population code. We performed two experiments - one while gradually increasing the percentage of neurons firing at baseline rate, and another while increasing the percentage of neurons firing at maximum rate. In both experiments, the maximum firing rate was fixed at $2$ Hz, the baseline firing rate as $0.25$ Hz, and the tuning width $\sigma$ was fixed at $0.2$. 

\begin{figure}[H]
\centering
\includegraphics[width=0.7\textwidth]{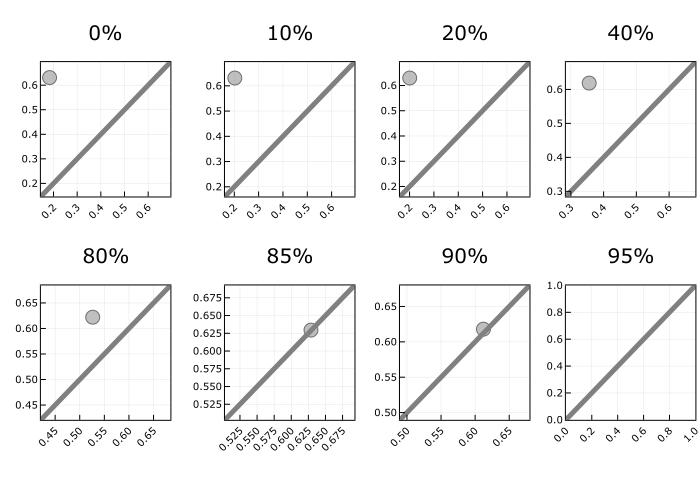}
\caption{Dimension-1 persistence diagrams as we varied the percentage of neurons firing at baseline. }
\label{fig:baseline_experiment}
\end{figure}

In the first experiment, we simulated a collection of 64 neurons. We then randomly selected a portion of the neurons and adjusted their firing rates to be $0.25$ Hz. Figure \ref{fig:baseline_experiment} summarizes how the persistence diagram changes as we increased the portion of such neurons firing at baseline rate. As one increases the portion of neurons firing at baseline, the point far from the diagonal decreases in lifetime until the signal disappears.

\begin{figure}[H]
\centering
\includegraphics[width=0.7\textwidth]{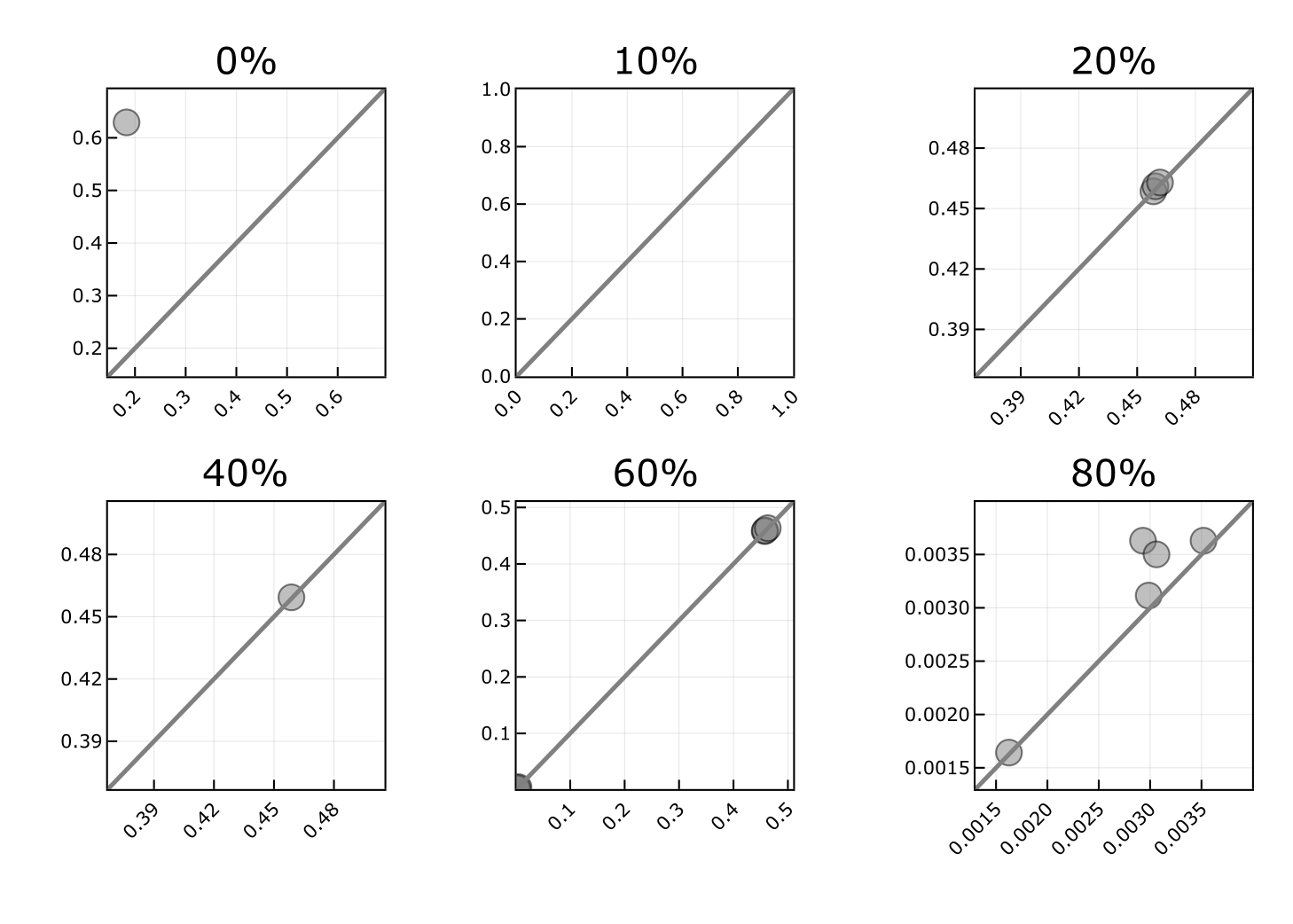}
\caption{Dimension-1 persistence diagrams as we varied the percentage of neurons firing at maximum rate. Even if 10\% of the population are firing at maximum rate, the collection of neurons fail to encode the single cyclic feature.}
\label{fig:maxrate_presence_experiment}
\end{figure}

In the second experiment, we performed a similar analysis by randomly selecting a portion of the simulated neurons and adjusting their firing rates to be at $2$ Hz. We examined how the persistence diagram changed as the percentage of such maximum firing neurons increased (Fig.~\ref{fig:maxrate_presence_experiment}). Even if $10$ \% of the neurons are replaced with neurons uniformly firing at maximum rate, the persistence diagram fails to encode the cyclic feature. 

In summary, the persistence diagram is robust against the presence of neurons that are firing at baseline, while it is quite sensitive to the presence of neurons that are uniformly firing at maximum rate. One possible explanation for such behavior is the choice of the spike train similarity in Equation \ref{eq:spiketrain_similarity}. Given a pair of spike trains, the windowed cross-correlation dissimilarity assigns a high dissimilarity value if one of the two spike trains is sparse, as illustrated in Figure \ref{fig:distance_comparison}. Such behavior can make the persistence diagram robust against the presence of sparse spike trains, as discussed in Section \ref{sec:spiketrain_dissimilarities}. 

\subsubsection*{Reliability of neurons} % maybe call the "reliability across trials" the "consistency" across trials. 

We added a novel type of noise, hereby referred to as the ``unreliability.'' When simulating a neuron via a tuning curve, it is usually assumed that that neuron will react according to its tuning curve with some variation coming from Gaussian noise. However, neurons do not fire the same way even if it is presented with the same stimulus repeatedly. In order to incorporate such variation in reaction, we decided to introduce intervals in which the simulated neurons fire at baseline firing rate. 

\begin{figure}[H]
\centering
\includegraphics[width=1\textwidth]{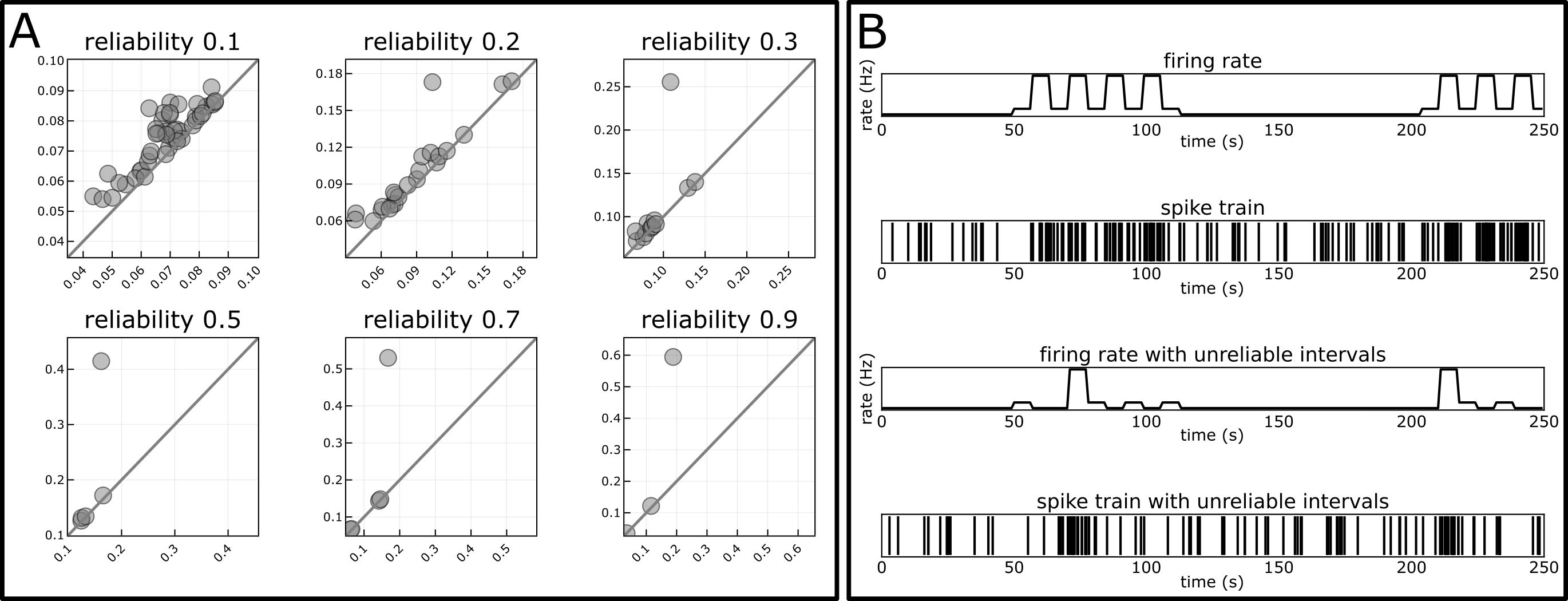}
\caption{Effect of reliability on persistence diagrams. \textbf{A.} Persistence diagrams from simulated neurons with varying levels of reliability parameters are shown. The reliability parameter $p$ is the probability at which a unit of intervals fires at the predicted rate as opposed to the baseline firing rate. A low $p$ value leads to neurons firing at baseline for more intervals, resulting in a persistence diagram without a clear signal. \textbf{B.} (top) Example firing rate and (2nd row) its corresponding spike train. (3rd row) Example firing rate with unreliable intervals (low $p$ value), and (bottom) its corresponding spike train.}
\label{fig:reliability_experiment}
\end{figure}

Recall that the stimulus video consisted of images of various orientations. Given an orientation, we presented 7 frames of the same orientations (with different centers). For each non-overlapping unit of intervals corresponding to the 7 frames, we adjusted the firing rate so that it would fire at the predicted rate $r_{\theta_p}(\theta)$ with probability $p$ and at baseline firing rate $C$ with probability $1-p$. Figure \ref{fig:reliability_experiment} summarizes the changes to the persistence diagram as we change the probability $p$. 

%---------------------------------------
\subsection{Analogous cycles matches involving multiple geometrically related cycles}

The analogous cycles method matches cycles across distinct stimulus spaces. In cases where the constituent cycles are geometrically independent, as in the toroidal spaces studied in the main text, this technique is generally effective. However, there are settings when a natural registration of the neural manifolds may result in "nesting" behaviors (as illustrated in {Fig.~\ref{fig:three_to_one}A} below) that may cause the analogous cycles method to produce unintuitive matches. In this section, we illustrate the output of the analogous cycles method on such data and describe a modification that selects a different parameter in Step 3(a) of Algorithm 1 (SI Section~{\ref{methods:analogous_cycles}}) that is better suited for such "nested" cycles. 

However, we caution the reader that this alternative method comes with two major drawbacks. First, as the authors demonstrate in \cite{analogous_bars}, the default choice of parameter comes with a theoretical guarantee that the result of the algorithm is independent of choice of cycle representative; for other choices of parameters, this guarantee fails and so internal algorithmic choices inside a persistence solver may give very different matches. Second, the choice we will make in this section is motivated by a shared, extrinsic geometric scale across systems\footnote{Here and throughout, by \emph{extrinsic} geometry we mean that there is an embedding of the geometric features of interest in some metric space. For example, thinking of place fields as explicitly embedded in a planar environment, rather than abstractly interacting via intersection patterns that may result from many candidate geometries.}. In any case where such a scale is not known to exist, which is the most likely setting for neural data, any choice of parameter besides the default, which is most permissive, will require substantial \emph{ad} \emph{hoc} justification. Finally, we note that in \cite{analogous_bars}, the authors develop a more general and complete algebraic framework for studying matches, which we choose not to present in full detail in this paper due to its mathematical density. We invite readers interested in edge cases to refer to that work for more information.

\medskip

While it is possible to describe this example in terms of spike trains and neural manifolds, we find it simpler to understand the example if we consider the underlying geometry directly. Let $P$ and $Q$ be the sets of points in the euclidean plane illustrated in (SI Fig.~{\ref{fig:three_to_one}}A), representing, for example, centers of hippocampal place fields. To apply the constructions described in this paper, we will use the euclidean distance between points as a measure of within-system dissimilarity, and if we overlay the two point clouds, we can compute cross-system dissimilarity using euclidean distance between points in the two populations\footnote{Observe that this cross-dissimilarity measure involves an \emph{extrinsic} planar geometry that may or may not be encoded by neural activity. As such, it is unclear whether real neural systems will exhibit the behavior described in this example. We nonetheless believe it is important to consider to illustrate how the method may fail.}. Intuitively, we might  expect the single significant point in $\pd(P)$ to be analogous to the combination of the three points in $\pd(Q)$. However, applying the method of analogous cycles does not recover any analogous cycles among the significant points of $\pd(P)$ and $\pd(Q)$ (SI Fig.~{\ref{fig:three_to_one}}B).

\begin{figure}[t!]
    \centering
    \includegraphics[width=1\linewidth]{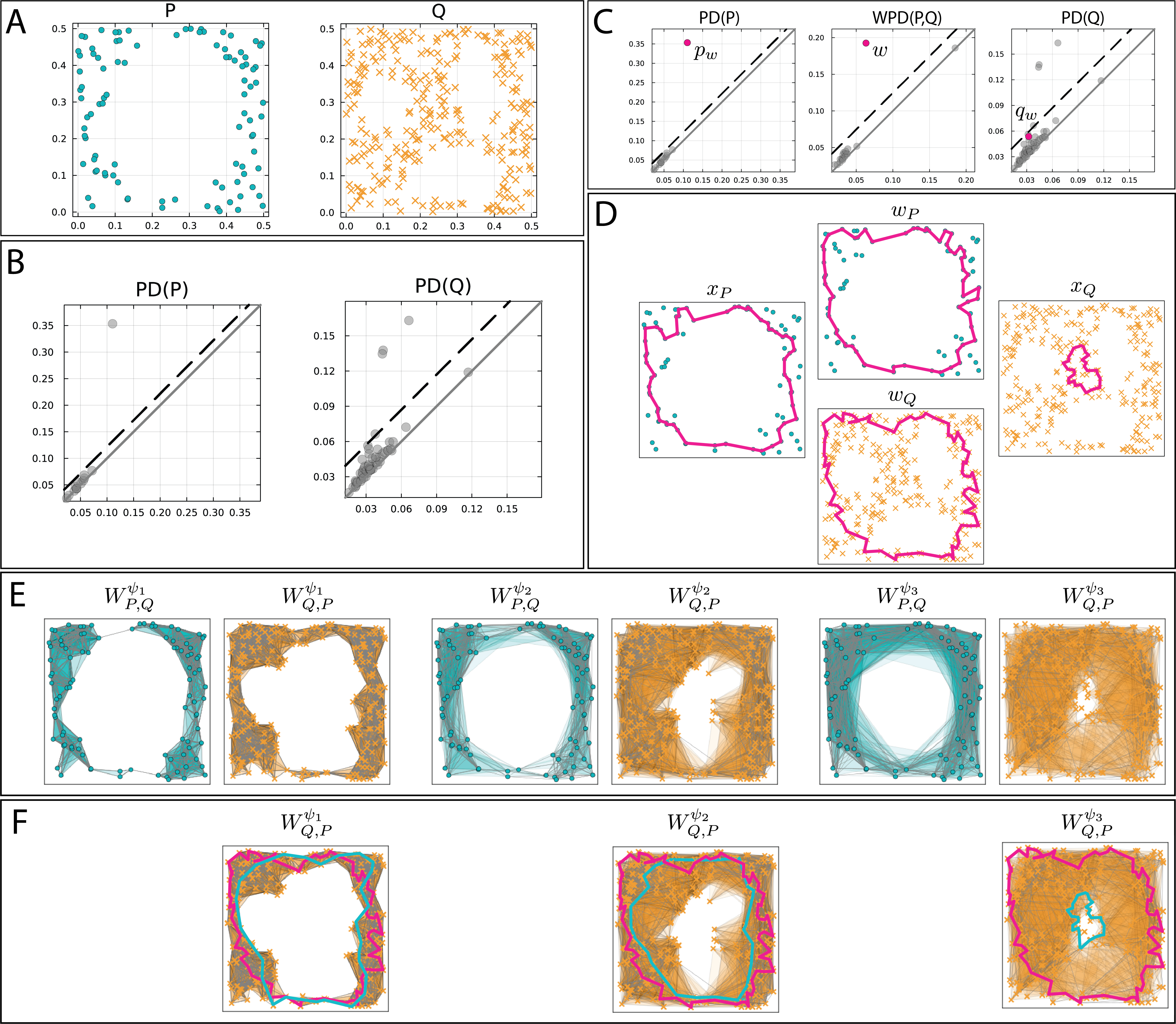}
    \caption{Example where a cycle in $P$ is analogous to a "nested" combination of multiple cycles in $Q$. \textbf{A.} Sets of points $P$ and $Q$ in the euclidean plane, corresponding to centers of hippocampal place fields. \textbf{B.} Applying the analogous cycles method produces no matches among the significant points in $\pd(P)$ and $\pd(Q)$. \textbf{C.} Applying analogous cycles without restricting to significant cycles matches the significant point in $\pd(P)$ to a non-significant point in $\pd(Q)$, as highlighted in pink. 
    \textbf{D.} Representatives of the cycles involved in the match in (C). (Center, top) Cycle representative of the selected Witness cycle in $\wpd(P,Q)$ in $P$. (Center, bottom) Cycle representative of the selected Witness cycle in $\wpd(P,Q)$ in $Q$. (Left) Cycle representative of the corresponding analogous cycle in $\pd(P)$. (Right) Cycle representative of the corresponding analogous cycle in $\pd(Q)$. The method produces these matches due to a conservative choice of parameter, which can be varied to reduce the number of potential matches in favor of geometric similarity. \textbf{E.} An illustration of the witness complexes on $P$ and $Q$ constructed using different choices of parameters $\psi_1 < \psi_2 < \psi_3 $. \textbf{F.} The witness cycle $w_Q$ (pink) and example homologous cycles (teal) in $W^{\psi_1}_{Q,P}$, $W^{\psi_2}_{Q,P}$, and $W^{\psi_3}_{Q,P}$. Observe that smaller values of $\psi_i$ produce cycles more geometrically similar to the target cycle in $P$. }
    \label{fig:three_to_one}
\end{figure} 

To understand why the output differs from the expectation, consider the structure of the witness complexes in this example. Given the significant point $w \in \wpd(P,Q)$ (SI Fig.~{\ref{fig:three_to_one}}C, center), in Step 3(a) of Algorithm 1 (SI Section~{\ref{methods:analogous_cycles}}), we fix the parameter $\psi(w)$ to be the largest parameter smaller than the death parameter of $w$. We then construct the witness complexes $W^{\psi}_{P,Q}$ and $W^{\psi}_{Q,P}$ and find the homology classes $[w_P] \in W^{\psi}_{P,Q}$ and $[w_Q] \in W^{\psi}_{Q,P}$ that correspond to $w$. The cycle representatives $w_P$ and $w_Q$ confirm that $w$ represents the shared circle between $P$ and $Q$ (SI Fig.~{\ref{fig:three_to_one}}D, center). 

The method then looks for homology classes $[x_P] \in H_1(X^{\varepsilon_P}_P)$ (respectively, $[x_Q] \in H_1(X^{\varepsilon_Q}_Q)$) that represent $[w_P]$ (respectively, $[w_Q]$) via persistent extension. The involved algebraic operations guarantee that $[x_P]$ and $[x_Q]$ are the persistent classes in $P$ and $Q$ with the smallest birth times that are homologous to $[w_P]$ and $[w_Q]$ (SI Section {\ref{methods:analogous_cycles}}). A visualization of representatives $x_P$ and $x_Q$ show that $x_Q$ is qualitatively similar to $w_Q$ (SI Fig.~{\ref{fig:three_to_one}}D). The problem is that the cycle representative $x_Q$ has a much smaller scale than $w_Q$. 

In its final step, the analogous cycles method returns collections of points $(p_w, q_w)$, with $p_w \subseteq \pd(P)$ and $q_w \subseteq \pd(Q)$, representing $[x_P]$ and $[x_Q]$. In this example, the output $(p_w, q_w)$ is highlighted in pink (SI Fig.~{\ref{fig:three_to_one}}C, left, right). As discussed in \emph{Materials} \emph{and} \emph{Methods} (main text), the method only reports collections of points $(p_w, q_w)$ involving the significant points of $\pd(P)$ and $\pd(Q)$. Here, $q_w$ is not a significant point, so $p_w$ is not analogous to any significant point in $\pd(Q)$. 

\medskip
% the effect of choosing psi the way we did
The key step in the algorithm that controls the geometric similarity between $w_Q$ and $x_Q$ (respectively, $w_P$ and $x_P$) is the construction of the witness complex $W^{\psi}_{Q,P}$ (respectively, $W^{\psi}_{P,Q}$), specifically the parameter $\psi$. By default, $\psi$ is chosen to be the largest parameter smaller than the death parameter of $w$ because the resulting witness complexes $W^{\psi}_{P,Q}$ and $W^{\psi}_{Q,P}$ contain all possible cycle representatives of $w$. This allows the analogous cycles method to find classes $[x_Q]$ which represent to $[w_Q]$ through persistent extension (respectively, $[x_P]$ and $[w_P]$), even if $x_Q$ and $w_Q$ are geometrically dissimilar. For example, SI Fig.~{\ref{fig:three_to_one}}F (right) illustrates an example witness complex $W^{\psi}_{Q,P}$ where $[w_Q]$ (pink) extends to $[x_Q]$ (teal), even if $w_Q$ and $x_Q$ are geometrically dissimilar. This $\psi$ is the most permissive choice of parameter, providing the largest number of possible matches and thus, in the absence of the significance test, a robust falsification when no matches appear. However, in some cases, it does allow matches of cycles across geometric scales that can interact with significance testing, which has the effect of rejecting potential matches with small geometric scale as noise, to create false negatives.

% A more sophisticated version of the method of analogous bars, described in \cite{analogous_bars}, provides a general framework for discovering such matches at the cost of substantial algebraic complexity. For purposes of this application, in the following section we describe a simpler heuristic modification to Algorithm 1 using an alternative parameter selection.}

\subsubsection{Witness-modified analogous cycles}
\label{sec:AC_variation_witness}

To find relations among cycles with extrinsic geometric relations, one can implement a variation of the analogous cycles method utilizing a different choice of parameter $\psi$ in Algorithm~{\ref{alg:analogous_cycles}} Step 3(a). Varying $\psi$ controls the extent to which the identified analogous cycles in $P$ and $Q$ geometrically resemble the witness cycles $w_P$ and $w_Q$. 

Continuing our example from above, SI Figure~{\ref{fig:three_to_one}}E
illustrates witness complexes built for three different choices of parameter $\psi_1 < \psi_2 < \psi_3$. As $\psi$ increases, the witness complex $W^\psi_{Q,P}$ expands from a relatively narrow ring to a much larger annulus. In $W^{\psi_3}_{Q,P}$, cycle $w_Q$ (pink) is homologous to the small teal cycle in $Q$ (SI Fig.~{\ref{fig:three_to_one}}F, right). In contrast, in $W^{\psi_1}_{Q,P}$ or $W^{\psi_2}_{Q,P}$, only cycles in $Q$ that are geometrically similar to $w_Q$ can be homologous to $[w_Q]$ (SI Fig.~{\ref{fig:three_to_one}}F, left, center). If we choose $\psi$ to be $\psi_1$ or $\psi_2$ in step 3(a) of Algorithm~{\ref{alg:analogous_cycles}} and implement the remaining steps, the output will be a collection of points $(p_w, q_w)$ in which $p_w$ and $q_w$ represent cycles that are geometrically similar to $w_P$ and $w_Q$. 

\begin{figure}[b!]
    \centering
    \includegraphics[width=0.7\linewidth]{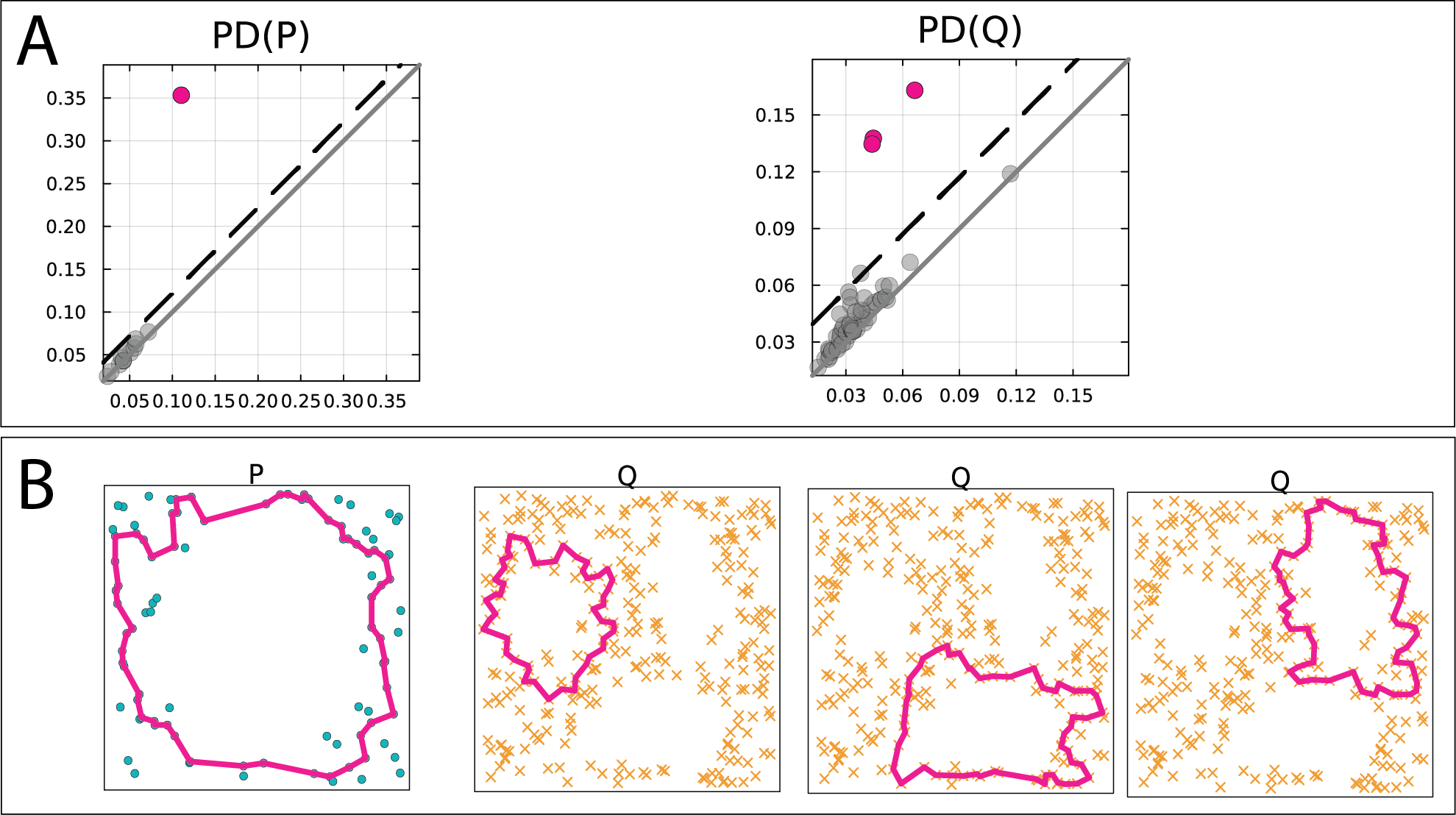}
    \caption{Output of witness-modified analogous cycles method captures one-to-many relations among cycles. \textbf{A.} The single significant feature in $\pd(P)$ is analogous to the combination of three highlighted points in $\pd(Q)$. \textbf{B.} Cycle representatives of the highlighted points in $\pd(P)$ and $\pd(Q)$ show that the combination of the three cycles in $Q$ form a cycle that is geometrically similar to the cycle in $P$.}
    \label{fig:three_to_one_alternative}
\end{figure}

Given $w \in \wpd(P,Q)$, the parameter $\psi$ can, in practice, be any value in \newline $[\text{birth}(w), \text{death}(w))$, where $\text{birth}(w)$ is the birth parameter of $w$ and $\text{death}(w)$ is the death parameter of $w$. We explicitly describe two options for $\psi$: 

\begin{itemize}
%\item $\psi$ is the birth parameter of the point $w$
\item $\psi$ is the largest parameter smaller than the death parameter of $w$. This is the default option in analogous cycles method (Algorithm~{\ref{alg:analogous_cycles}}).
\item $\psi$ is the largest parameter smaller than all death parameters of the significant points in $\pd(P)$ and $\pd(Q)$. That is, $\psi$ is the largest parameter smaller than $\min_{p \in \pd_*(P) \cup \pd_*(Q)} \{ \text{death}(p) \}$, where $\pd_*(P)$ and $\pd_*(Q)$ are the significant points on $\pd(P)$ and $\pd(Q)$. If $\min_{p \in \pd_*(P) \cup \pd_*(Q)} \{ \text{death}(p) \}$ is greater than $\text{death}(w)$, then we use the default choice of $\psi$ above. 
\end{itemize}

We refer to the modified analogous cycle in which the witness complex is built using the minimum death parameter of significant points as the \emph{witness-modified} \emph{analogous} \emph{cycles} method. When we apply witness-modified analogous cycles to the example in SI Fig.~{\ref{fig:three_to_one}}A, the algorithm returns a match between the single significant point in $\pd(P)$ and a linear combination of the three significant points in three points in $\pd(Q)$ (SI Fig.~{\ref{fig:three_to_one_alternative}}A). An inspection of the cycle representatives illustrates that the combination of the three highlighted cycles in $Q$ form a large cycle that is geometrically similar to the one in $P$ (SI Fig.~{\ref{fig:three_to_one_alternative}}B).

\begin{figure}[h!]
    \centering
    \includegraphics[width=0.8\linewidth]{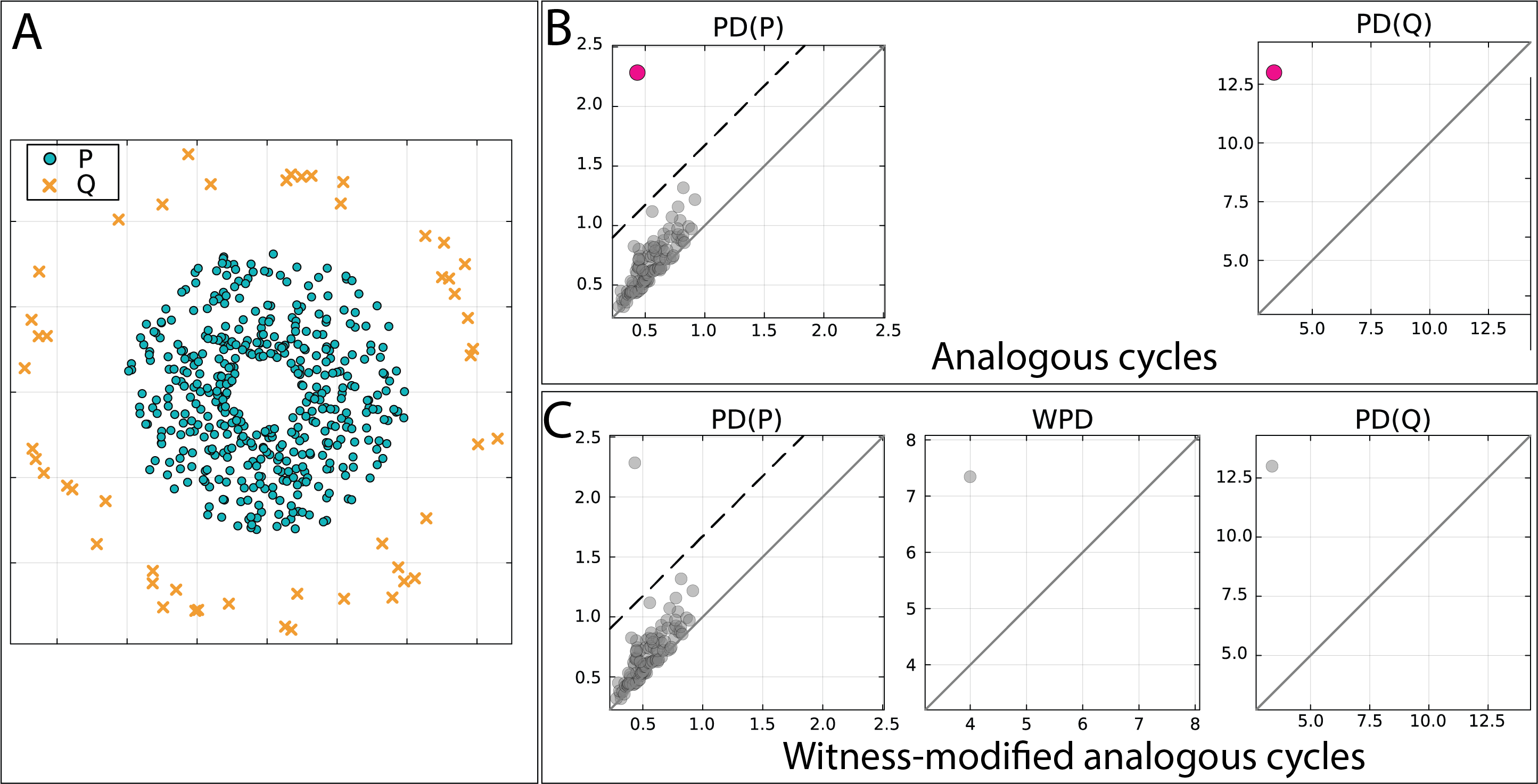}
    \caption{Example point cloud illustrating the potential problem of incomparable parameters in the witness-modified analogous cycle method. \textbf{A.} Example point clouds $P$ and $Q$. \textbf{B.} The default analogous cycles method finds a pair of analogous cycles. \textbf{C.} The witness-modified analogous cycles doesn't find any analogous cycles. The minimum death parameter of significant points in $\pd(P)$ and $\pd(Q)$, which is around 2.3, is smaller than the birth parameter of the point in $\wpd$.}
    \label{fig:witness_varied_AC}
\end{figure}

\begin{figure}[h!]
    \centering
    \includegraphics[width=0.8\linewidth]{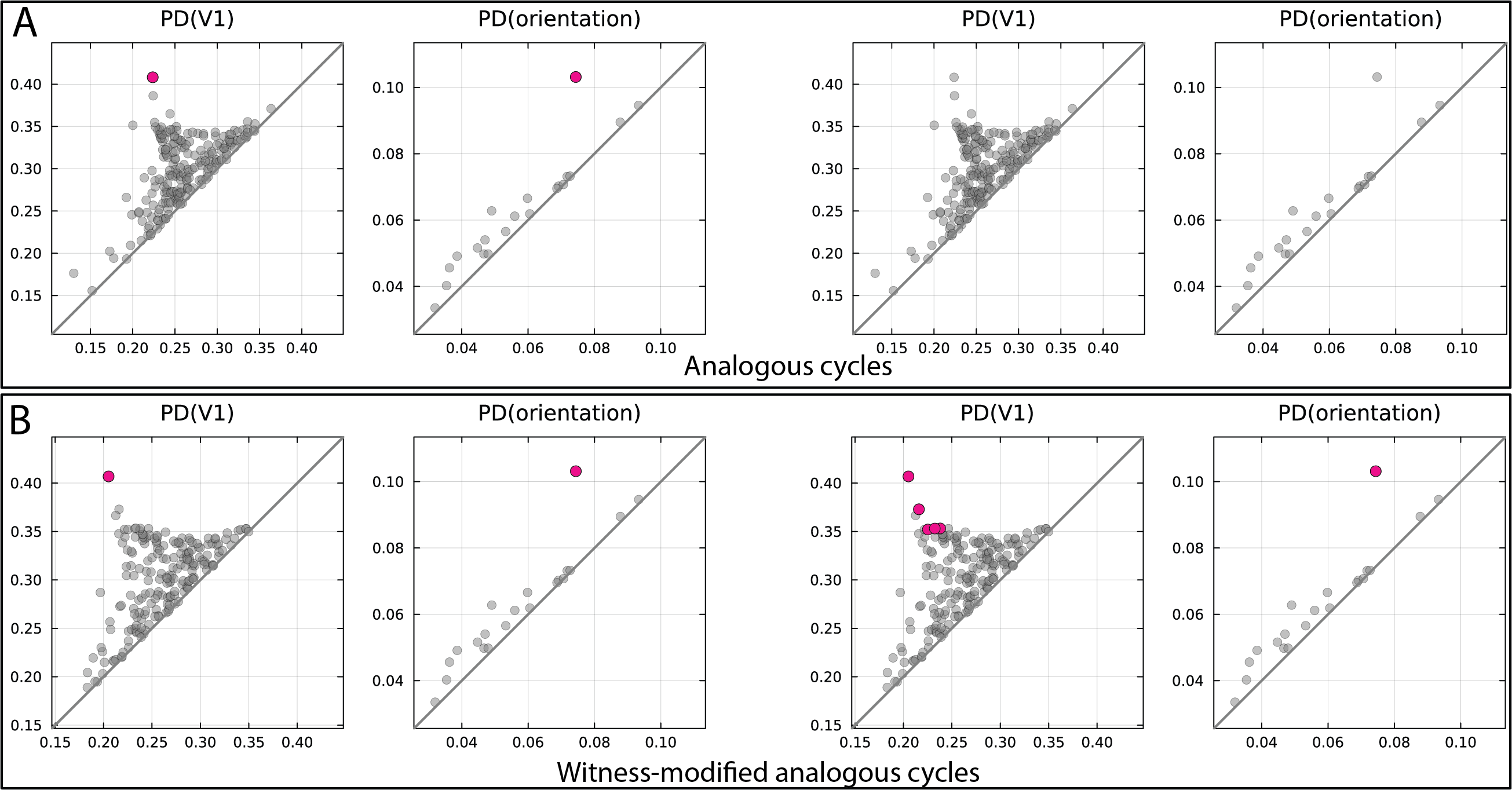}
    \caption{The witness-modified analogous cycles method may not always find homologous cycles. Here, we present a comparison of the default analogous cycles and the witness-modified analogous cycles method on a random subsample of the simulated V1 simple cells and orientation cells. The witness-modified analogous cycles method fixes the witness parameter $\psi$ at the birth time of the significant point in $\wpd$. We took the $800$ simulated V1 simple cells and $64$ simulated orientation cells from Fig.~2 of the main text, randomly sampled $300$ simulated V1 simple cells, and ran the default analogous cycles method (without significance thresholding) and the witness-modified analogous cycles. \textbf{A.} Output of the two methods on a $300$ subsample of the simulated V1 simple cells and $64$ orientation cells. (Left) The default analogous cycles identifies a pair of points in $\pd(V1)$ and $\pd(\text{orientation})$ highlighted in pink. The identification is consistent with the full output in Fig.~2 of main text. (Right) The witness-modified analogous cycles method doesn't find any analogous cycles.
    \textbf{B.} Outputs in another $300$ random selection of V1 simple cells and $64$ orientation neurons. (Left) Again, the default analogous cycles method identifies a pair of points in $\pd(V1)$ and $\pd(\text{orientation})$ in a manner consistent with Fig.~2 of main text. (Right) The witness-modified analogous cycles method finds some analogous cycles.}
    \label{fig:witness_varied_AC_V1_ori}
\end{figure}

\medskip
% caution & possibility for future work 
While the witness-modified analogous cycles method finds matches that are geometrically more similar than the default method, there are potential issues with this choice of parameter. First, the strategy of using the minimum death parameter of the significant points in $\pd(P)$ and $\pd(Q)$ to fix the witness complex parameter $\psi$ works only if the dissimilarity matrices $D_P$, $D_Q$, and the cross-system dissimilarity matrix $D_{P,Q}$ have comparable scales. When $P$ and $Q$ represent vastly different systems, such as in the simulated visual stimuli and neural activity analyzed in Fig.~2 in the main text, the death parameters of points in $\pd(P)$ and $\pd(Q)$ may not be comparable to the parameters in $\wpd$ (see SI Fig.~{\ref{fig:witness_varied_AC}}). In such situations, one may need to employ a different strategy for systematically exploring smaller parameters of $\psi$. Secondly, by decreasing the default parameter $\psi$ of the witness complex, one makes a trade-off between finding homologous cycles and geometrically similar cycles (see SI Fig.~{\ref{fig:witness_varied_AC_V1_ori}}). As there can be advantages to both approaches, in settings where complicated interactions in neural manifold geometry across populations are suspected, we recommend that users explore a variety of different parameters $\psi$. A systematic exploration of the optimal parameter $\psi$ is left for future research. 

% ---------- example involving multiple cycle matches in stimulus space
\subsubsection{An example of using both versions of analogous cycles to disentangle relations between neural manifolds}
\label{sec:complicated_stimulus_space}

\begin{figure}[h!]
    \centering
    \includegraphics[width=1\linewidth]{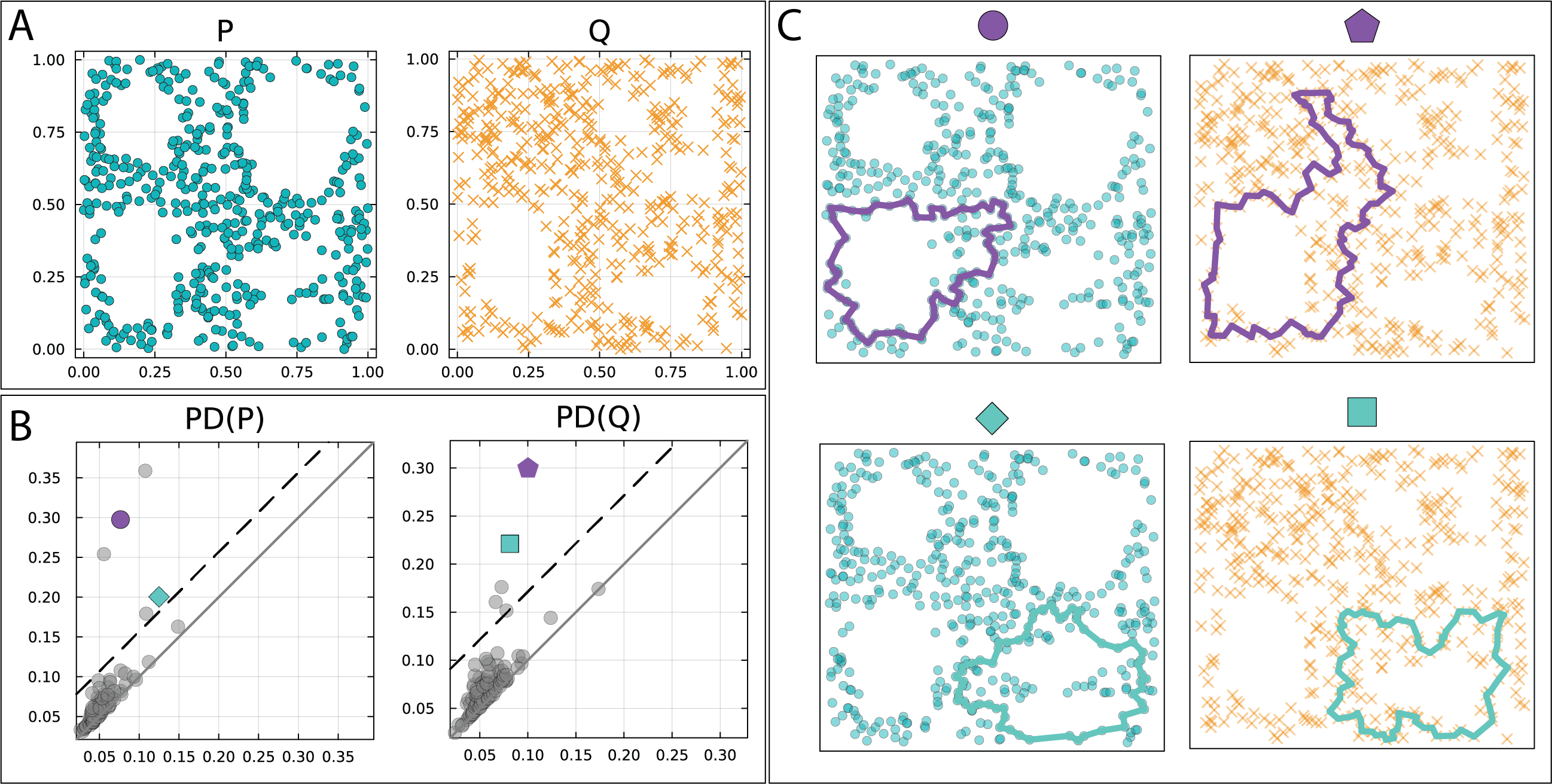}
    \caption{The default analogous cycles method on example point clouds with multiple types of matches. \textbf{A.} Point clouds $P$ and $Q$. \textbf{B.} The default analogous cycles method finds two analogous pairs, shown in purple and teal. \textbf{C.} Cycle representatives of the purple analogous points (top) and teal analogous points (bottom).}
    \label{fig:complicated_stimulus}
\end{figure}

We illustrate the use of both the default analogous cycles method and the witness-modified analogous cycle method to compare structure in spaces with more intricate cycle relations. Once again, we restrict our attention to sets of points $P$ and $Q$, with euclidean distance thought of as a proxy for spike train dissimilarity, as illustrated in SI Figure~{\ref{fig:complicated_stimulus}}A. From a visual inspection, both $P$ and $Q$ have five intrinsic cycles. The cycle in the top left corner of $P$ isn't analogous to any cycles in $Q$. The cycle in the top right corner of $P$ roughly matches to a combination of the three cycles in the similar region in $Q$. The cycles in the bottom left corner of $P$ and $Q$ are well-matched, and in the bottom right corner, the point cloud $P$ has two cycles that match to a cycle larger in $Q$ occupying the similar region.

The default analogous cycles method finds two analogous pairs indicated by the purple and teal points in SI Figure~{\ref{fig:complicated_stimulus}}B. By plotting the cycle representatives, we see that the default analogous cycles method finds the analogous cycles in the bottom left corners (purple) and the bottom right corners (teal). However, it failed to identify the cycle in $P$ that resembles a combination of three smaller cycles of $Q$ in the upper right region (SI Fig.~{\ref{fig:complicated_stimulus}}C).

\begin{figure}[h!]
    \centering
    \includegraphics[width=0.9\linewidth]{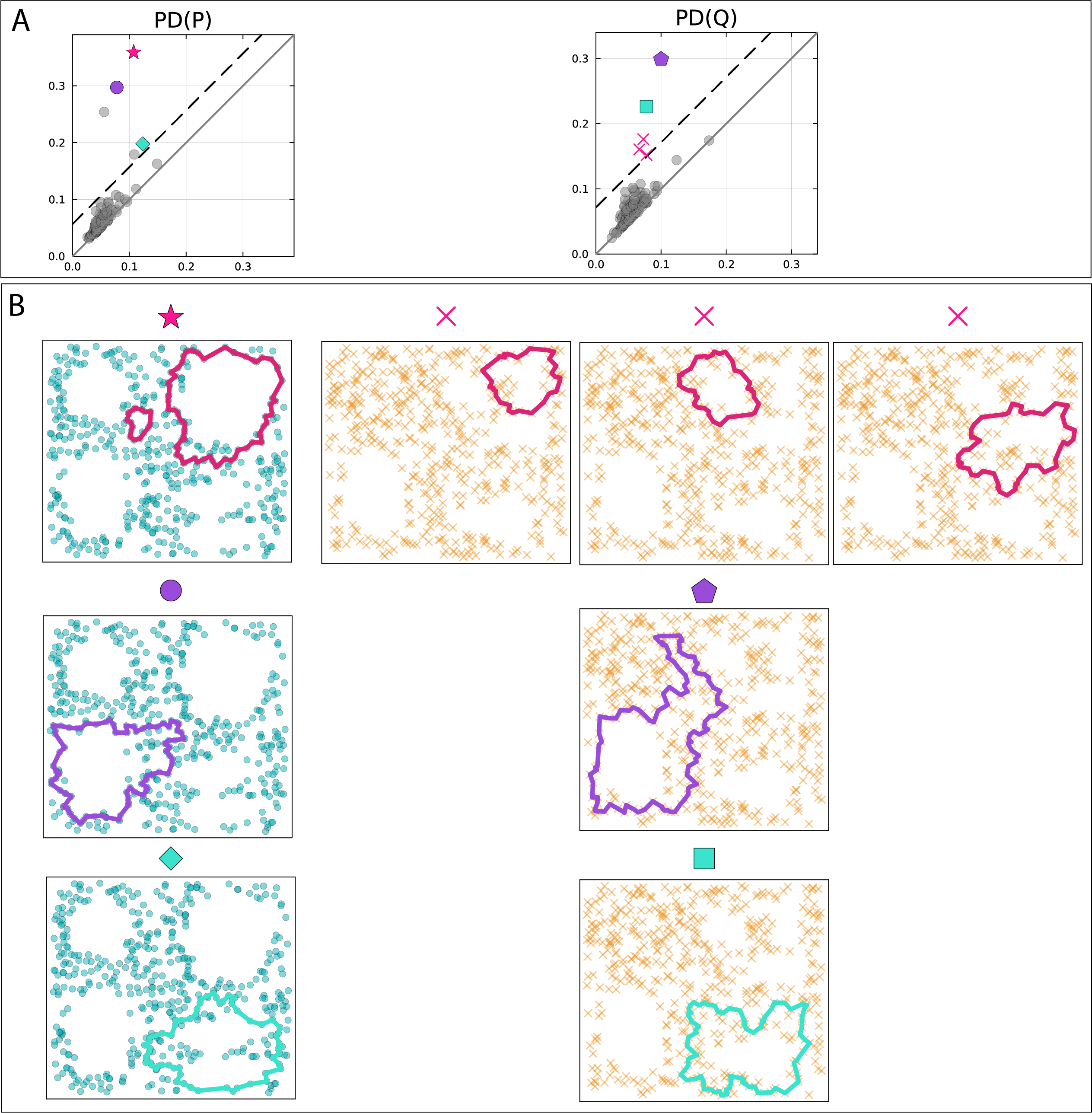}
    \caption{Output of the witness-modified analogous cycles on point clouds $P$ and $Q$ in SI Fig.~\ref{fig:complicated_stimulus}A. \textbf{A.} Output of the witness-modified analogous cycles finds three analogous pairs, shown in pink, purple, and teal. \textbf{B.} Cycle representatives of the analogous pairs show that the witness-modified analogous cycles identified the cycle in the top right corner of $P$ as a combination of three smaller cycles in a similar region in $Q$. It also identified the cycles in the bottom left corner and the cycles in the bottom right corner. }
\label{fig:complicated_stimuli_witness_variation}
\end{figure}

When we apply the witness-varied analogous cycles method, we find all expected analogous pairs, including the cycle in $P$ that is analogous to a combination of three cycles in $Q$ (SI Fig.~{\ref{fig:complicated_stimuli_witness_variation}}).

%---------------------------------------

\section{Supplementary Figures}

\begin{figure}[h!]
\centering
\includegraphics[width=0.55\textwidth]{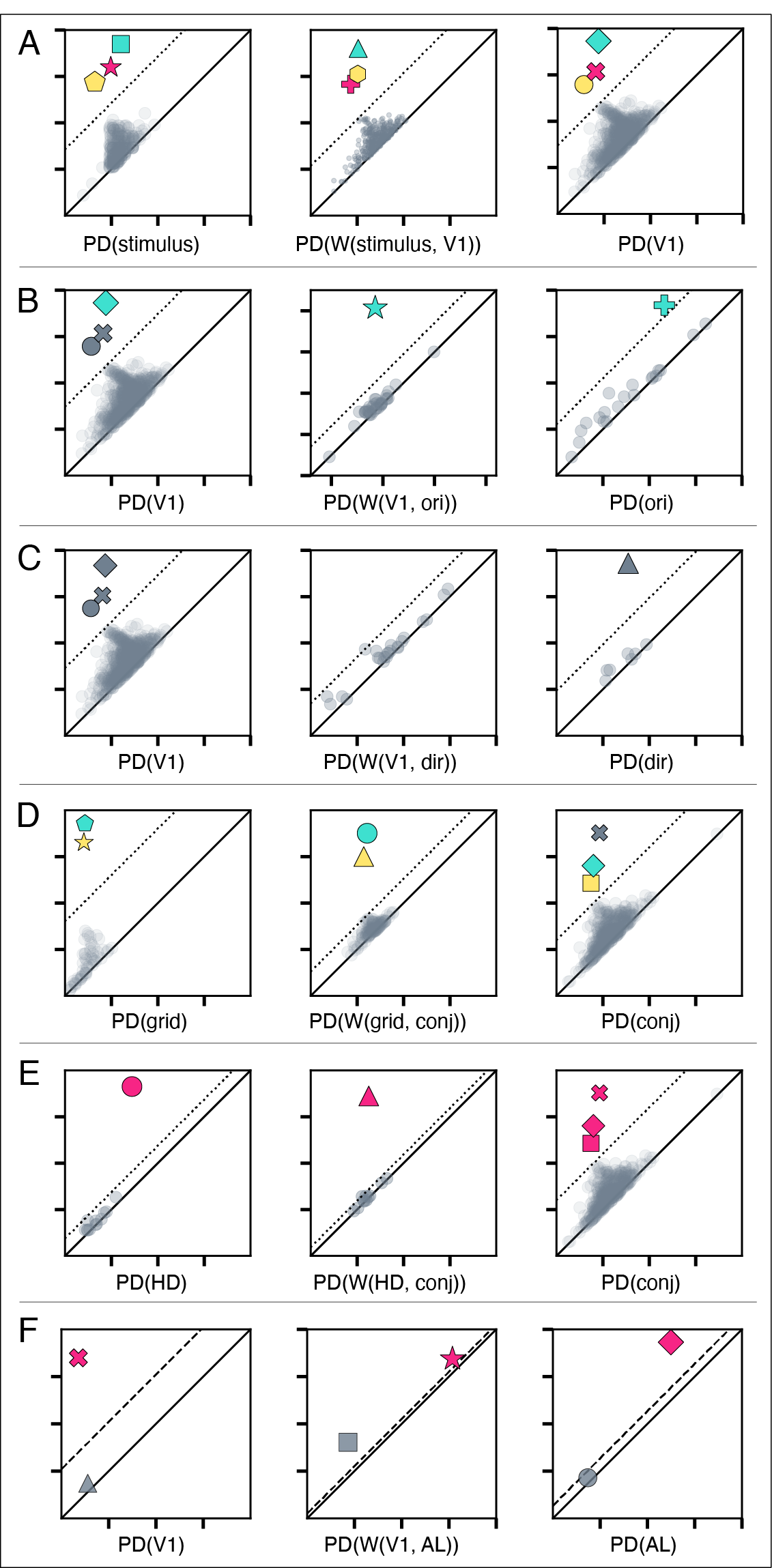}
\caption{Analogous cycles for all experiments, including the Witness persistence diagrams. (Left, right) Persistence diagrams of two systems. (Center) Witness persistence diagrams. \textbf{A.} Simulated stimulus and V1 simple cells. \textbf{B.} Simulated V1 simple cells and orientation cells. \textbf{C.} Simulated V1 simple cells and direction cells. \textbf{D.} Simulated grid and conjunctive cells. \textbf{E.} Simulated head-direction and conjunctive cells. \textbf{F.} Experimental V1 and AL cells. Since the persistence diagrams are sparse, the significance thresholds were computed using random spike trains.}
\label{fig:analogous_with_Witness}
\end{figure}
\quad

\newpage

\begin{figure}[h!]
\centering
\includegraphics[width=0.8\textwidth]{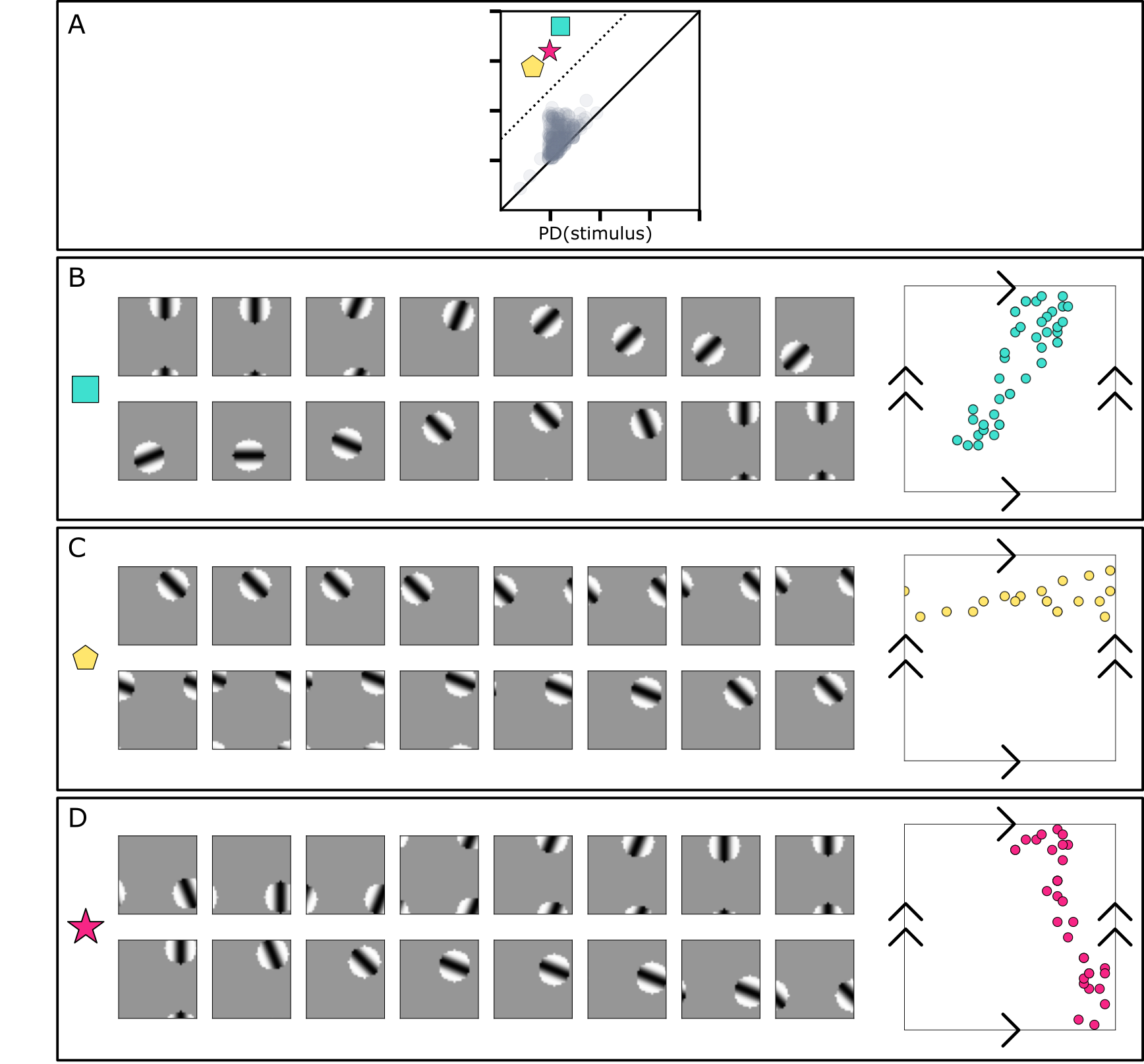}
\caption{Visualizations of cycle representatives of the three significant points in $\pd(\text{stim})$. (Left) Down-sampled cycle representatives of each significant point. (Right) Locations of the circular masks of the cycle representative. \textbf{A.} Persistence diagram $\pd(\text{stim})$. \textbf{B.} The teal square in the persistence diagram represents circular feature stemming from the orientations. \textbf{C. } The yellow pentagon in the persistence diagram represents the circular feature among the $x$-coordinates. \textbf{D. }The pink star in the persistence diagram represents the circular feature among the $y$-coordinates.}
\label{fig:stimulus_cyclereps}
\end{figure}
\quad

\newpage

\begin{figure}[h!]
\centering
\includegraphics[width=0.8\textwidth]{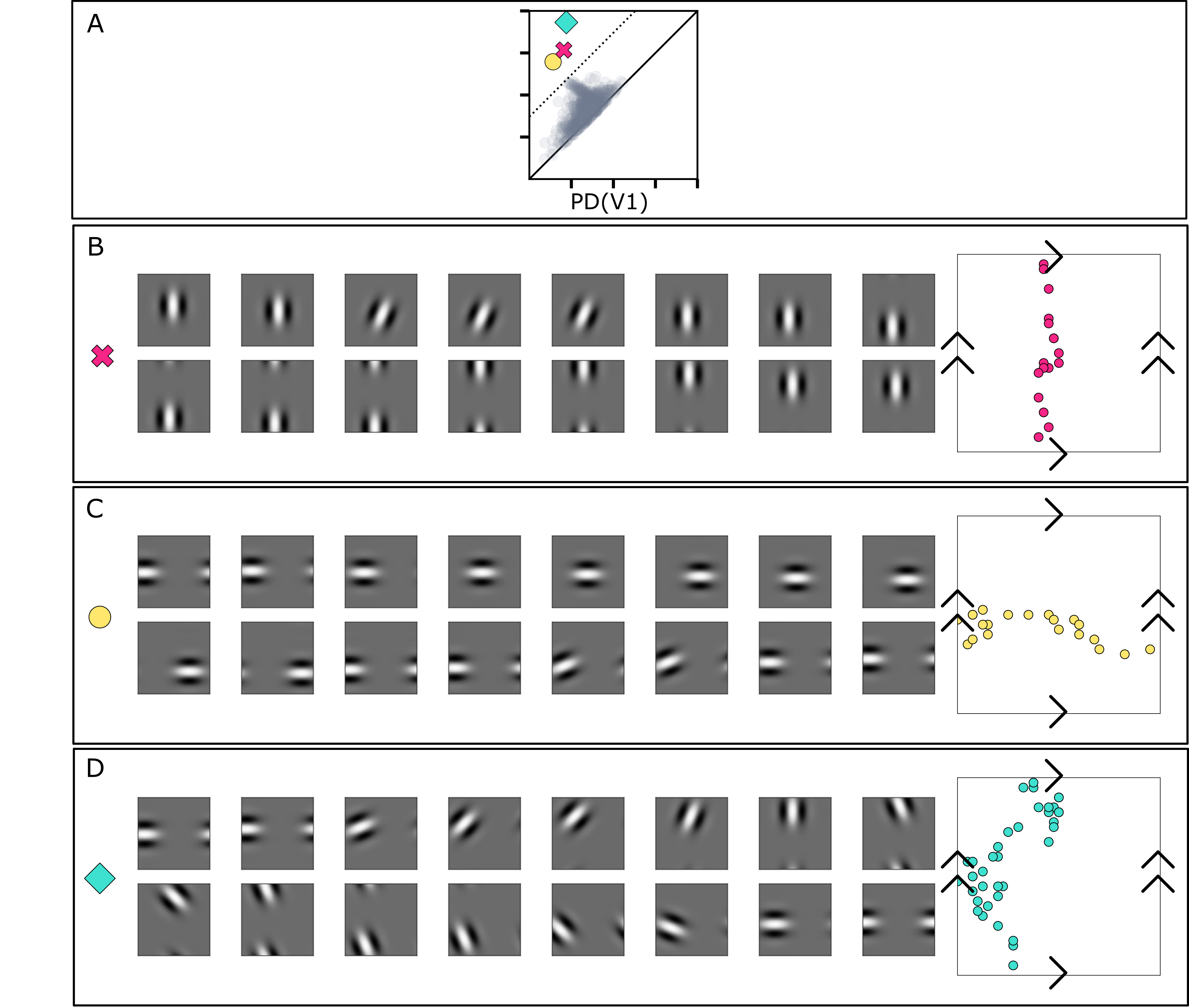}
\caption{Visualizations of cycle representatives of the three significant points in $\pd(V1)$.}
\label{fig:V1_reps}
\end{figure}
\quad

\newpage

\begin{figure}[h!]
\centering
\includegraphics[width=0.9\textwidth]{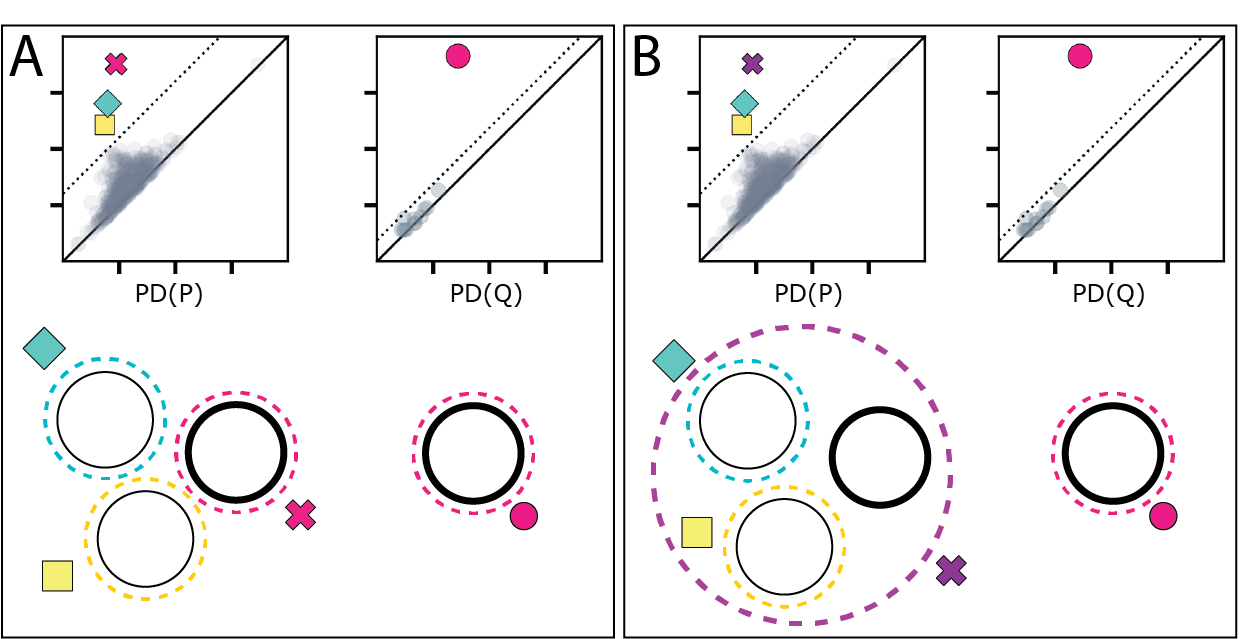}
\caption{Illustration of different choices of basis for the cycles in persistence diagrams and their effect on the analogous cycles. Let $P$ and $Q$ be two systems with three and one cycles each. Assume that the single cycle in $Q$ is related to the bold cycle in $P$. \textbf{A} A specific cycle basis for $\pd(P)$ assigns a single circle to each significant point on the persistence diagram as indicated by the colors. Using such assignment, the analogous cycles method will identify the pink "x" point in $\pd(P)$ to the pink "o" point in $\pd(Q)$. \textbf{B} An alternative cycle basis for $\pd(P)$ assigns a single circle to the teal diamond and the yellow square points on $\pd(P)$ and assigns the union of the three circles to the purple "x" point in $\pd(P)$. Using such basis, the analogous cycles method will match the pink "o" point in $\pd(Q)$ to the union of the three significant points on $\pd(P)$. Recall that the homology is computed with $\mathbb{Z}_2$ coefficients. Considering the formal sum of the three cycles (the "x", square, and diamond), the bold circle appears once, whereas the circles corresponding to the square and the diamond appear twice. In $\mathbb{Z}_2$ coefficients, such formal sum of the three cycles is equivalent to the single bold cycle. Under this basis, the analogous cycle matches the pink "o" point in $\pd(Q)$ to the union of the three significant points in $\pd(P)$ because such union represents the single bold cycle in $P$.}
\label{SIfig:change_of_basis}
\end{figure}
\quad

\newpage
\begin{figure}[h!]
    \centering
    \includegraphics[width=0.6\linewidth]{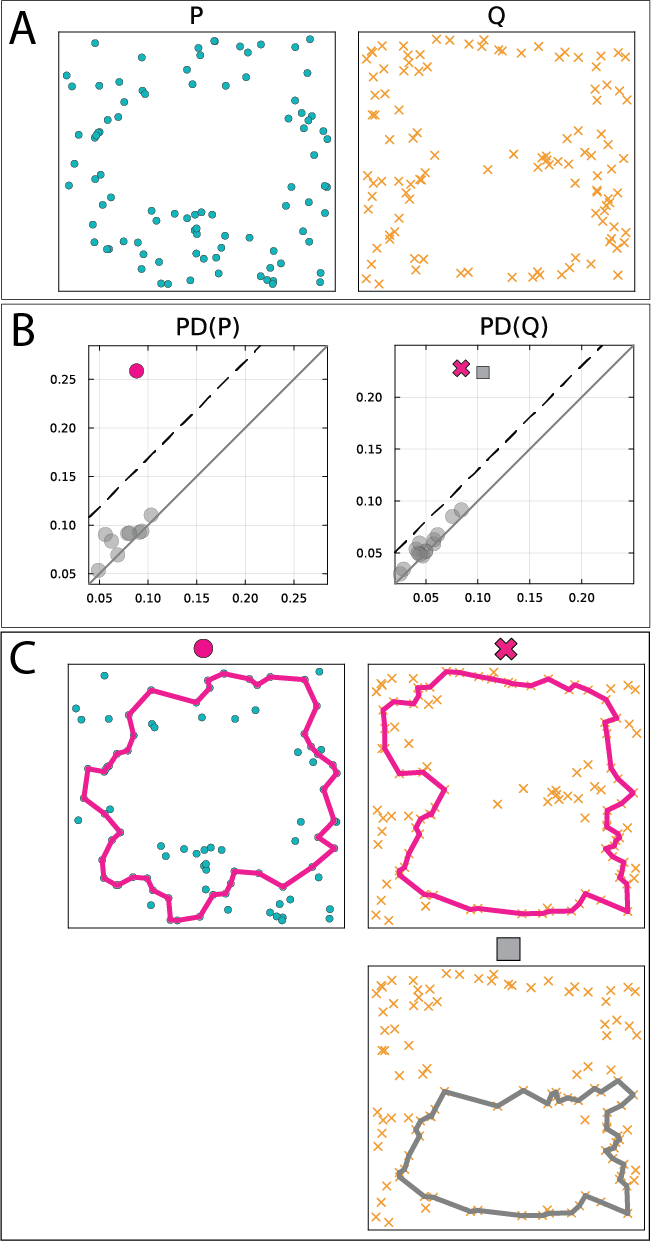}
    \caption{Example illustrating the dependence of analogous cycles on the choice of basis for cycles in the persistence diagrams. \textbf{A.} Example point clouds $P$ and $Q$. \textbf{B.} Output of the analogous cycles identifies the single significant point in $\pd(P)$ with one of the two significant points in $\pd(Q)$. The analogous cycles are marked in pink. From a visual inspection of panel A, one might expect the pink circle in $\pd(P)$ to be analogous to a combination of the cross and the square points in $\pd(Q)$ to reflect the fact that the large circle in $P$ looks like a combination of the two circles in $Q$. Such discrepancy is explained in panel C. \textbf{C.} Plots of the cycle representatives of significant points in $\pd(P)$ and $\pd(Q)$ show that the pink circle in $\pd(P)$ and the pink cross in $\pd(Q)$ are analogous because they do, indeed, represent cycles that are similar. The choice of basis for cycles in $\pd(Q)$ assigns the "larger" circle to the cross and the bottom circle to the square.}
    \label{fig:ambiguous_pair_basis}
\end{figure}
\quad
\newpage

\begin{figure}[h!]
\centering
\includegraphics[width=0.6\textwidth]{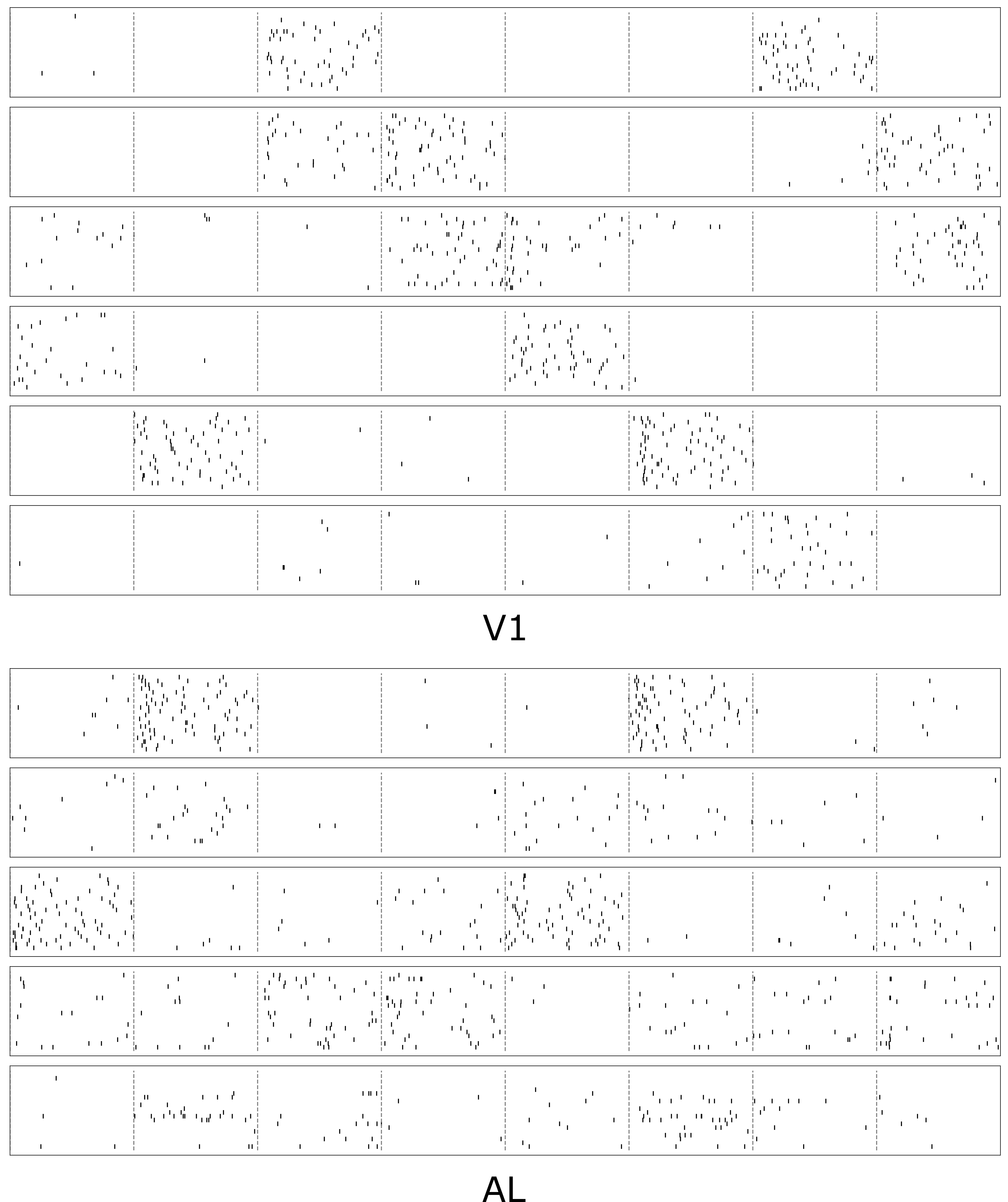}
\caption{Cycle representatives of the analogous cycles in V1 and AL. Both cycle representatives show collections of neurons that form circular features. }
\label{fig:UCSB_cr}
\end{figure}
\quad

% Examples of offset extensions. I don't know if this is necessary to include. 
% \begin{figure}
% \centering
% \includegraphics[width=1\textwidth]{simulation_offset_examples.png}
% \caption{Example offset extensions. Example offset bar extensions. For this paper, our main figures only report the baseline extensions \IY{Hard to see. The purple colors are difficult to see.} \IY{need a home (section) for this figure}}
% \label{fig:offset_examples}
% \end{figure}

%%% Add this line AFTER all your figures and tables
%\FloatBarrier

%\movie{Stimulus video presented in simualted visual system.}

%\movie{Stimulus video presented in in vivo experiment.}
\quad 
\newpage
\quad

\printbibliography